\pdfoutput=1
\documentclass[aps,pre,twocolumn,groupedaddress]{revtex4-2}

\usepackage{amsmath}
\usepackage{amssymb}
\usepackage{amsfonts}

\usepackage{graphicx}

\usepackage{hyperref}

\begin{document}


\title{Nonergodicity of $d$-dimensional generalized L\'evy walks\\and their relation to other space-time coupled models}


\author{Tony Albers}
\email[]{tony.albers@physik.tu-chemnitz.de}
\affiliation{Institute of Physics, Chemnitz University of Technology, 09107 Chemnitz, Germany}
\author{G\"unter Radons}
\email[]{guenter.radons@physik.tu-chemnitz.de}
\affiliation{Institute of Physics, Chemnitz University of Technology, 09107 Chemnitz, Germany}
\affiliation{Institute of Mechatronics, 09126 Chemnitz, Germany}


\date{\today}

\begin{abstract}
We investigate the nonergodicity of the generalized L\'evy walk introduced by Shlesinger et al. [Phys. Rev. Lett. \textbf{58}, 1100 (1987)] with respect to the squared displacements.
We present detailed analytical derivations of our previous findings outlined in a recent Letter [Phys. Rev. Lett. \textbf{120}, 104501 (2018)], give profound interpretations, and especially emphasize three surprising results:
First, we find that the mean-squared displacements can diverge for a certain range of parameter values.
Second, we show that an ensemble of trajectories can spread subdiffusively, whereas individual time-averaged squared displacements show superdiffusion.
Third, we recognize that the fluctuations of the time-averaged squared displacements can become so large that the ergodicity breaking parameter diverges, what we call infinitely strong ergodicity breaking.
The latter phenomenon can also occur for paramter values where the lag-time dependence of the mean-squared displacements is linear indicating normal diffusion.
In order to numerically determine the full distribution of time-averaged squared displacements, we use importance sampling.
For an embedding of our new findings into existing results in the literature, we define a more general model which we call \textit{variable speed generalized L\'evy walk}
and which includes well known models from the literature as special cases such as the space-time coupled L\'evy flight or the anomalous Drude model.
We discuss and interpret our findings regarding the generalized L\'evy walk in detail and compare them with the nonergodicity of the other space-time coupled models following from the more general model.
\end{abstract}


\maketitle

\section{\label{sec:1}Introduction}

L\'evy walks \cite{shlesinger1987} are a special class of continuous time random walks with a spatio-temporal coupling.
In contrast to the standard L\'evy flight \cite{shlesinger1982}, where random jumps with infinite second moment and waiting times with finite mean between them are completely independent from each other,
the spatio-temporal coupling of L\'evy walks prevents a divergence of the mean-squared displacement \cite{shlesinger1993,klafter1996}.
For the standard L\'evy walk \cite{zumofen1993_1,zumofen1993_2}, this coupling is achieved by assigning a constant velocity to the random walker meaning that the resulting motion consists of randomly oriented flight episodes,
where heavy-tail distributed travel distances are linearly coupled to corresponding flight durations.
For moving particles with mass, the finite velocity brings L\'evy walks closer to reality compared to L\'evy flights with their instantaneous jumps \cite{zaburdaev2015,zaburdaev2016}.
In addition to this linear coupling, nonlinear couplings between flight durations and covered distances have been investigated \cite{shlesinger1987}.
Moreover, also continuous time random walks with a linear or nonlinear coupling between jumps and waiting times are sometimes referred to as L\'evy walks \cite{dentz2015}.
In the next section of this article, we show how these models are related to each other and give a detailed overview over previous studies.
L\'evy walks including its variations and modifications \cite{barkai1998,friedrich2006,eule2008,taylor-king2016} can model all kinds of anomalous diffusion ranging from subdiffusion and normal diffusion
to superdiffusion, ballistic diffusion, and even superballistic diffusion \cite{zaburdaev2015}.
Therefore, L\'evy walk models have wide applications.
They have been used to model the dynamics of one-dimensional iterated maps \cite{zumofen1993_1,zumofen1993_2} and nonintegrable Hamiltonian systems \cite{klafter1994},
where the former result from higher-dimensional dissipative dynamics and the latter have applications in plasma physics \cite{del-castillo-negrete2000} and turbulence \cite{solomon1993,del-castillo-negrete1998}.
In addition, also photon counting statistics of blinking quantum dots \cite{jung2002} and perturbation spreading in many-particle systems \cite{cipriani2005,zaburdaev2011} can be modeled by L\'evy walks.
Furthermore, human travel behavior \cite{brockmann2006} or search strategies of predators \cite{sims2008} follow L\'evy walk patterns which are also applied to the target search of robots \cite{krivonosov2016}.
Moreover, diffusion of cold atoms in optical lattices \cite{marksteiner1996,kessler2012,barkai2014}, fluid stretching in two-dimensional heterogeneous media \cite{dentz2016_1,dentz2016_2},
and turbulent pair dispersion \cite{thalabard2014} are related to the generalized L\'evy walk model introduced by Shlesinger et. al. \cite{shlesinger1987}, which is in the focus of the present article.
This model, where flight velocities and flight durations are nonlinearly coupled, was developed to reproduce the Richardson-Obukhov law of turbulence \cite{richardson1926,obukhov1959},
i.e., a cubic increase of the mean-squared displacement.
Unexpectedly and despite of a well-defined finite velocity of the random walker at any instant of time, in full contrast to the standard L\'evy flight, we will show that for this more general coupling,
there is a certain parameter range where the mean-squared displacement diverges and, therefore, prevents a cubic increase.
A very important aspect of anomalous diffusion is nonergodicity, i.e., the nonequivalence of ensemble and time averages.
Interestingly, even if the underlying state or phase space of the process is fully accessible for each trajectory, ensemble and time averages may not coincide,
what in the physical literature is called weak ergodicity breaking \cite{bouchaud1992} or weak nonergodicity \cite{fulinski2011}.
This phenomenon has attracted much attention in the last years due to the progress in single-particle tracking experiments \cite{saxton1997},
where in contrast to classical ensemble-based methods such as pulsed field gradient nuclear magnetic resonance \cite{kaerger1983}, time averages are evaluated \cite{feil2012,barkai2012,metzler2012,hoefling2013}.
Weak nonergodicity has been observed in several experiments on different processes such as the fluorescence of single nanocrystals \cite{brokmann2003},
diffusion of lipid granules in living fission yeast cells \cite{jeon2011}, and diffusion of proteins in the plasma membrane of living cells \cite{weigel2011,manzo2015}.
Since the discovery of weak nonergodicty of subdiffusive continuous time random walks with respect to its squared displacements \cite{lubelski2008,he2008},
where it was recognized that the time-averaged squared displacement shows a linear increase indicating normal diffusion and remains random even for long trajectories,
many theoretical models of anomalous diffusion known in the literature have been investigated \cite{metzler2014}.
Among others, fractional Brownian motion \cite{deng2009}, diffusion on fractals \cite{meroz2010}, geometric Brownian motion \cite{peters2013}, scaled Brownian motion \cite{thiel2014},
heterogeneous diffusion processes \cite{cherstvy2013}, integrated Brownian motion \cite{albers2014}, and globally correlated random walks \cite{budini2017} were investigated with respect to their ergodic behavior.
Of course, also L\'evy walks were studied.
While many investigations focused on the standard model \cite{godec2013,froemberg2013_1,froemberg2013_2} with a constant flight velocity independent of the flight durations,
the generalized L\'evy walk introduced by Shlesinger et. al. \cite{shlesinger1987} was investigated in a previous publication of the authors \cite{albers2018}, where many surprising results were found.
In this article, we show connections of generalized L\'evy walks to other space-time coupled models of anomalous diffusion, recall our findings from our previous article,
present the details of the calculations in the appendices, and give profound interpretations of the findings.

\section{\label{sec:2}Variable speed generalized L\'evy walk}

\begin{figure}
\includegraphics[width=\linewidth]{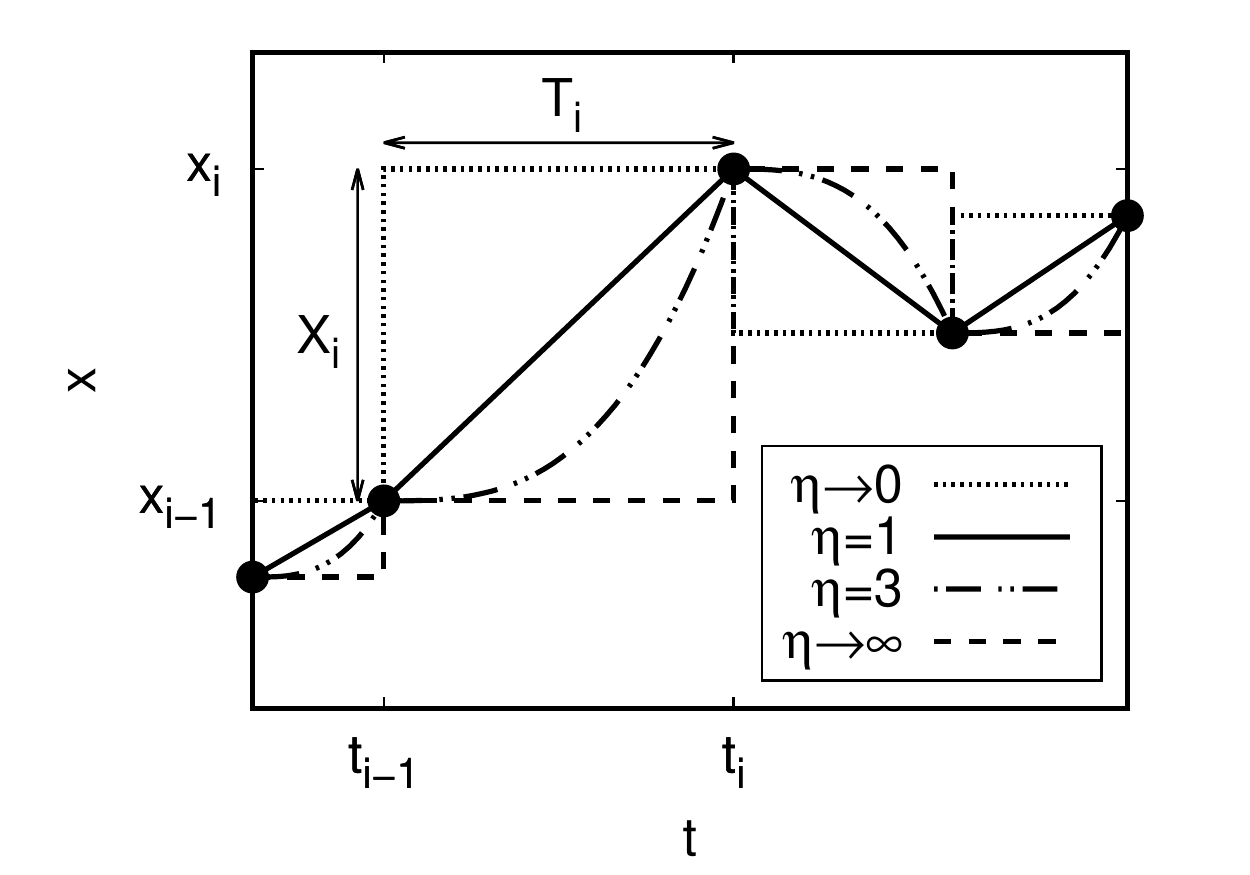}
\caption{\label{fig:trajectory}
Schematic representation of one realization of the variable speed generalized L\'evy walk for different values of the parameter $\eta$ in one spatial dimension.
The black dots represent a sequence of turning points in time and space, where at each turning point, a new elementary event characterized by a duration $T_i$ and a spatial displacement $X_i$ is initiated.
The statistics of the elementary events is described by the distribution $\psi(x,t)$ of Eq.~(\ref{eq:psi_x_t}).
The turning points can be connected in different ways according to different values of the exponent $\eta$ in Eq.~(\ref{eq:x_t}).
The case $\eta=1$ (straight lines) corresponds to the generalized L\'evy walk to be discussed in the subsequent sections,
and the cases $\eta\rightarrow0$ and $\eta\rightarrow\infty$ correspond to the space-time coupled L\'evy flight in the jump-first and wait-first interpretation, respectively.
Note that for general values of $\eta$, the velocity is time-dependent also between the turning events.}
\end{figure}

\begin{figure}
\includegraphics[width=\linewidth]{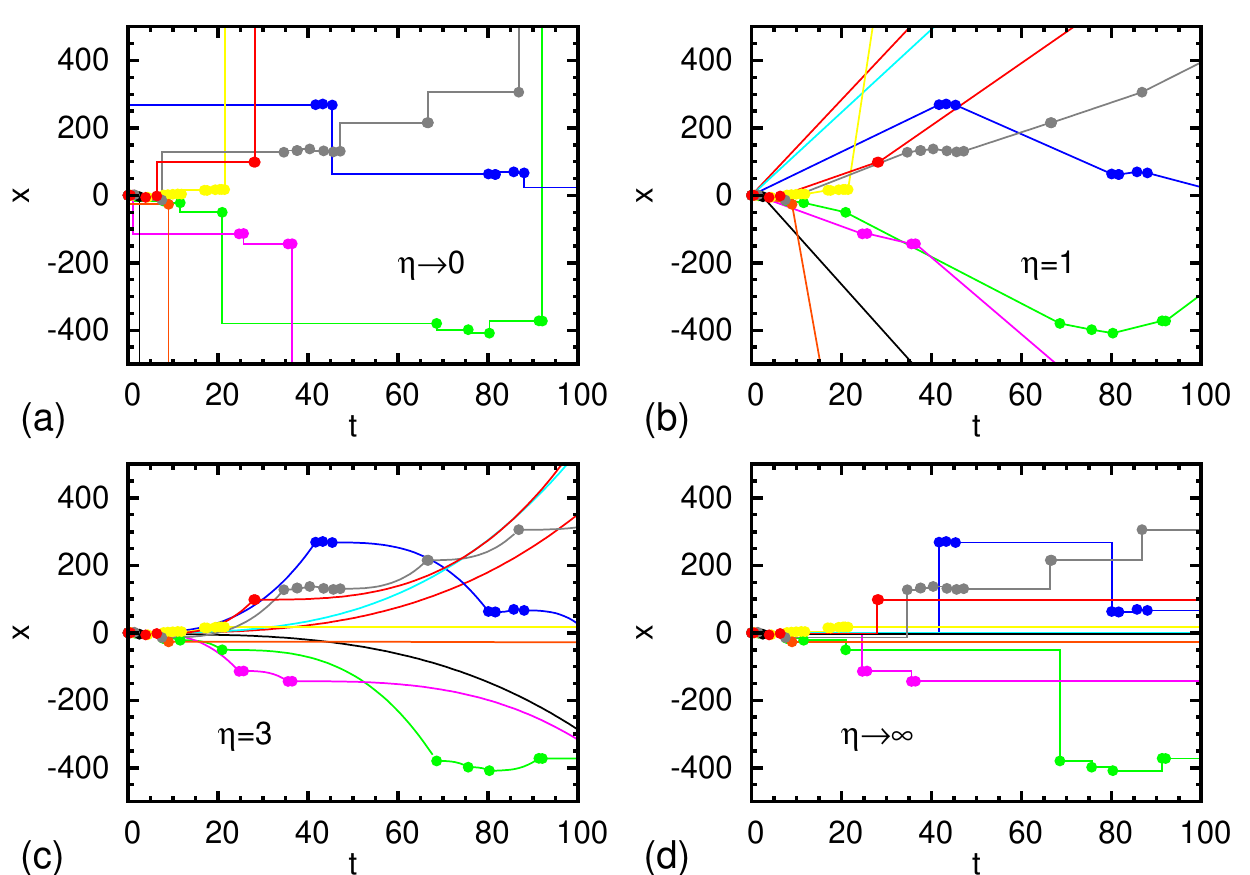}
\caption{\label{fig:trajectories}
Ten numerically generated realizations of the variable speed generalized L\'evy walk for different values of the parameter $\eta$ in one spatial dimension ($\gamma=0.5$, $\nu=1.5$, $t_0=1.0$, $c=1.0$).
All four figures show the same ten sequences of turning points (colored dots), but the connections between them are very different:
(a) $\eta\rightarrow0$, space-time coupled L\'evy flight, jump-first interpretation (b) $\eta=1$, generalized L\'evy walk (c) $\eta=3$ (d) $\eta\rightarrow\infty$, space-time coupled L\'evy flight, wait-first interpretation.}
\end{figure}

We consider a general space-time coupled model of anomalous diffusion which is characterized by three exponents ($\gamma$, $\nu$, and $\eta$)
and was first introduced by the authors in the supplemental material of a previous publication \cite{albers2018}.
This model, for general parameters $\gamma$, $\nu$, and $\eta$, is called in the following \textit{variable speed generalized L\'evy walk}.
It consists of a sequence of independent and identically distributed space-time coupled elementary events.
The duration $T_i$, $i=1,2,...$, of each event is drawn randomly from a heavy-tailed probability density function, which we choose explicitly as
\begin{equation}
\label{eq:psi_t}
\psi(t)=\frac{\gamma}{t_0}\left(\frac{t}{t_0}+1\right)^{-\gamma-1},\quad\gamma>0,\,t_0>0.
\end{equation}
We note, however, that the essential results of this paper depend only on the tail exponent $\gamma$.
The distance $|\mathbf{X}_i|$, which is covered during a complete elementary event, is connected with the random duration $T_i$ in a deterministic way,
\begin{equation}
\label{eq:X_T}
|\mathbf{X}_i|=c\,T_i^{\nu},\quad\nu>0,\,c>0.
\end{equation}
The spatial direction of each event is uniformly chosen at random such that the process is isotropic.
Therefore, the statistics of an elementary event is captured by the multivariate probability density
\begin{equation}
\label{eq:psi_x_t}
\psi(\mathbf{x},t)=\frac{1}{S_d(|\mathbf{x}|)}\,\delta(|\mathbf{x}|-ct^{\nu})\,\psi(t),
\end{equation}
where the surface $S_d(|\mathbf{x}|)=\left(2\pi^{d/2}/\Gamma(d/2)\right)|\mathbf{x}|^{d-1}$ of the $d$-dimensional sphere with radius $|\mathbf{x}|$
accounts for the correct normalization of $\psi(\mathbf{x},t)$ in $d$ Euclidean dimensions.
$\psi(\mathbf{x},t)\,\text{d}^d\mathbf{x}\,\text{d}t$ is the probability that a distance $\mathbf{X}_i$ lying in the infinitesimal volume $\text{d}^d\mathbf{x}$ around $\mathbf{x}$
is covered during a complete elementary event of duration $T_i\in[t,t+\text{d}t]$.
A sequence of such events leads to a series of turning points in time and space (the black dots in Fig.~\ref{fig:trajectory}), where each turning point marks the beginning of a new elementary event.
While the distribution $\psi(\mathbf{x},t)$ determines the statistics of the turning points, the paths between them have to be specified.
Let $(t_i,\mathbf{x}_i)$ be the coordinate of the $i^{\text{th}}$ turning point in time and space
and $\mathbf{e}_i=(\mathbf{x}_i-\mathbf{x}_{i-1})/|\mathbf{x}_i-\mathbf{x}_{i-1}|$ the spatial unit vector on the straight line connecting the $(i-1)^{\text{th}}$ turning point and the $i^{\text{th}}$ one,
the position $\mathbf{x}(t)$ of a random walker at time $t$ with $t\in[t_{i-1},t_i]$ between these two turning points is given by
\begin{equation}
\label{eq:x_t}
\mathbf{x}(t)=\mathbf{x}_{i-1}+v_{i,\eta}\,(t-t_{i-1})^{\eta}\,\mathbf{e}_i,\quad v_{i,\eta}=c\,T_i^{\nu-\eta}.
\end{equation}
Possible paths between turning points in time and space are illustrated in Fig.~\ref{fig:trajectory} for the simplest case of only one spatial dimension.
Obviously, the new exponent $\eta>0$ controlls the temporal progress between two turning points and, therefore, interpolates between different models of anomalous diffusion that are known in the literature.
For instance, for $\eta=1$, there is a constant velocity $|\mathbf{V}_i|=c\,T_i^{\nu-1}$ during each elementary event that depends on the duration $T_i$ of the event.
This case corresponds to the straight lines in Fig.~\ref{fig:trajectory} and defines the generalized L\'evy walk introduced by Shlesinger et al. \cite{shlesinger1987} which we will investigate in detail in the next sections.
If in addition $\nu=1$, the flight velocity $|\mathbf{V}_i|=c$ does not depend on the flight duration leading to the standard L\'evy walk often encountered in the literature \cite{zumofen1993_1,zumofen1993_2}.
The special case $\eta=\nu$ of the variable speed generalized L\'evy walk is known in the literature as anomalous Drude model \cite{schulz-baldes1997} or L\'evy walk collision process \cite{barkai1998}.
For $\eta\rightarrow0$ or $\eta\rightarrow\infty$, successive turning points are connected by instantaneous jumps and waiting times between them (dotted line in Fig.~\ref{fig:trajectory}).
Due to the deterministic coupling of waiting times and jumps according to Eq.~(\ref{eq:X_T}), these processes are sometimes also called L\'evy walks (see for instance \cite{dentz2015}).
Because of the instantaneous jumps and in order to distinguish these processes from the ``real" L\'evy walks, we call them space-time coupled L\'evy flights \cite{klafter1987}.
Furthermore, they are also known as stored-energy-driven L\'evy flights \cite{akimoto2013,akimoto2014}.
Ten realizations of the generalized L\'evy walk ($\eta=1$), the space-time coupled L\'evy flight in the jump-first ($\eta\rightarrow0$) and wait-first ($\eta\rightarrow\infty$) interpretation
as well as the special case ($\eta=3$) are shown in Fig.~\ref{fig:trajectories}.
They share the same sequence of turning points but the paths between them are different.
Note that for general values of $\eta$, the velocity is time-dependent also between the turning events.
Our variable speed generalized L\'evy walk is related to a large number of special cases that were thoroughly studied in the literature.

The weakly nonergodic behavior of the standard L\'evy walk ($\nu=1$, $\eta=1$) with respect to the squared displacements was investigated in \cite{godec2013,froemberg2013_1,froemberg2013_2}.
Furthermore, it was shown that the standard L\'evy walk can also be described by a set of coupled Langevin equations using a subordination technique \cite{magdziarz2012,eule2012}.
From that, the time-lag dependence of the ensemble-averaged squared displacement and the ensemble average of the time-averaged squared displacement was recovered \cite{wang2019}.
The propagator and its moments of the standard L\'evy walk ($\nu=1$, $\eta=1$) and the standard space-time coupled L\'evy flight ($\nu=1$, $\eta\rightarrow\infty$) were discussed in detail in \cite{rebenshtok2014_1,rebenshtok2014_2,froemberg2015}
using L\'evy's central limit theorem and the so-called infinite covariant density, i.e., a formally non-normalizable density describing the outer tails of the propagator.
Interestingly, while the infinite covariant density was derived using a moment generating function approach, it can also be obtained by the single big jump approach \cite{vezzani2019,wang2020}
with the idea that the tails of the propagator are determined by the occurence of a very long flight event during the observation time that solely influences the statistics of the whole trajectory.
The generalized space-time coupled L\'evy flight ($\eta\rightarrow\infty$) for all relevant values of the parameters $\gamma$ and $\nu$ was investigated in detail in the literature
including weak nonergodicity with respect to the squared displacements \cite{akimoto2013,akimoto2014} as well as the propagator and its moments \cite{dentz2015}.
The ensemble-averaged squared displacement and the ensemble average of the time-averaged squared displacement for a special case ($\nu=\eta$) of the variable speed generalized L\'evy walk
were derived in \cite{meyer2017} using a scaling Green-Kubo relation.
Our model is also related to the intermittent dynamics of one-dimensional iterated maps with infinite invariant measure caused by the existence of marginally unstable fixed points.
These fixed points lead to heavy-tail distributed sojourn times in their vicinity according to Eq.~(\ref{eq:psi_t}) \cite{geisel1984}.
A famous example is the Pomeau-Manneville map \cite{pomeau1980}.
Interpreting the time series or a function of the time series of such an iterated map as the increments of a random walk, this leads to a L\'evy-walk type of motion where consecutive visits of the fixed points' vicinity,
which are interupted by short chaotic bursts, can be identified as the elementary events, which follow the statistics of Eq.~(\ref{eq:psi_x_t}), but the turning points are slightly differently connected in comparison with Eq.~(\ref{eq:x_t}).
Such an analysis was done for instance in \cite{meyer2017}, whereas in \cite{meyer2018}, the reason for the anomalous diffusion was additionally decomposed into its constitutive causes known as the Joseph, Noah, and Moses effect.
In contrast, in \cite{akimoto2015}, an iterated map defined on the unit interval was used meaning that the increments are strictly positive.
In this case, the elementary events follow the statistics in Eq.~(\ref{eq:psi_x_t}) without the prefactor of $1/2$ on the right hand side.
For this process, the propagator, which is related to the distribution of the time integral of the absolute value of the velocity process of our model, and its moments were derived \cite{akimoto2015}.
Other dynamical behavior related to our variable speed generalized L\'evy walk is found for the diffusion of cold atoms in optical lattices.
These processes can theoretically be described by a pair of coupled Langevin equations determining the time evolution of position and velocity of the atoms.
The velocity process can be regarded as a sequence of random excursions from the origin that are independently and identically distributed because of the Markovian nature of the Langevin process.
Due to the coupling of position and velocity, these excursions lead to elementary events following the space-time scaling in Eq.~(\ref{eq:psi_x_t}) for the special case $\nu=3/2$ \cite{kessler2012,barkai2014}.
The propagator and its moments (showing strong anomalous diffusion) for this case and also for general values of $\nu$ were investigated in detail in \cite{aghion2017,aghion2018} using again the concept of the infinite covariant density.
Moreover, this infinite density was also recovered using the single big jump approach in \cite{vezzani2019}.
A general study of renewal processes with heavy-tail distributed sojourn times including the statistics of rare events described by non-normalizable densities can be found in \cite{wang2018}.

As already mentioned, our variable speed generalized L\'evy walk was first introduced in the supplemental material of a previous publication \cite{albers2018} and initiated further investigations.
In \cite{bothe2019}, it was shown that the parameter $\eta$ only controls the prefactor of the time-lag dependence of the mean-squared displacements.
However, for certain ranges of the parameters, the mean-squared displacements can diverge what we discuss later in the article.
Therefore, in the following, we concentrate our investigation on the generalized L\'evy walk ($\eta=1$) and refer to the other space-time coupled models only if there are some major differences between these models.
The propagator and its moments of the variable speed generalized L\'evy walk were derived in \cite{vezzani2020} using again the single big jump approach.
Moreover, the velocity process of our model was studied in \cite{akimoto2020}.
There, an analytical expression for the propagator featuring an infinite invariant density as well as the connection of the latter to the distributional behavior of certain time averages was found.
The Moses, Noah, and Joseph effects in the variable speed generalized L\'evy walk were investigated in \cite{aghion2021}.

In the literature, there are further studies on weak ergodicity breaking that go beyond the time-lag dependencies of the mean-squared displacements.
Especially, for continuous time random walks on a lattice with heavy-tail distributed waiting times and the corresponding nonlinear iterated maps with marginally unstable fixed points,
the distribution of the fraction of occupation time of a certain state and its connection to the equilibrium distribution of an ensemble of random walkers was derived \cite{bel2005,bel2006_1,bel2006_2}.
Because the fraction of occupation time can be interpreted as the probability to be in a specific state obtained from a time average,
its distribution is key in understanding the distribution of other time averages such as the time-averaged position of the random walker \cite{rebenshtok2007}.
The influence of infinite invariant densities on the distribution of some time-averaged observation functions was also investigated \cite{akimoto2008,korabel2012}.

The aims for the rest of the present article are the following:
First, we want to recall our findings from a previous short publication \cite{albers2018} regarding the weak nonergodicity of the generalized L\'evy walk model ($\eta=1$) with respect to the squared displacements.
Second, we want to present detailed derivations of these findings.
Finally and most importantly, we want to give profound interpretations leading to a deeper understanding of weak nonergodicity in the generalized L\'evy walk and the other space-time coupled models
that follow from the general model defined in this section.

The rest of the paper is organized as follows.
In section \ref{sec:3}, we define the generalized L\'evy walk as special case of our variable speed generalized L\'evy walk.
Exact analytical results concerning the ensemble-averaged squared displacement, the ensemble average of the time-averaged squared displacement, and the randomness of the time-averaged squared displacements
for the generalized L\'evy walk in the full two-dimensional parameter space are discussed in sections \ref{sec:4}, \ref{sec:5}, and \ref{sec:6}, respectively.
Corresponding derivations of the analytical results can be found in the appendices.
A summary of our findings and a final discussion are presented in section \ref{sec:7}.

\section{\label{sec:3}The generalized L\'evy walk}

\begin{figure}
\includegraphics[width=\linewidth]{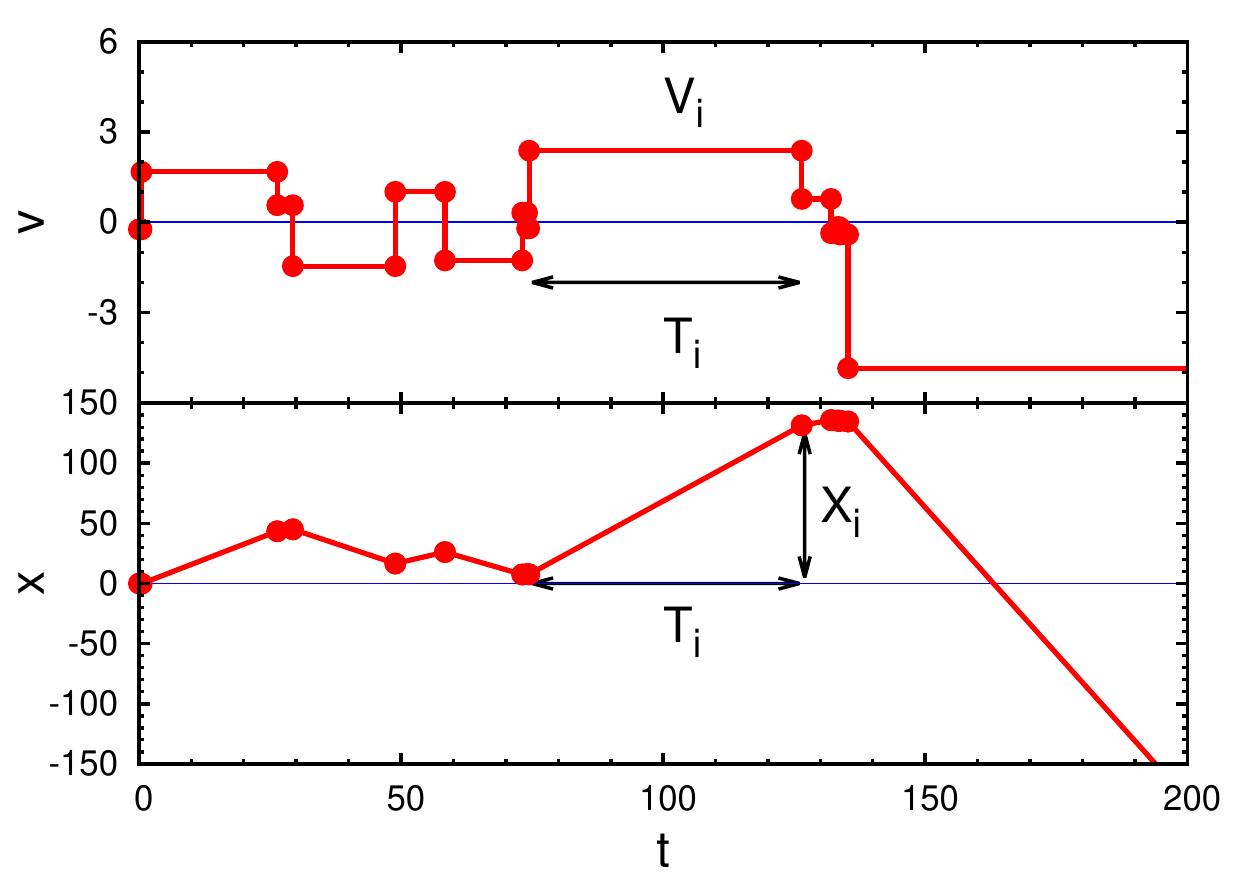}
\caption{\label{fig:Levy_walk}
One numerically generated realization of the generalized L\'evy walk ($\gamma=0.5$, $\nu=1.5$, $t_0=1.0$, $c=0.33$),
where the upper panel shows the velocity process in dependence on time, and the lower panel shows the corresponding integrated process, i.e., the position in dependence on time.
The trajectory consists of a sequence of independent and identically distributed flights, where for each flight of random duration $T_i$ distributed according to $\psi(t)$ in Eq.~(\ref{eq:psi_t}),
the constant velocity $V_i=\pm c\,T_i^{\nu-1}$ leads to a spatial displacement $X_i=V_i\,T_i=c\,T_i^{\nu}$.}
\end{figure}

The generalized L\'evy walk was first introduced by Shlesinger et. al. in \cite{shlesinger1987} and follows from our variable speed generalized L\'evy walk introduced in Sec.~\ref{sec:2} as special case for $\eta=1$.
Therefore, it consists of a sequence of independent and identically distributed flights (see Fig.~\ref{fig:Levy_walk}),
where the random durations $T_i$ of the flights are drawn from the heavy-tailed probability density $\psi(t)$ in Eq.~(\ref{eq:psi_t}).
Due to normalization, the characteristic exponent $\gamma$ must be positive.
For $0<\gamma\leq1$, the mean flight duration $\langle T\rangle=\int t\,\psi(t)\,\text{d}t$ diverges, whereas it is finite for $\gamma>1$.
In each flight event, a random walker moves with a constant speed $|\mathbf{V}_i|$ that depends on the random duration $T_i$ of the flight in a deterministic way that follows from Eq.~(\ref{eq:X_T}),
\begin{equation}
\label{eq:V_T}
|\mathbf{V}_i|=\frac{|\mathbf{X}_i|}{T_i}=c\,T_i^{\nu-1},\quad\nu>0,\,c>0.
\end{equation}
The characteristic exponent $\nu$ is also positive in order to guarantee that larger flight durations imply larger covered distances $|\mathbf{X}_i|$.
All possible flight directions have the same probability such that the L\'evy walk is isotropic.
By a simpe change of variables, we can calculate the probability density $p(|\mathbf{x}|)$ of the distances $|\mathbf{X}_i|$ from Eq.~(\ref{eq:psi_t}) and Eq.~(\ref{eq:X_T}),
\begin{equation}
\label{eq:p_x}
p(\mathbf{|x|})=\int_0^{\infty}\delta(|\mathbf{x}|-ct^{\nu})\,\psi(t)\,\text{d}t\sim|\mathbf{x}|^{-\frac{\gamma}{\nu}-1}.
\end{equation}
For $2\nu<\gamma$, the second moment $\langle|\mathbf{X}|^2\rangle=\int|\mathbf{x}|^2\,p(|\mathbf{x}|)\,\text{d}|\mathbf{x}|$ of this distribution is finite, whereas it diverges for $2\nu\geq\gamma$.
Note that in \cite{shlesinger1987}, the generalized L\'evy walk was introduced in a slightly different way.
There, the distances $|\mathbf{X}_i|$ were drawn randomly from the probability density $p(|\mathbf{x}|)$ in Eq.~(\ref{eq:p_x}) and the durations $T_i$ followed deterministically,
$T_i=|\mathbf{X}_i|/|\mathbf{V}_i|\propto|\mathbf{X}_i|^{\frac{1}{\nu}}$.
Of course, this defintion of the generalized L\'evy walk is completely equivalent to our definition.

The statistics of the flights is described by the multivariate probability density $\psi(\mathbf{x},t)$ in Eq.~(\ref{eq:psi_x_t}).
Another important statistics is the distribution $W(\mathbf{x},t)$
that captures the probability $W(\mathbf{x},t)\,\text{d}^d\mathbf{x}$ of walking a distance ending in the infinitesimal volume $\text{d}^d\mathbf{x}$ around $\mathbf{x}$ in time $t$ with a single flight whose duration is larger than $t$,
\begin{equation}
\label{eq:W_x_t}
\begin{split}
W(\mathbf{x},t)&=\frac{1}{S_d(|\mathbf{x}|)}\int_t^{\infty}\delta(|\mathbf{x}|-ct'^{\nu-1}t)\,\psi(t')\,\text{d}t'\\
&=\int_1^{\infty}\lambda^dt\,\psi(\lambda\mathbf{x},\lambda t)\,\text{d}\lambda.
\end{split}
\end{equation}
Here, we used the notation introduced in Eq.~(\ref{eq:psi_x_t}).
The distribution $W(\mathbf{x},t)$ is remarkable in two senses.
First, it is a multivariate probability density with respect to the first argument but a cumulative distribution with respect to the second one.
Therefore, its normalization reads $\lim\limits_{t\rightarrow0}\int W(\mathbf{x},t)\,\text{d}^d\mathbf{x}=1$.
Second, and most importantly, the prefactor $\lambda^dt$ in the second line of Eq.~(\ref{eq:W_x_t}) does not appear in the corresponding expression of the distribution $W(\mathbf{x},t)$ in \cite{shlesinger1987}.
This seemingly little difference leads to big consequences.
In Sec.~\ref{sec:4}, we obtain different time-lag dependencies of the ensemble-averaged squared displacement compared to the ones in \cite{shlesinger1987} what is additionally confirmed numerically.
As a consequence, we show that the generalized L\'evy walk is not able to produce a cubic increase of the ensemble-averaged squared displacement.
Moreover, the latter diverges in a certain region of the two-dimensional parameter space, a fact that was not realized in \cite{shlesinger1987}.
These discrepancies have already been pointed out in a previous publication of the authors \cite{albers2018} and, additionally, were later confirmed by other authors in \cite{bothe2019}.

The distributions $\psi(\mathbf{x},t)$ and $W(\mathbf{x},t)$ can be used to express the propagator $p(\mathbf{x},t)$,
the probability density of finding a random walker at position $\mathbf{x}$ at time $t$ with initial condition $p(\mathbf{x},t=0)=\delta(\mathbf{x})$, in terms of convolutions.
The latter can be simplified by applying Fourier ($\mathbf{k}$) and Laplace ($s$) transforms.
Following the derivation in \cite{zumofen1993_2} for the one-dimensional case, we obtain for $d$ dimensions an analogous result (see Appendix \ref{sec:0}),
\begin{equation}
\label{eq:p_k_s}
p(\mathbf{k},s)=\frac{W(\mathbf{k},s)}{1-\psi(\mathbf{k},s)},
\end{equation}
where the arguments $(\mathbf{x},t)$ or $(\mathbf{k},s)$ indicate the space we are working in.
Diffusion processes are typically characterized by the time-lag dependence of the mean-squared displacement (MSD) that can be determined either as ensemble average or as time average.
In the next section, we investigate the ensemble-averaged squared displacement while the time-averaged squared displacement is studied in sections \ref{sec:5} and \ref{sec:6}.

\section{\label{sec:4}Ensemble-averaged squared displacement (EASD)}

\begin{figure*}
\includegraphics[width=\linewidth]{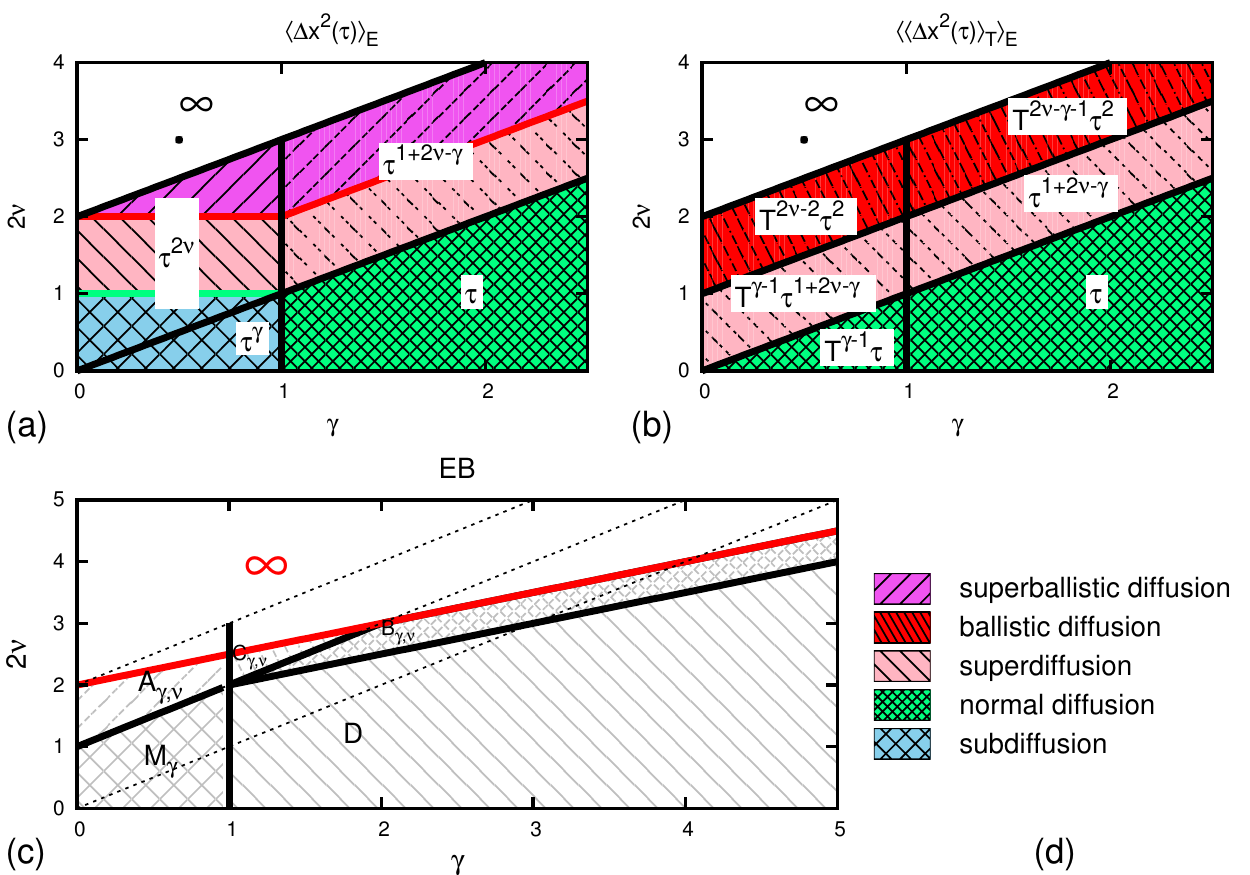}
\caption{\label{fig:phase_diagrams}
From \cite{albers2018}.
Phase diagram for the ensemble-averaged squared displacement (a), the ensemble average of the time-averaged squared displacement (b), and the ergodicity breaking parameter (c).
Different ranges of validity of the analytical results are separated by thick black lines in the two-dimensional parameter space.
Different kinds of diffusion are color-coded as indicated in the key.
The dotted lines in Figure (c) serve as a guide to the eye for a better comparison with the phase diagram in Figure (b).}
\end{figure*}

\begin{figure}
\includegraphics[width=\linewidth]{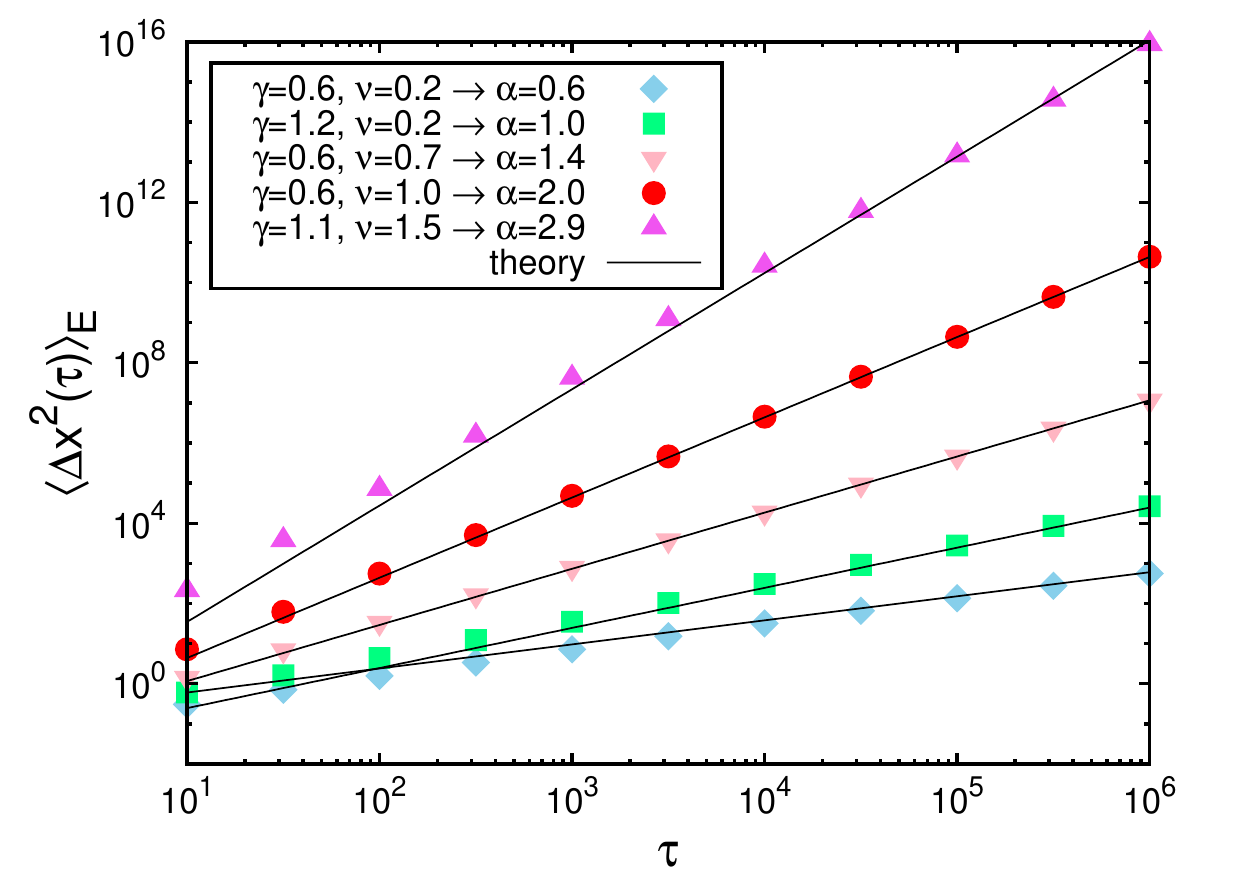}
\caption{\label{fig:EASD_analytical_numerical}
Ensemble-averaged squared displacement numerically determined from $N=10^6$ realizations of the generalized L\'evy walk for different values of the parameters $\gamma$ and $\nu$ ($t_0=1.0$, $c=0.33$) as specified in the key.
The colors of the symbols indicate resulting types of anomalous diffusion according to Fig.~\ref{fig:phase_diagrams} (d).
The black lines are the corresponding theoretical curves obtained from Eq.~(\ref{eq:LW_MSD_E_tau_1}) and Eq.~(\ref{eq:LW_MSD_E_tau_2}) and are in good agreement with the numerical results.}
\end{figure}

The ensemble-averaged squared displacement (EASD) for an ensemble of $N$ trajectories $\mathbf{x}_i(t)$ is defined as
\begin{equation}
\label{eq:EASD_definition}
\begin{split}
\langle\Delta\mathbf{x}^2(\tau)\rangle_{\text{E}}&=\langle[\mathbf{x}(\tau)-\mathbf{x}(0)]^2\rangle_{\text{E}}\\
&=\lim\limits_{N\rightarrow\infty}\frac{1}{N}\sum\limits_{i=1}^N[\mathbf{x}_i(\tau)-\mathbf{x}_i(0)]^2.
\end{split}
\end{equation}
In Eq.~(\ref{eq:EASD_definition}) and in the following, the symbols $\langle\dots\rangle_{\text{E}}$ and $\langle\dots\rangle_{\text{T}}$ denote ensemble and time averages, respectively.
Of course, in experiments and simulations, the limit in the second line of Eq.~(\ref{eq:EASD_definition}) cannot be performed, so one approximates the EASD by its sample mean for a finite ensemble.
Typically for anomalous diffusion, the EASD increases asymptotically according to a power law \cite{bouchaud1990,metzler2000,klages2008},
\begin{equation}
\label{eq:EASD_asymptotic}
\langle\Delta\mathbf{x}^2(\tau)\rangle_{\text{E}}\simeq\langle D_{\alpha}\rangle_{\text{E}}\,\tau^{\alpha}\quad(\tau\rightarrow\infty),
\end{equation}
where $\alpha$ is the diffusion exponent and the symbol $\langle D_{\alpha}\rangle_{\text{E}}$ denotes the generalized diffusion coefficient.
Depending on the value of $\alpha$, one distinguishes different kinds of anomalous diffusion.
For $\alpha<1$, one has subdiffusion, while the case $\alpha>1$ is called superdiffusion.
The special case $\alpha=1$ corresponds to normal diffusion, and $\alpha=2$ is referred to as ballistic diffusion.
The case $\alpha>2$ is often called superballistic diffusion.
Because of the initial condition $p(\mathbf{x},t=0)=\delta(\mathbf{x})$ of the propagator, which implies that $\mathbf{x}_i(0)=\mathbf{0}$, the EASD is identical to the second moment of the propagator.
Because the Fourier transform of the propagator is a moment-generating function (also see Eq.~(\ref{eq:MSD_E_s}) of Appendix \ref{sec:A}), the EASD can be calculated via
\begin{equation}
\label{eq:EASD_propagator}
\begin{split}
\langle\Delta\mathbf{x}^2(\tau)\rangle_{\text{E}}&=\int_{\mathbb{R}^d}\mathbf{x}^2\,p(\mathbf{x},\tau)\,\text{d}^d\mathbf{x}\\
&=\mathcal{L}^{-1}\left\{-\left.\frac{\partial^2}{\partial\mathbf{k}^2}p(\mathbf{k},s)\right|_{\mathbf{k}=\mathbf{0}};s,\tau\right\},
\end{split}
\end{equation}
where we use the symbol $\mathcal{L}^{-1}$ for the inverse Laplace transform.
The arguments $\tau$ and $s$ refer to the time lag and its Laplace conjugated variable, respectively.
Applying Abelian and Tauberian theorems \cite{feller1991,hughes1995} to Eq.~(\ref{eq:EASD_propagator}) together with Eq.~(\ref{eq:p_k_s}) shows that the long-time behavior ($\tau\rightarrow\infty$) of the EASD
is determined by the small-$s$ behavior of the transformed distributions $\psi(\mathbf{k},s)$ and $W(\mathbf{k},s)$.
Calculating the corresponding asymptotics and inserting the results into Eq.~(\ref{eq:EASD_propagator}), we obtain the asymptic behavior of the time-lag dependence of the EASD.
The details of the calculations as well as the final equations are presented in Appendix \ref{sec:A}.
For the following discussion, it is only important to realize that the character of the diffusive behavior is only determined by the two positive exponents $\gamma$ and $\nu$.
Therefore, the effective parameter space of the generalized L\'evy walk is two-dimensional.
The analytical results of the EASD for the full two-dimensional parameter space are summarized in the phase diagram in Fig.~\ref{fig:phase_diagrams} (a).
Furthermore, a comparison of the analytical results with numerical simulations of the generalized L\'evy walk is shown in Fig.~\ref{fig:EASD_analytical_numerical}.
We can see a very good agreement.

In the following, we want to discuss the phase diagram for the EASD of the generalized L\'evy walk in Fig.~\ref{fig:phase_diagrams} (a).
A first interesting observation is that the dependence of the diffusion exponent on the two characteristic parameters $\gamma$ and $\nu$
is identical to that of the corresponding generalized space-time coupled L\'evy flight ($\eta\rightarrow\infty$) \cite{klafter1987,zumofen1989,akimoto2013,akimoto2014},
where the flights of the generalized L\'evy walk are replaced by waiting times and jumps of corresponding duration and length, respectively.
The same dependence is found for the special case $\eta=\nu$ of the variable speed generalized L\'evy walk \cite{barkai1998}.
This indicates that the diffusion exponent is only determined by the statistics of the turning points but not by their connection in time and space what was later confirmed in \cite{bothe2019}.
As a consequence, as long as the EASD is finite, our phase diagram in Fig.~\ref{fig:phase_diagrams} (a) is valid for all values of $\eta$.
However, the divergence of the EASD of the generalized L\'evy walk for $2\nu\geq\gamma+2$ in the two-dimensional parameter space (white in Fig.~\ref{fig:phase_diagrams} (a))
does not occur for the space-time coupled L\'evy flight and the special case $\eta=\nu$ and has not been recognized in \cite{shlesinger1987}.
As a consequence, and in full contrast to \cite{shlesinger1987}, an asymptotic time-lag dependence of the EASD which is equal or faster than a cubic increase cannot be found for the generalized L\'evy walk.
We want to give a simple explanation for the divergence of the EASD of the generalized L\'evy walk and the convergence of the EASD for the other models.
To do so, we calculate a lower bound of the EASD by considering the contribution to the latter coming from all realizations of the process whose duration $T_1=t_1$ of the first elementary event is larger than $\tau$.
Furthermore, we use that for these trajectories according to Eq.~(\ref{eq:x_t}), the squared displacement after time lag $\tau$ is given by $(ct_1^{\nu-\eta}\tau^{\eta})^2$.
Therefore, we obtain the following bound,
\begin{equation}
\label{eq:EASD_divergence}
\begin{split}
\langle\Delta\mathbf{x}^2&(\tau)\rangle_{\text{E}}>\int_{\tau}^{\infty}(ct_1^{\nu-\eta}\tau^{\eta})^2\,\psi(t_1)\,\text{d}t_1\\
&\sim\int_{\tau}^{\infty}t_1^{2\nu-2\eta-\gamma-1}\,\text{d}t_1=\infty\quad\text{if}\quad2\nu\geq\gamma+2\eta.
\end{split}
\end{equation}
For the special case $\eta=1$ of our variable speed generalized L\'evy walk, we get from Eq.~(\ref{eq:EASD_divergence}) the correct condition for the divergence of the EASD of the generalized L\'evy walk.
Furthermore, we can see that for the generalized space-time coupled L\'evy flight ($\eta\rightarrow\infty$) and the special case $\eta=\nu$ of the variable speed generalized L\'evy walk,
the condition for the divergence of the EASD cannot be fulfilled, which explains the finiteness of the EASD for these models.
Therefore, we can conclude that while the diffusion exponents are determined by the statistics of the turning points,
the divergence or the convergence of the EASD depends on the specific connections between the turning points in time and space.

\begin{figure}
\includegraphics[width=\linewidth]{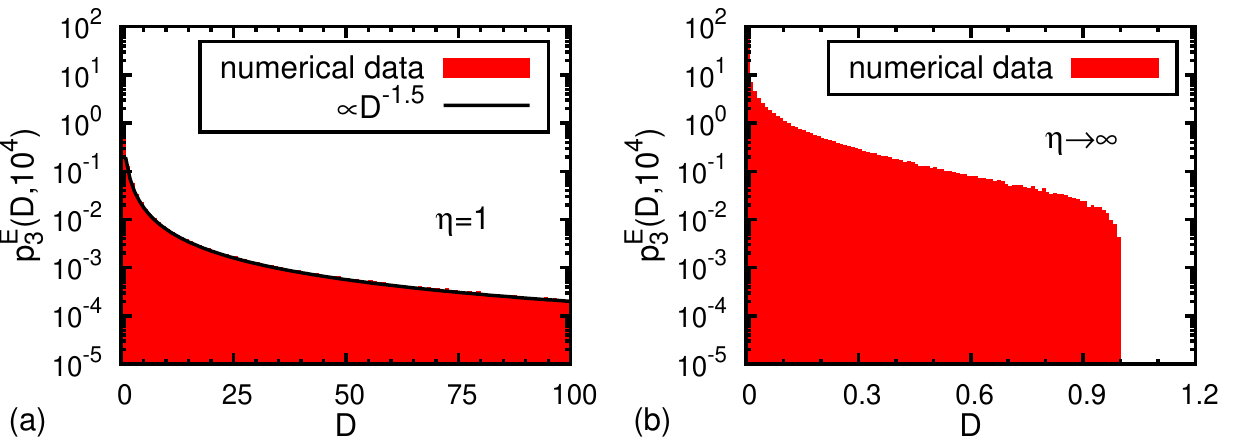}
\caption{\label{fig:DOGD_LW_LF}
Distribution of generalized diffusivities, Eq.~(\ref{eq:DOGD}), numerically determined from $N=10^6$ trajectories of the generalized L\'evy walk ($\eta=1$) (a)
and the space-time coupled L\'evy flight ($\eta\rightarrow\infty$) (b) for $\gamma=0.5$ and $\nu=1.5$ ($t_0=1.0$, $c=1.0$).
The black line in figure (a) describes the asymptotic behavior of the distribution for the generalized L\'evy walk leading to a divergence of the first moment
and, therefore, to a divergence of the ensemble-averaged squared displacement.
The distribution of generalized diffusivities for the space-time coupled L\'evy flight, however, has a cutoff at $D_{\text{max}}=c^2$ leading to a finite ensemble-averaged squared displacement.}
\end{figure}

The finiteness or divergence of the EASD for the different models can be visualized with the distribution of generalized diffusivities (DOGD),
which was first introduced for inhomogeneous and anisotropic normal diffusion processes in \cite{bauer2011,heidernaetsch2013} and later extended to anomalous diffusion in \cite{albers2013}.
We define a generalized diffusivity $D_{\alpha}(\tau)$ as a single squared displacement rescaled by the asymptotic time-lag dependence of the EASD, $D_{\alpha}(\tau)=\Delta\mathbf{x}^2(\tau)/\tau^{\alpha}$.
From an ensemble of trajectories, we can obtain the DOGD,
\begin{equation}
\label{eq:DOGD}
p_{\alpha}(D,\tau)=\langle\delta(D-D_{\alpha}(\tau))\rangle_{\text{E}}.
\end{equation}
Of course, the DOGD could also be obtained as time average from a long single-particle trajectory.
The advantage of the rescaling of the squared displacements is that the distribution may become stationary.
The first moment of the DOGD is asymptotically equal to the generalized diffusion coefficient, $\int D\,p_{\alpha}(D,\tau)\,\text{d}D\simeq\langle D_{\alpha}\rangle_{\text{E}}$ for $\tau\rightarrow\infty$.
The DOGD that is numerically obtained for the generalized L\'evy walk and the generalized space-time coupled L\'evy flight for $\gamma=1/2$ and $\nu=3/2$ is shown in Fig.~\ref{fig:DOGD_LW_LF}.
According to Shlesinger et. al. \cite{shlesinger1987}, this parameter choice corresponds to the Richardson case, where a cubic increase of the EASD of the generalized L\'evy walk was expected.
This cubic increase can actually be found for the generalized space-time coupled L\'evy flight but not for the generalized L\'evy walk, because for the latter, the EASD diverges,
see the black dot in the phase diagram in Fig.~\ref{fig:phase_diagrams} (a).
However, the DOGDs for both models actually become stationary if the scaling exponent $\alpha$ is equal to three.
The DOGD for the generalized L\'evy walk has a heavy tail, which leads to a divergence of the first moment of the DOGD and, therefore, also to a divergence of the EASD.
The DOGD for the generalized space-time coupled L\'evy flight, however, has a cutoff leading to a finite first moment and a finite EASD.
The heavy tail of the DOGD for the generalized L\'evy walk can be obtained by a simple estimation.
Large diffusivities are obtained by large displacements which are caused by large flight velocities that are connected to large flight durations $t_f$ for $\nu>1$.
In this case, we can write $D_{\alpha}(\tau)=(ct_f^{\nu-1}\tau)^2/\tau^{\alpha}\propto t_f^{2\nu-2}$.
A simple change of variables from flight durations $t_f$ distributed according to $\psi(t_f)$ to generalized diffusivities leads to the asymptotic behavior of the DOGD,
$p_{\alpha}(D,\tau)\simeq\int\delta(D-D_{\alpha}(\tau))\,\psi(t_f)\,\text{d}t_f\sim D^{-\gamma/(2\nu-2)-1}$ for $D\rightarrow\infty$.
The first moment of this distribution diverges for $2\nu\geq\gamma+2$ meaning that we recover the condition for the divergence of the EASD of the generalized L\'evy walk.
The cutoff of the DOGD for the generalized space-time coupled L\'evy flight is caused by the fact that there is a maximal jump length that can occur in the time interval $[0,\tau]$
generated by a preceding waiting time of duration $\tau$.
Therefore, the maximal squared displacement reads $\Delta\mathbf{x}^2(\tau)_{\text{max}}=(c\tau^{\nu})^2$.
For $0<\gamma<1$ and $2\nu>\gamma$, the corresponding asymptotic time-lag dependence of the EASD is proportional to $\tau^{2\nu}$.
This leads to the cutoff of the DOGD at $D_{\text{max}}=c^2$.

The kind of observed diffusion ranging from subdiffusion to superdiffusion in the phase diagram for the EASD of the generalized L\'evy walk in Fig.~\ref{fig:phase_diagrams} (a) can also be explained.
Keep in mind that for a normal diffusion process, the diffusion coefficient $\langle D_1\rangle_{\text{E}}$, i.e., the prefactor of the linear increase of the EASD,
is given by $\langle D_1\rangle_{\text{E}}=\langle\mathbf{X}^2\rangle/\langle T\rangle$,
where $\langle\mathbf{X}^2\rangle$ is the second moment of the covered distances per flight, and $\langle T\rangle$ is the mean flight duration.
The former is finite for $2\nu<\gamma$ and diverges for $2\nu\geq\gamma$.
$\langle T\rangle$ is finite for $\gamma>1$ and diverges for $\gamma\leq1$.
This means, for $\gamma>1$ and $2\nu<\gamma$, the normal diffusion coefficient is finite, and, therefore, we observe a linear increase of the EASD.
For $\gamma>1$ and $2\nu\geq\gamma$, the normal diffusion coefficient diverges, which means superdiffusion.
For $\gamma\leq1$ and $2\nu<\gamma$, the normal diffusion coefficient vanishes, which means subdiffusion.
For $\gamma\leq1$ and $2\nu\geq\gamma$, both characteristic quantities of the generalized L\'evy walk, $\langle\mathbf{X}^2\rangle$ and $\langle T\rangle$, diverge,
and so, we can observe in this region of the two-dimensional parameter space subdiffusion as well as superdiffusion.
Because $\langle\mathbf{X}^2\rangle\propto\langle T^{2\nu}\rangle$, subdiffusion is found for $2\nu<1$ and superdiffusion for $2\nu>1$.

In the next section, we contrast the asymptotic time-lag dependence of the EASD with the one for the time-averaged squared displacement.

\section{\label{sec:5}Time-averaged squared displacement (TASD)}

\begin{figure}
\includegraphics[width=\linewidth]{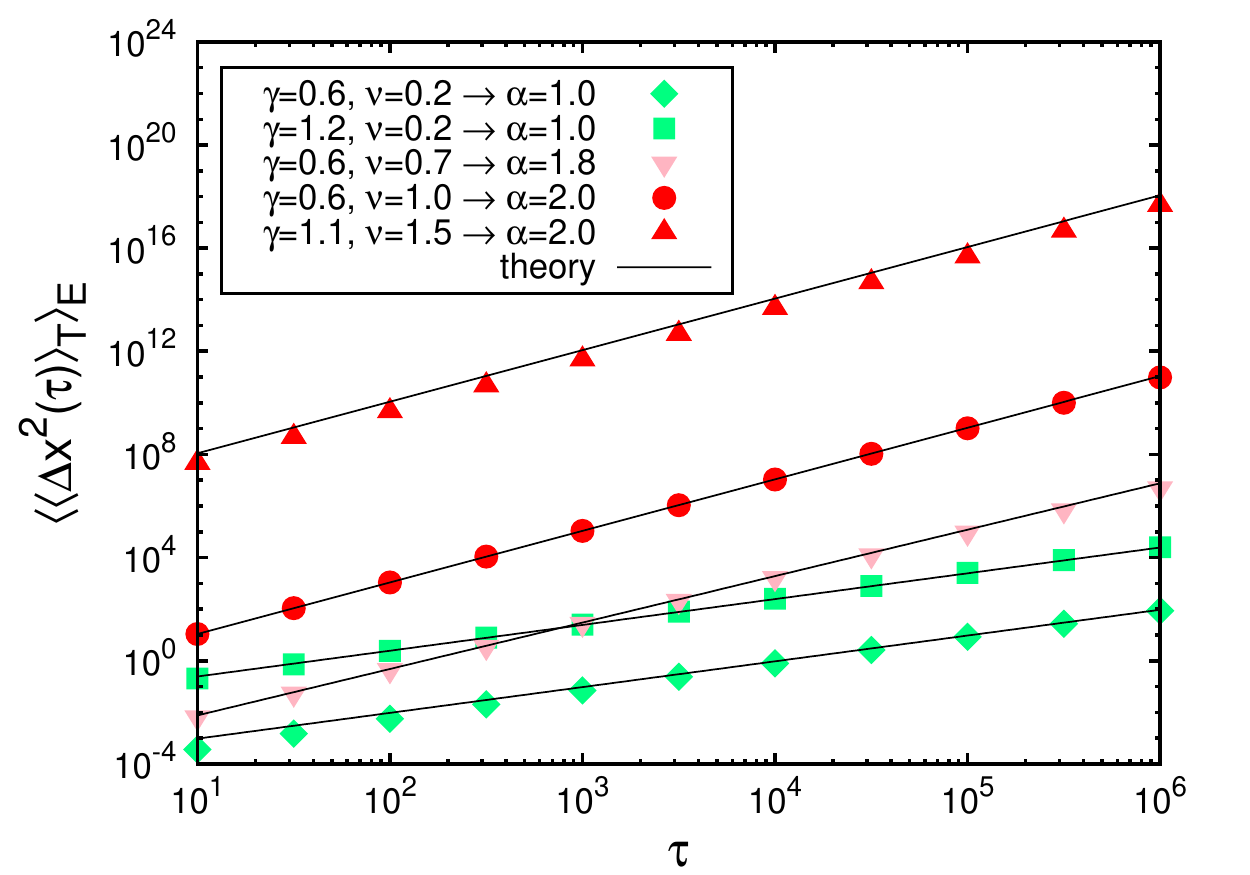}
\caption{\label{fig:EATASD_analytical_numerical}
Ensemble average of the time-averaged squared displacement numerically determined from $N=10^4$ realizations of length $T=10^8$ of the generalized L\'evy walk
for different values of the parameters $\gamma$ and $\nu$ ($t_0=1.0$, $c=0.33$) as specified in the key.
The colors of the symbols indicate resulting kinds of anomalous diffusion according to Fig.~\ref{fig:phase_diagrams} (d).
The black lines are the corresponding theoretical curves obtained from Eq.~(\ref{eq:LW_MSD_T_E_tau_1}) and Eq.~(\ref{eq:LW_MSD_T_E_tau_2}) and are in good agreement with the numerical results.}
\end{figure}

The time-averaged squared displacement (TASD) for a long stochastic or chaotic trajectory $\mathbf{x}(t)$ of length $T$ is defined as
\begin{equation}
\label{eq:TASD_definition}
\langle\Delta\mathbf{x}^2(\tau)\rangle_{\text{T}}=\frac{1}{T-\tau}\int_0^{T-\tau}[\mathbf{x}(t+\tau)-\mathbf{x}(t)]^2\,\text{d}t.
\end{equation}
Because the TASD is a random variable for every finite $T$ and in some cases even for $T\rightarrow\infty$, we first investigate its mean, i.e., the time-lag dependence of the ensemble average of the TASD (EATASD),
\begin{equation}
\label{eq:EATASD_definition}
\langle\langle\Delta\mathbf{x}^2(\tau)\rangle_{\text{T}}\rangle_{\text{E}}=\lim_{N\rightarrow\infty}\frac{1}{N}\sum\limits_{i=1}^N\langle\Delta\mathbf{x}_i^2(\tau)\rangle_{\text{T}},
\end{equation}
where, of course, the limit $N\rightarrow\infty$ cannot be performed in experiments and simulations.
The distribution of the TASD is the topic of the next section.
The details of the analytical derivation of the asymptotic time-lag dependence of the EATASD can be found in Appendix \ref{sec:B}.
Here, we focus on the discussion of the analytical results, which are again illustrated in the form of a phase diagram now in Fig.~\ref{fig:phase_diagrams} (b).
A comparison of the analytical results with numerical simulations can be found in Fig.~\ref{fig:EATASD_analytical_numerical}, where we can see again a very good agreement.

First of all, we recognize that also the EATASD of the generalized L\'evy walk diverges for $2\nu\geq\gamma+2$.
Again, the possible divergence of the EATASD of our variable speed generalized L\'evy walk defined in Eqs.~(\ref{eq:psi_t}), (\ref{eq:psi_x_t}), and (\ref{eq:x_t}) can be explained by considering a lower bound.
For all trajectories whose duration $T_1$ of the first elementary event is longer than the measurement time $T$, the TASD according to the definition in Eq.~(\ref{eq:TASD_definition}) and taking account of Eq.~(\ref{eq:x_t}) is given by
\begin{equation}
\label{eq:TASD_estimation}
\begin{split}
\langle\Delta\mathbf{x}^2(\tau)\rangle_{\text{T}}^{T_1>T}&=\frac{1}{T-\tau}\int_0^{T-\tau}[cT_1^{\nu-\eta}((t+\tau)^{\eta}-t^{\eta})]^2\,\text{d}t\\[1ex]
&\overset{T\gg\tau}{\sim}T_1^{2\nu-2\eta}\tau^2T^{2\eta-2}.
\end{split}
\end{equation}
If we only consider the contribution to the EATASD coming from these trajectories, we obtain
\begin{equation}
\label{eq:EATASD_divergence}
\begin{split}
\langle\langle\Delta\mathbf{x}^2&(\tau)\rangle_{\text{T}}\rangle_{\text{E}}\geq\int_T^{\infty}\langle\Delta\mathbf{x}^2(\tau)\rangle_{\text{T}}^{T_1>T}\,\psi(T_1)\,\text{d}T_1\\
&\sim\int_T^{\infty}T_1^{2\nu-2\eta-\gamma-1}\,\text{d}T_1=\infty\quad\text{if}\quad2\nu\geq\gamma+2\eta.
\end{split}
\end{equation}
When we compare this result with Eq.~(\ref{eq:EASD_divergence}), we see that the estimated conditions for the divergence of the EASD and the EATASD are equal.
Therefore, this simple estimation explains why the EATASD can diverge for the generalized L\'evy walk ($\eta=1$)
but not for the space-time coupled L\'evy flight in the wait first interpretation ($\eta\rightarrow\infty$) or the special case $\eta=\nu$.

\begin{figure}
\includegraphics[width=\linewidth]{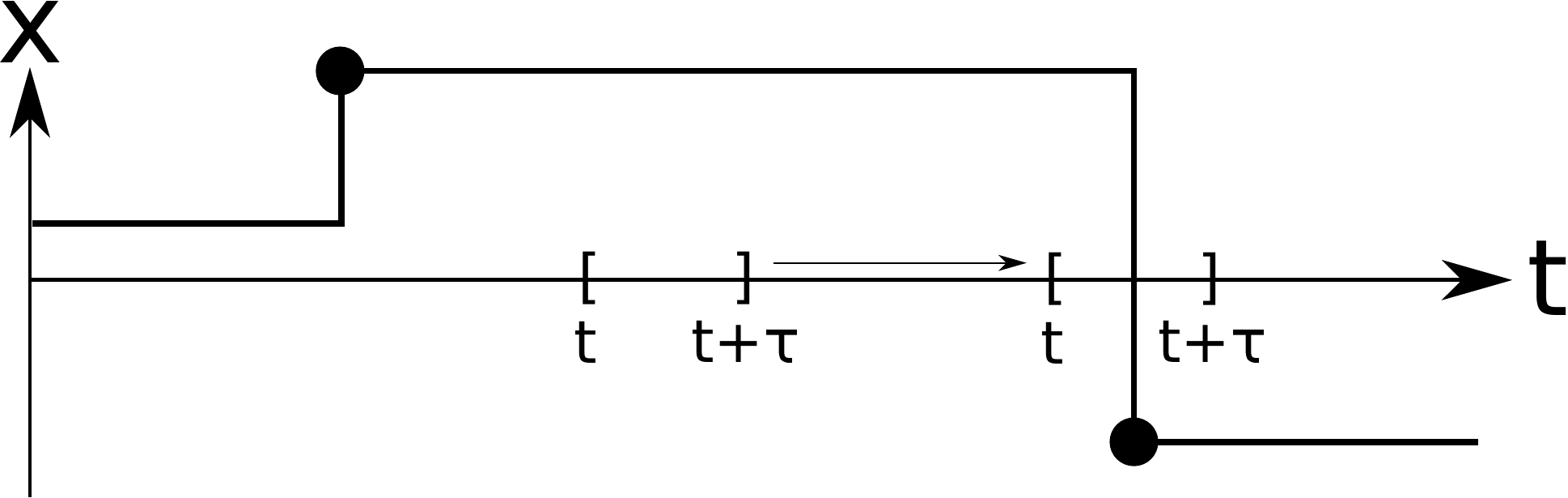}
\caption{\label{fig:explanation1}
Schematic representation of one realization of a L\'evy flight, where the squared displacement $[x(t+\tau)-x(t)]^2$ is only non-vanishing if at least one jump occurs in the interval $[t,t+\tau]$.}
\end{figure}

In the following, we want to discuss in detail the phase diagram for the EATASD of the generalized L\'evy walk in Fig.~\ref{fig:phase_diagrams} (b) in the section of the two-dimensional parameter space where the EATASD is finite.
At first sight, we can see that the colors indicating the kind of anomalous diffusion are in general different from the ones appearing in the phase diagram for the EASD in Fig.~\ref{fig:phase_diagrams} (a).
Therefore, we can say that the generalized L\'evy walk is nonergodic with respect to the squared displacements.
Interestingly, and in full contrast to the generalized L\'evy walk, the EATASD of the space-time coupled L\'evy flight in the wait-first interpretation ($\eta\rightarrow\infty$)
always shows a linear time-lag dependence \cite{akimoto2013,akimoto2014} although both models have the same turning point statistics and the same diffusion exponents with respect to the EASD.
The reason is the different connection of the turning points in both models.
For the space-time coupled L\'evy flight, a squared displacement during the time interval $[t,t+\tau]$ is only nonzero if at least one jump occurs during this time window.
This can be inferred from Fig.~\ref{fig:explanation1}.
The non-vanishing contribution to the TASD coming from a single jump is proportional to the length of the time window, i.e., to the time lag $\tau$.
This contribution is asymptotically dominant for $T\gg\tau$ over the contributions where more than one jump occurs during the time interval $[t,t+\tau]$ \cite{miyaguchi2013}.
In \cite{akimoto2013}, it was shown that this relation holds for all values of the parameters $\gamma$ and $\nu$.
Therefore, we can conclude that the linear time-lag dependence of the EATASD of the space-time coupled L\'evy flight is a consequence of the special structure of the trajectories consisting of jumps and waiting times.
For the generalized L\'evy walk with its ballistic connections of the turning points, however, several time-lag dependencies of the EATASD arise.

For $2\nu<\gamma$, a linear time-lag dependence of the EATASD is also found for the generalized L\'evy walk.
For $\gamma>1$, this result is hardly surprising because in this case, both the second moment $\langle\mathbf{X}^2\rangle$ of the covered distances per flight as well as the mean flight duration $\langle T\rangle$ are finite,
and we recover normal and ergodic diffusion with respect to the squared displacements.
For $\gamma<1$, however, the mean flight duration diverges leading to ergodicity breaking because the total measurement time $T$ cannot be much longer than the typical time scale of the system.
Interestingly, in this case, the nonergodicity resembles the one for the subdiffusive continuous time random walk \cite{lubelski2008,he2008,albers2013}, i.e., we obtain subdiffusion from the EASD and ``normal diffusion'' from the EATASD.
For $\gamma<2\nu<\gamma+1$ and $\gamma>1$, the EASD and the EATASD coincide with respect to the diffusion exponent but not with respect to the diffusion coefficient.
This kind of nonergodicity was observed earlier in \cite{zumofen1993_3,godec2013} for the special case of the standard L\'evy walk ($\nu=1$) and was called ultraweak ergodicity breaking.
The reason is that the statistics of the squared diplacements is non-stationary in the sense that it depends on the so-called aging time $t_a$, the elapsed time between the beginning of the process and the beginning of the measurement.
For $t_a\rightarrow\infty$, however, stationarity and also ergodicity is recovered in the sense that also the diffusion coefficients obtained from the EASD and the EATASD coincide.
For $\gamma+1<2\nu<\gamma+2$ and $\gamma>1$, we obtain ballistic diffusion from the EATASD but superballistic diffusion from the EASD.
For the complete section of the two-dimensional parameter space where $\gamma<1$, we observe differences in the obtained diffusion exponents and diffusion coefficients in general as expected from the diverging time scale of the system,
i.e., the diverging mean flight duration $\langle T\rangle$.
Especially interesting is the triangular region $\gamma<1$ and $\gamma<2\nu<1$, where the EASD shows subdiffusion but the EATASD indicates superdiffusion.
To our knowledge, such kind of ergodicity breaking, which we call ``subdiffusion appearing as superdiffusion'', has not been recognized before in any model of anomalous diffusion,
but it has significant impact on the interpretation of data coming from experiments or simulations.
For example, it means that one could measure subdiffusion in a pulsed field gradient nuclear magnetic resonance experiment, where ensemble averages are measured,
but superdiffusion in a single-particle tracking experiment on the same system, where time averages are measured.
Whereas the observed types of anomalous diffusion from the EASD have already been explained in the previous section with the divergence or finiteness of the characteristic quantities $\langle\mathbf{X}^2\rangle$ and $\langle T\rangle$,
we want to give in the following descriptive explanations for the kinds of anomalous diffusion appearing in the EATASD.

To do so, we first decompose the time integral in the definition of the TASD in Eq.~(\ref{eq:TASD_definition}) into two contributions,
where the first one captures the $N_T$ completed flights until measurement time $T$ and the second one describes the contribution from the backward recurrence time \cite{godreche2001},
i.e., from the time interval $[t_{N_T},T]$ containing the last incomplete flight,
\begin{widetext}
\begin{equation}
\label{eq:TASD_contributions}
\begin{split}
\langle\Delta\mathbf{x}^2(\tau)\rangle_{\text{T}}&=\frac{1}{T-\tau}\int_0^{T-\tau}[\mathbf{x}(t+\tau)-\mathbf{x}(t)]^2\,\text{d}t\overset{T\gg\tau}{\simeq}\frac{1}{T}\int_0^T[\mathbf{x}(t+\tau)-\mathbf{x}(t)]^2\,\text{d}t\\[1ex]
&=\frac{1}{T}\left[\sum\limits_{i=1}^{N_T}\int_{t_{i-1}}^{t_i}[\mathbf{x}(t+\tau)-\mathbf{x}(t)]^2\,\text{d}t+\int_{t_{N_T}}^T[\mathbf{x}(t+\tau)-\mathbf{x}(t)]^2\,\text{d}t\right].
\end{split}
\end{equation}
If we perform the ensemble average of the TASD in Eq.~(\ref{eq:TASD_contributions}), we obtain for the EATASD the following decomposition,
\begin{equation}
\label{eq:EATASD_contributions}
\langle\langle\Delta\mathbf{x}^2(\tau)\rangle_{\text{T}}\rangle_{\text{E}}=\frac{1}{T}\left[\left\langle\sum\limits_{i=1}^{N_T}\int_{t_{i-1}}^{t_i}[\mathbf{x}(t+\tau)-\mathbf{x}(t)]^2\,\text{d}t\right\rangle_{\text{E}}+\left\langle\int_{t_{N_T}}^T[\mathbf{x}(t+\tau)-\mathbf{x}(t)]^2\,\text{d}t\right\rangle_{\text{E}}\right].
\end{equation}
\end{widetext}
These two contributions to the EATASD correspond to the completed flights and to the last incomplete flight and are studied in detail in the appendices \ref{sec:C} and \ref{sec:D}, respectively.
Here, we summarize the results.
In Appendix \ref{sec:C}, we show that long completed flights lead to superdiffusion for $\gamma<2\nu<\gamma+1$ and ballistic diffusion for $2\nu>\gamma+1$ in accordance with the phase diagram in Fig.~\ref{fig:phase_diagrams} (b).
Interestingly, for $2\nu<\gamma$, long completed flights lead to subdiffusion, but this contribution is dominated by the one of short completed flights that cause normal diffusion as expected.
This can be understood as follows.
For $\nu\rightarrow0$, the coupling between flight duration $T_i$ and travelled distance $|\mathbf{X}_i|\propto T_i^{\nu}$ is weak.
This means that travelled distances of a certain order of magnitude can be caused by long flights or short flights, i.e., these distances are covered in a long or a short time, respectively.
Long flights lead then to subdiffusion similar to the long waiting times in the subdiffusive continuous time random walk.
Short flights, however, lead to normal diffusion.
In contrast, in appendix \ref{sec:D}, we show that the contribution from the last incomplete flight always leads to ballistic diffusion with the asymptotics shown in Fig.~\ref{fig:phase_diagrams} (b) for $2\nu>\gamma+1$.
For $2\nu<\gamma+1$, this contribution is dominated by the superdiffusive contribution of the long completed flights.
Note that also the dependence on the total measurement time $T$ and the condition $T\gg\tau$ for meaningful time averaging has thereby to be taken into account.
In this way, the full phase diagram for the EATASD of the generalized L\'evy walk in Fig.~\ref{fig:phase_diagrams} (b) is explained.
Furthermore, these considerations explain why in the triangular region $\gamma<1$ and $\gamma<2\nu<1$ of the two-dimensional parameter space,
where the EASD shows subdiffusion because the second moment $\langle\mathbf{X}^2\rangle\propto\langle T^{2\nu}\rangle$ of the covered distances per flight diverges weaker than the mean flight duration $\langle T\rangle$,
the EATASD indicates superdiffusion.
However, such kind of ergodicity breaking cannot be found for the space-time coupled L\'evy flight in the wait-first interpretation ($\eta\rightarrow\infty$)
although the turning point statistics as well as the diffusion exponents obtained from the EASD are identical to that of the generalized L\'evy walk.
For wait and jump models with heavy-tailed distributed waiting times such as the subdiffusive continuous time random walk or the space-time coupled L\'evy flight,
linear time-lag dependencies of the EATASD were found \cite{lubelski2008,he2008,akimoto2013,akimoto2014} caused by the geometry of the trajectories as explained previously in this section.

\section{\label{sec:6}Ergodicity breaking (EB) parameter}

\begin{figure}
\includegraphics[width=\linewidth]{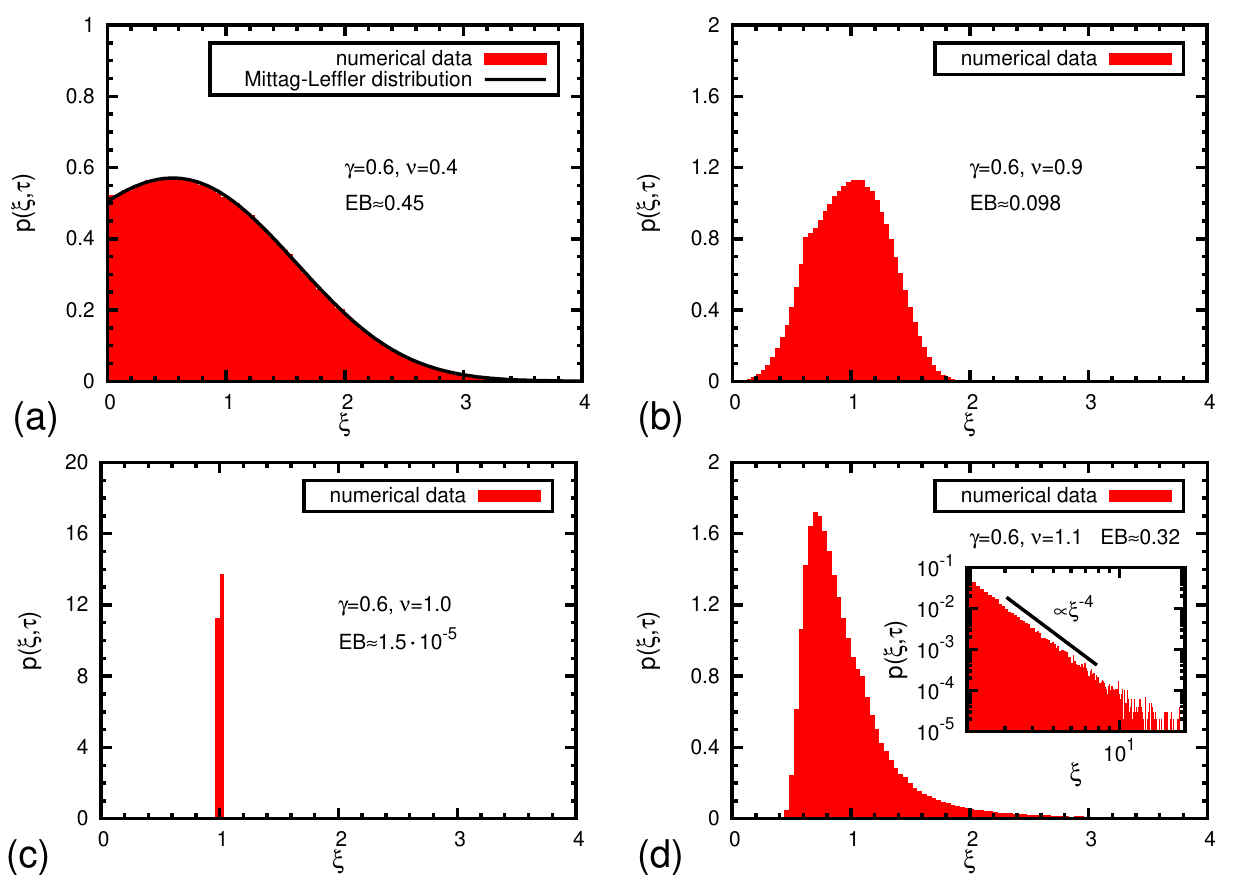}
\caption{\label{fig:DOXI_1}
Distribution $p(\xi,\tau)$ of rescaled time-averaged squared displacements numerically determined from $N=10^6$ realizations of length $T=10^7$ of the generalized L\'evy walk for $\gamma=0.6$ (infinite mean flight duration)
and increasing values of the parameter $\nu$ ($t_0=1.0$, $c=0.33$, $\tau=100$) belonging to different kinds of anomalous diffusion according to the phase diagram of the EASD in Fig.~\ref{fig:phase_diagrams} (a):
(a) $\nu=0.4$, subdiffusion (b) $\nu=0.9$, superdiffusion (c) $\nu=1.0$, ballistic diffusion (d) $\nu=1.1$, superballistic diffusion.
In figure (a), the distribution is compared with the Mittag-Leffler distribution known for instance from the distribution $p(\xi,\tau)$ of the subdiffusive continuous time random walk,
where the same temporal behavior of the mean-squared displacements was found, i.e., subdiffusion with respect to the EASD and normal diffusion regarding the EATASD \cite{he2008}.
The case $\nu=1.0$ in figure (c) belongs to the standard L\'evy walk, where according to \cite{froemberg2013_1,froemberg2013_2}, a delta distribution is expected.
The inset in figure (d) for $\nu=1.1$ shows the distribution $p(\xi,\tau)$ in a double-logarithmic plot on a larger scale in order to pronounce the heavy tail of the distribution.
Figure (a) belongs to the sector $M_{\gamma}$ in the phase diagram of the EB parameter in Fig.~\ref{fig:phase_diagrams} (c).
All other figures belong to sector $A_{\gamma,\nu}$.}
\end{figure}

\begin{figure}
\includegraphics[width=\linewidth]{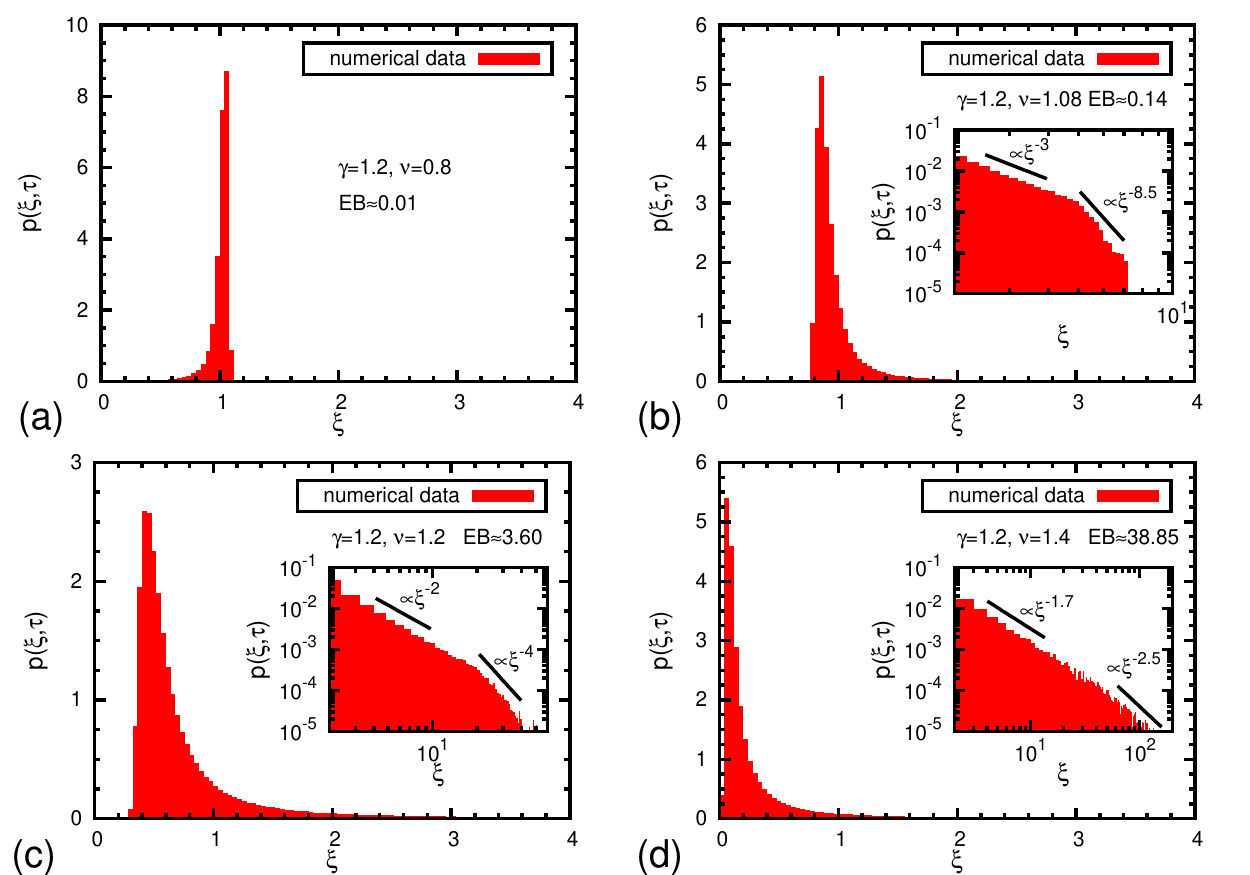}
\caption{\label{fig:DOXI_2}
Distribution $p(\xi,\tau)$ of rescaled time-averaged squared displacements numerically determined from $N=10^6$ realizations of length $T=10^7$ of the generalized L\'evy walk for $\gamma=1.2$ (finite mean flight duration)
and increasing values of the parameter $\nu$ ($t_0=1.0$, $c=0.33$, $\tau=100$) belonging to different kinds of anomalous diffusion according to the phase diagram of the EASD in Fig.~\ref{fig:phase_diagrams} (a)
and belonging to different sectors in the phase diagram of the EB parameter in Fig.~\ref{fig:phase_diagrams} (c):
(a) $\nu=0.8$, superdiffusion, sector D (b) $\nu=1.08$, superdiffusion, sector $B_{\gamma,\nu}$ (c) $\nu=1.2$, superballistic diffusion, sector $C_{\gamma,\nu}$ (d) $\nu=1.4$, superballistic diffusion, sector $\infty$.
The insets show the distributions $p(\xi,\tau)$ in a double-logarithmic plot on a larger scale in order to pronounce the heavy tails of the distributions.}
\end{figure}

\begin{figure}
\includegraphics[width=\linewidth]{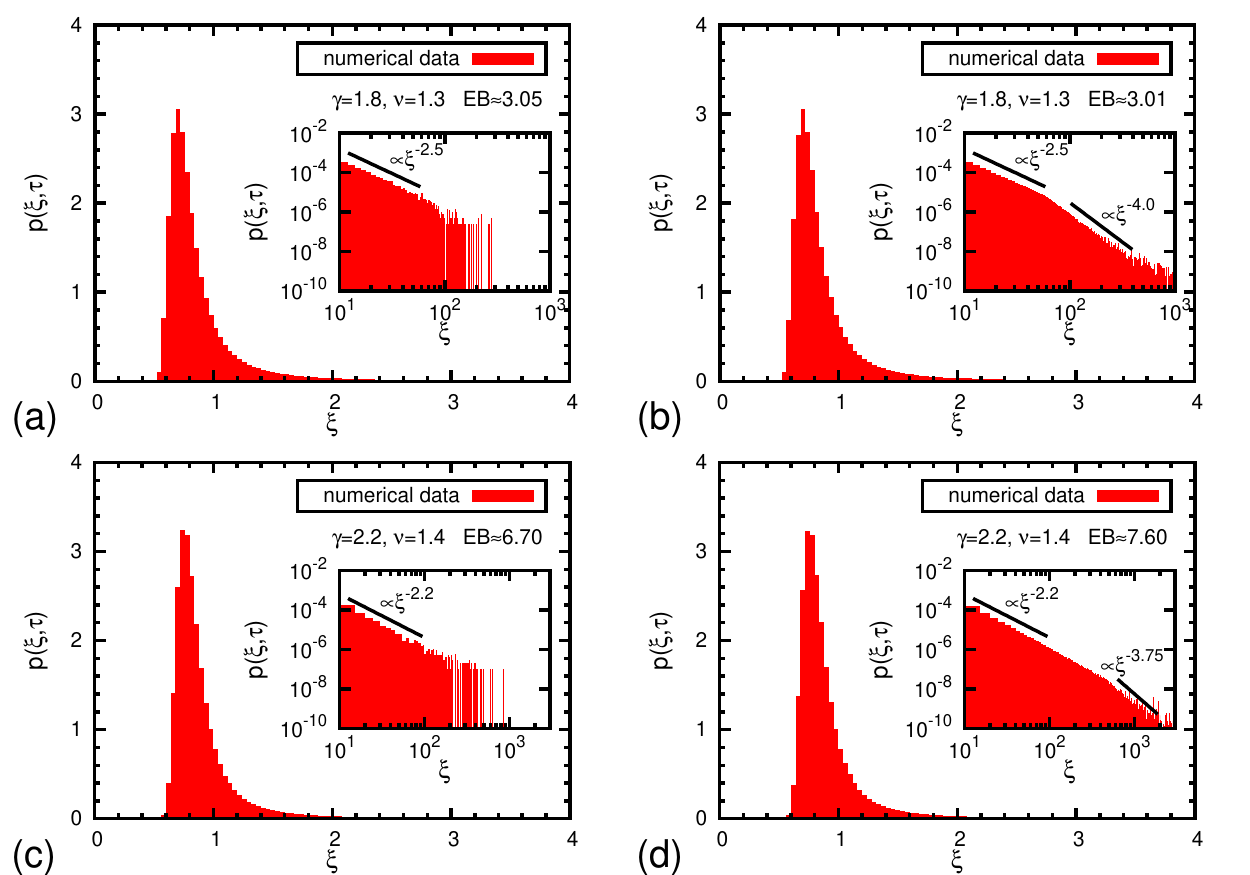}
\caption{\label{fig:DOXI_3}
Comparison of the distributions $p(\xi,\tau)$ of rescaled time-averaged squared displacements numerically determined with simple sampling (left figures (a) and (c)) and hybrid sampling (simple sampling + importance sampling, right figures (b) and (d))
for two parameter combinations both belonging to sector $B_{\gamma,\nu}$ in the phase diagram of the EB parameter in Fig.~\ref{fig:phase_diagrams} (c): (a) and (b) $\gamma=1.8$, $\nu=1.3$ (c) and (d) $\gamma=2.2$, $\nu=1.4$.
$N=2\cdot10^6$ realizations of duration $T=10^4$ of the generalized L\'evy walk were used ($t_0=1.0$, $c=0.33$, $\tau=1.0$).
The insets show the distributions $p(\xi,\tau)$ in a double-logarithmic plot on a larger scale in order to show the advantage of hybrid sampling versus simple sampling.
Parameters for the hybrid sampling: $\xi^*=10.0$, $p=0.000625$, $t^*=3000$ (b) and $\xi^*=10.0$, $p=0.0005$, $t^*=1000$ (d).}
\end{figure}

Last but not least, in this section, we want to investigate the fluctuations of the TASD with respect to different realizations of the generalized L\'evy walk.
To do so, we consider the rescaled random variable
\begin{equation}
\label{eq:xi_tau}
\widehat{\xi}(\tau)=\frac{\langle\Delta\mathbf{x}^2(\tau)\rangle_{\text{T}}}{\langle\langle\Delta\mathbf{x}^2(\tau)\rangle_{\text{T}}\rangle_{\text{E}}},
\end{equation}
whose mean value is equal to unity due to the rescaling, i.e., $\langle\widehat{\xi}(\tau)\rangle_{\text{E}}=1$.
The variance of the random variable $\widehat{\xi}(\tau)$, which is the square of the relative fluctuations of the TASD, is known in the literature as ergodicity breaking (EB) parameter \cite{he2008},
\begin{equation}
\label{eq:EB_tau}
\text{EB}(\tau)=\text{Var}\left(\widehat{\xi}(\tau)\right)=\left\langle\widehat{\xi}^2(\tau)\right\rangle_{\text{E}}-\left\langle\widehat{\xi}(\tau)\right\rangle_{\text{E}}^2.
\end{equation}
The distribution $p(\xi,\tau)=\langle\delta(\xi-\widehat{\xi}(\tau))\rangle_{\text{E}}$ of the random variable $\widehat{\xi}(\tau)$ fully captures the random nature of the TASD.
For an ergodic process, the TASDs determined from different realizations of the process coincide if the measurement time $T$ goes to infinity.
As a consequence, the EB parameter goes asymptotically to zero and the distribution $p(\xi,\tau)$ becomes a delta distribution, i.e., $\lim_{T\rightarrow\infty}p(\xi,\tau)=\delta(\xi-1)$.
Often for anomalous diffusion processes, the TASDs obtained from different realizations of the process coincide with respect to the diffusion exponent but not with respect to the diffusion coefficient, i.e.,
in a double-logarithmic plot, different TASDs correspond to parallel lines with different absolute terms.
In this case, the random variable $\widehat{\xi}(\tau)$ does not depend on the time lag $\tau$ and is equal in distribution to the random variable $\xi^*$ defined as rescaled time-averaged squared velocity,
\begin{equation}
\label{eq:xi_xi*}
\widehat{\xi}\overset{\text{d}}{=}\xi^*,\quad\xi^*=\frac{\int_0^T\mathbf{v}^2(t)\,\text{d}t}{\left\langle\int_0^T\mathbf{v}^2(t)\,\text{d}t\right\rangle_{\text{E}}}.
\end{equation}
This was shown in previous publications of the authors \cite{albers2014,albers2016} by using the Green-Kubo formula \cite{kubo1966,hansen2006,godec2013} (see Eq.~(\ref{eq:Green_Kubo_formula}),
which connects the TASD with the autocorrelation function of the velocity process defined as time average (see Eq.~(\ref{eq:C_v_t}).
Moreover, because $[\mathbf{x}(t+\tau)-\mathbf{x}(t)]^2\simeq\mathbf{v}^2(t)\,\tau^2$ for $\tau\rightarrow0$, this equality in distribution generally holds in the limit $\tau\rightarrow0$.
An analytical treatment of the random variable $\xi^*$ is much simpler than the original problem because the velocity process $v(t)=|\mathbf{v}(t)|$ is piecewise constant.
In Appendix \ref{sec:E}, an analytical derivation of the EB parameter, which also allows conclusions on the full distribution $p(\xi,\tau)$ of the generalized L\'evy walk, is presented.
Our analytical results for the EB parameter are again illustrated in the form of a phase diagram in Fig.~\ref{fig:phase_diagrams} (c).
Furthermore, numerically determined distributions $p(\xi,\tau)$ for several values of the parameters $\gamma$ and $\nu$ are shown in Figs.~\ref{fig:DOXI_1}, \ref{fig:DOXI_2}, and \ref{fig:DOXI_3}.
In the following, we want to discuss the phase diagram in Fig.~\ref{fig:phase_diagrams} (c) in detail.

Similar to the MSDs, also the EB parameter of the generalized L\'evy walk can diverge, what we call ``infinitely strong ergodicity breaking'', but contrary to the MSDs, the condition for the divergence reads $2\nu\geq\gamma/2+2$.
As a consequence, the EB parameter can even diverge when the MSDs are finite.
Again, the possible divergence of the EB parameter of the models following from the variable speed generalized L\'evy walk can be understood by considering a lower bound
coming from all trajectories whose duration $T_1$ of the first elementary event is longer than the measurement time $T$.
Because the EB parameter is essentially determined by the second moment of the random TASD and using Eq.~(\ref{eq:TASD_estimation}), we obtain
\begin{equation}
\label{eq:EB_divergence}
\begin{split}
\text{EB}&\geq\int_T^{\infty}\left(\langle\Delta\mathbf{x}^2(\tau)\rangle_{\text{T}}^{T_1>T}\right)^2\,\psi(T_1)\,\text{d}T_1\\
&\sim\int_T^{\infty}T_1^{4\nu-4\eta-\gamma-1}\,\text{d}T_1=\infty\quad\text{if}\quad2\nu\geq\frac{\gamma}{2}+2\eta.
\end{split}
\end{equation}
For the generalized L\'evy walk ($\eta=1$), we obtain with this simple estimation the correct condition for the divergence of the EB parameter.
Furthermore, we can see that, for instance, the EB parameter does not diverge for the space-time coupled L\'evy flight in the wait first interpretation ($\eta\rightarrow\infty$) or the spacial case $\eta=\nu$.

We continue our discussion with the sections of the phase diagram in Fig.~\ref{fig:phase_diagrams} (c) where the EB parameter is finite.
For parameter choices from the section $\gamma<1$, where the mean flight duration diverges, TASDs obtained from different realizations of the generalized L\'evy walk show the same temporal scaling, i.e.,
the distribution of rescaled TASDs and the EB parameter do not depend on the time lag $\tau$.
In sector $\text{M}_{\gamma}$ ($\gamma<1$ and $2\nu<\gamma+1$), the EB parameter is given by Eq.~(\ref{eq:EB_1}) and is equal to the variance of the Mittag-Leffler distribution.
Therefore, we conclude that the distribution of rescaled TASDs is given by the Mittag-Leffler distribution,
\begin{equation}
\label{eq:Mittag_Leffler_distribution}
p(\xi)=\frac{\Gamma^{1/\alpha}(1+\alpha)}{\alpha\xi^{1+1/\alpha}}\,l_{\alpha}\left[\frac{\Gamma^{1/\alpha}(1+\alpha)}{\xi^{1/\alpha}}\right],
\end{equation}
where $l_{\alpha}(t)$ is the one-sided L\'evy stable probability density function whose Laplace transform is $\exp(-s^{\alpha})$ \cite{bouchaud1990}.
This analytical result is confirmed numerically, see Fig.~\ref{fig:DOXI_1} (a).
Furthermore, there is a simple explanation for this finding.
According to Eq.~(\ref{eq:xi_xi*}), the random variable $\widehat{\xi}$ is equal in distribution to the random variable $\xi^*$.
The time integral of the squared velocity appearing in the numerator of $\xi^*$ can be approximated as follows
\begin{equation}
\label{eq:time_integral_approximation}
\int_0^T\mathbf{v}^2(t)\,\text{d}t\approx\sum\limits_{i=1}^{N_T}\mathbf{V}_i^2T_i\approx N_T\,\langle\mathbf{V}_i^2T_i\rangle,
\end{equation}
where again $N_T$ is the number of completed flights up to measurement time T.
Therefore, the random variable $\xi^*$ is equal in distribution to the random variable $N_T/\langle N_T\rangle_{\text{E}}$.
From renewal theory \cite{bouchaud1990} it is known that this random variable follows the Mittag-Leffler distribution.
Because this consideration only holds if the mean value $\langle\mathbf{V}_i^2T_i\rangle\propto\langle T_i^{2\nu-1}\rangle=\int_0^{\infty}t^{2\nu-1}\psi(t)\,\text{d}t$ is finite,
which is the case for $2\nu<\gamma+1$, this explains the occurence of the Mittag-Leffler distribution in sector $\text{M}_{\gamma}$.

In sector $\text{A}_{\gamma,\nu}$ ($\gamma<1$ and $\gamma+1<2\nu<\gamma/2+2$), the EB parameter is given by a complicated formula that depends on both parameters $\gamma$ and $\nu$, see Eq.~(\ref{eq:EB_2}).
Note that this formula, which was initially derived in \cite{albers2016}, was reproduced recently in \cite{akimoto2020}, where it was also shown that the transition line between sector $\text{M}_{\gamma}$ and $\text{A}_{\gamma,\nu}$
is associated with the observable changing from being integrable to becoming non-integrable with respect to an associated infinite density.
For a fixed value of the parameter $\gamma$ and for increasing values of the parameter $\nu$ starting from the boundary between the sectors $\text{M}_{\gamma}$ and $\text{A}_{\gamma,\nu}$,
the EB parameter first decreases until it becomes zero for $\nu=1$, and then it increases until it even diverges for $2\nu\geq\gamma/2+2$.
Correspondingly, the distribution of recaled TASDs first becomes narrower until it converges to a delta distribution for $\nu=1$,
and then it becomes a heavy-tailed distribution that is responsible for the divergence of the EB parameter for $2\nu\geq\gamma/2+2$.
This evolution of the distribution of rescaled TASDs for a fixed value of the parameter $\gamma$ and increasing values of the parameter $\nu$ is shown in Fig.~\ref{fig:DOXI_1}.
The asymptotic behavior of the distribution $p(\xi)$ for $\nu>1$ can be estimated by a simple consideration.
Large values of $\xi$ are related to large TASDs which are connected to the occurrence of long flights because large flight durations lead to large flight velocities for $\nu>1$ and, therefore, to large displacements.
According to Eq.~(\ref{eq:TASD_estimation}), for large flight durations $t_f$, the TASD is proportional to $t_f^{2\nu-2}$.
By a change of variables from long flight durations $t_f$ distributed according to $\psi(t_f)$ to the random variable $\widehat{\xi}\sim t_f^{2\nu-2}$, we obtain
\begin{equation}
\label{eq:p_xi_asymptotic}
p(\xi)\sim\int\delta\left(\xi-t_f^{2\nu-2}\right)\,\psi(t_f)\,\text{d}t_f\sim\xi^{-1-\frac{\gamma}{2\nu-2}}.
\end{equation}
This asymptotic behavior of the distribution of rescaled TASDs is confirmed numerically, see the insets in Figs.~\ref{fig:DOXI_1}, \ref{fig:DOXI_2}, and \ref{fig:DOXI_3}.
Furthermore, from this asymptotic behavior, we can infer that the EB parameter, which is essentially determined by the second moment of $p(\xi)$,
diverges for $2\nu\geq\gamma/2+2$ in full agreement with the phase diagram in Fig.~\ref{fig:phase_diagrams} (c).
Note that the vanishing EB parameter for $\nu=1$ correpsonds to the standard L\'evy walk with constant flight velocities independent on the flight durations.
This case was investigated in detail in \cite{froemberg2013_1,froemberg2013_2}.

In the section $\gamma>1$, where the mean flight duration is finite, the EB parameter increases and the distribution of rescaled TASDs becomes broader for increasing values of the time lag $\tau$.
In this case, our analytical results for the EB parameter are only valid in the limit $\tau\rightarrow0$.
In sector D ($\gamma>1$ and $2\nu<\gamma/2+3/2$), the EB parameter goes to zero for $T\rightarrow\infty$, see Eq.~(\ref{eq:EB_3}), and the distribution $p(\xi,\tau)$ converges to a delta distribution.
This is what we expect for the normal diffusion sector in the two-dimensional parameter space and is in agreement with the previous findings for the standard L\'evy walk ($\nu=1$) in \cite{godec2013,froemberg2013_1,froemberg2013_2}.
In sectors $\text{B}_{\gamma,\nu}$ and $\text{C}_{\gamma,\nu}$ ($\gamma>1$ and $\gamma/2+3/2<2\nu<\gamma/2+2$),
the process becomes nonergodic in the sense that the EB parameter becomes larger for increasing values of $T$ and is given by a complicated $T$ dependent expression, see Eq.~(\ref{eq:EB_4}) and Eq.~(\ref{eq:EB_5}).
This transition from an ergodic to a nonergodic behavior can be understood with the help of the Khinchin theorem \cite{khinchin1949,lee2007,lapas2008,weron2010},
which states that the EB parameter goes to zero for $T\rightarrow\infty$ if the covariance function of the squared displacements $\Delta\mathbf{x}^2(t',\tau)=[\mathbf{x}(t'+\tau)-\mathbf{x}(t')]^2$
and $\Delta\mathbf{x}^2(t'+t,\tau)$ goes to zero for $t\rightarrow\infty$,
\begin{equation}
\label{eq:Khinchin_theorem}
\text{Cov}\left(\Delta\mathbf{x}^2(t',\tau),\Delta\mathbf{x}^2(t'+t,\tau)\right)\overset{t\rightarrow\infty}{\longrightarrow}0.
\end{equation}
For a simple estimation of the covariance function, we use the relation $\Delta\mathbf{x}^2(t',\tau)\simeq\mathbf{v}^2(t')\tau^2\,(\tau\rightarrow0)$.
Therefore, for $\tau\rightarrow0$, the covariance function is essentially determined by $\langle\mathbf{v}^2(t')\mathbf{v}^2(t'+t)\rangle_{\text{E}}$.
We only consider the contributions where the instants of time $t'$ and $t'+t$ belong to the same flight of duration $t_f>t$.
This only occurs if the forward recurrence time (FRT) $t_f'$, i.e., the remaining duration of a flight at time $t'$, is longer than $t$.
From renewal theory \cite{godreche2001}, it is well known that for $\gamma>1$ (finite mean flight duration),
the corresponding distribution of the FRT becomes stationary for $t'\rightarrow\infty$, $\lim_{t'\rightarrow\infty}\psi^{\text{FRT}}_{t'}(t_f')\sim t_f'^{-\gamma}\,(t_f'\rightarrow\infty)$.
For $\nu>1$, we underestimate the velocity at time $t'$ and $t'+t$ by calculating the velocity from the FRT $t_f'$ instead from the real flight duration $t_f>t_f'$.
Therefore,
\begin{equation}
\label{eq:Khinchin_theorem_estimation}
\begin{split}
\lim\limits_{t'\rightarrow\infty}&\langle\mathbf{v}^2(t')\mathbf{v}^2(t'+t)\rangle_{\text{E}}>\int_t^{\infty}(ct_f'^{\nu-1})^4\lim\limits_{t'\rightarrow\infty}\psi^{\text{FRT}}_{t'}(t_f')\,\text{d}t_f'\\
&\sim\int_t^{\infty}t_f'^{4\nu-\gamma-4}\,\text{d}t_f'=\infty\quad\text{if}\quad2\nu\geq\frac{\gamma}{2}+\frac{3}{2}.
\end{split}
\end{equation}
We conclude that for $2\nu\geq\gamma/2+3/2$, the Khinchin theorem is violated which explains the transition from ergodic to nonergodic behavior.
The evolution of the distribution of rescaled TASDs for a fixed value $\gamma>1$ and increasing values of $\nu$ is shown in Fig.~\ref{fig:DOXI_2}.
Note that for $\nu>1$, the asymptotic behavior of the distribution is again described by Eq.~(\ref{eq:p_xi_asymptotic}) explaining the divergence of the EB parameter for $2\nu\geq\gamma/2+2$.

A very interesting observation from the phase diagram in Fig.~\ref{fig:phase_diagrams} (c) is that the EB parameter can even diverge in the section of the two-dimensional parameter space where normal diffusion occurs.
This may happen for $\gamma>4$.
However, it is very difficult to observe this new kind of ergodicity breaking numerically.
The reason is that the divergence of the EB parameter is caused by very long flights that become rare
because the mean flight duration $\langle T\rangle=t_0/(\gamma-1)$ decreases for increasing values of the parameter $\gamma$ and the tail of the distribution $\psi(t)$ of flight durations decays faster.
If the flights become shorter, more flights have to be generated in order to produce trajctories of a certain length $T$.
In other words, for increasing values of the parameter $\gamma$, computer simulations become more time-consuming.
In order to limit the time exposure for the numerical simulations, shorter trajectories have to be used.
For these trajectories, numerically determined distributions $p(\xi,\tau)$ are depicted in Fig.~\ref{fig:DOXI_3} (a) and (c).
We can see that the second algebraic decay of the distribution, which is described by Eq.~(\ref{eq:p_xi_asymptotic}) and is responsible for the divergence of the EB parameter, is not visible even for $\gamma=1.8$ and $\gamma=2.2$.

In order to numerically determine this second tail of the distribution of rescaled TASDs, one can use importance sampling \cite{bucklew2004,press2007}.
In order to get the full distribution, we use a method that we call hybrid sampling.
The basic idea is that we determine the bulk of the distribution with simple sampling and the tail with importance sampling.
In the following, we briefly describe the hybrid sampling method for a general case and then specify the method for the determination of the distribution of rescaled TASDs of the generalized L\'evy walk.
Let us consider a general random variable $X\in\mathbb{R}^+$ with probability density $p(x)$.
The random variable $X$ could be identified, for instance, with the TASD or the random variable $\widehat{\xi}(\tau)$.
By introducing some threshold value $x^*$, the distribution $p(x)$ can be divided in two parts, the bulk for $x<x^*$ and the tail for $x>x^*$.
We want to calculate the average of an arbitrary function $f(X)$ with respect to the probability density $p(x)$.
We can write
\begin{equation}
\label{eq:importance_sampling}
\begin{split}
\langle f(X)\rangle_p&=\int_0^{\infty}f(x)\,p(x)\,\text{d}x\\
&=\underbrace{\int_0^{x^*}f(x)\,p(x)\,\text{d}x}_{=I_1}+\underbrace{\int_{x^*}^{\infty}f(x)\,\frac{p(x)}{\tilde{p}(x)}\,\tilde{p}(x)\,\text{d}x}_{=I_2},
\end{split}
\end{equation}
where the new probability density $\tilde{p}(x)$ is chosen such that large values of the random variable $X$ are more likely compared with the original distribution $p(x)$.
The integrals $I_1$ and $I_2$ can be interpreted as expectation values $\langle\Theta(x^*-X)f(X)\rangle_p$ and $\langle\Theta(X-x^*)f(X)p(X)/\tilde{p}(X)\rangle_{\tilde{p}}$
with respect to the probability densities $p(x)$ and $\tilde{p}(x)$, respectively.
Of course, these mean values can be estimated by corresponding sample means.
Therefore, an unbiased estimator for $I_1$ is
\begin{equation}
\label{eq:estimator1}
\widehat{I}_1=\frac{1}{N}\sum\limits_{i=1}^N\Theta(x^*-X_i)\,f(X_i),\quad X_i\overset{\text{IID}}{\sim}p,
\end{equation}
and an unbiased estimator for $I_2$ is
\begin{equation}
\label{eq:estimator2}
\tilde{I}_2=\frac{1}{N}\sum\limits_{i=1}^N\Theta(X_i-x^*)\,f(X_i)\,\frac{p(X_i)}{\tilde{p}(X_i)},\quad X_i\overset{\text{IID}}{\sim}\tilde{p}.
\end{equation}
In order to determine the distribution of the random variable $\widehat{\xi}(\tau)$ defined in Eq.~(\ref{eq:xi_tau}), we first have to calculate the EATASD.
For this first step, the random variable $X$ can be identified with the TASD, and we have $f(X)=X$.
After estimating the EATASD, we can determine the distribution of rescaled TASDs.
In this second step, the random variable $X$ can be identified with $\widehat{\xi}(\tau)$, and for the estimation of the probability $P(X\in[a,b])$, we have $f(X)=\Theta(X-a)\Theta(b-X)$.
Additionally, for both steps, we have to specify the likelihood ratio $p(x)/\tilde{p}(x)$.
The probability for the occurrence of a certain TASD is connected with the probability for the corresponding realization of the generalized L\'evy walk
that is essentially determined by the probability of the associated sequence of flight durations.
For the simple sampling, we use the original distribution $\psi(t)$ of flight durations defined in Eq.~(\ref{eq:psi_t}).
For the importance sampling, we use a new distribution of flight durations,
\begin{equation}
\label{eq:psi_t_new}
\begin{split}
\tilde{\psi}(t)&=(1-p)\,\psi(t)+p\,\psi(t|t>t^*),\\
\psi(t|t>t^*)&=\Theta(t-t^*)\,\frac{\psi(t)}{\int_{t^*}^{\infty}\psi(t)\,\text{d}t},
\end{split}
\end{equation}
which for $0<p\leq1$ guarantees that long flights of duration $t>t^*$ are more likely.
The advantage of this choice for the new distribution $\tilde{\psi}(t)$ is that the likelihood ratio of a single flight only depends on whether the flight duration $t$ is smaller or larger than the threshold value $t^*$.
The likelihood ratio for one single flight is given by
\begin{equation}
\label{eq:likelihood_ratio1}
L^{\text{SF}}(t)=\frac{\psi(t)}{\tilde{\psi}(t)}=\begin{cases}\frac{1}{1-p},\,&t<t^*\\[1ex]\frac{1}{1-p+p/\lambda},\,&t>t^*\end{cases},\,\lambda=\int_{t^*}^{\infty}\psi(t)\,\text{d}t,
\end{equation}
and, therefore, the likelihood ratio for a complete trajectory reads
\begin{equation}
\label{eq:likelihood_ratio2}
L^{\text{Traj}}=\left(\frac{1}{1-p}\right)^{N_{t<t^*}}\,\left(\frac{1}{1-p+p/\lambda}\right)^{N_{t>t^*}},
\end{equation}
where $N_{t<t^*}$ and $N_{t>t^*}$ is the number of flights in one complete trajectory with flight durations smaller and larger than $t^*$, respectively.
Using hybrid sampling, we obtain the numerically determined distributions of rescaled TASDs for $\gamma=1.8$ and $\gamma=2.2$ shown in Fig.~\ref{fig:DOXI_3} (b) and (d).
A comparison of the left part of Fig.~\ref{fig:DOXI_3} with the right part of Fig.~\ref{fig:DOXI_3} shows the advantage of hybrid sampling over simple sampling.
Now we can see the second algebraic decay of the distribution of rescaled TASDs which is described by Eq.~(\ref{eq:p_xi_asymptotic}) and is responsible for the divergence of the EB parameter for $2\nu\geq\gamma/2+2$.
In order to get the full distribution of rescaled TASDs for $\gamma>4$, where long flights are even more unlikely,
one would have to apply a hybrid sampling method which also divides the tail of the distribution in several parts in order to resolve the complete tail.

\section{\label{sec:7}Summary and discussion}

In this article, we defined the variable speed generalized L\'evy walk consisting of a sequence of independent and identically distributed space-time coupled elementary events.
This general model includes several special cases which have been investigated previously in the literature.
All these models have in common that they possess the same statistics of turning points, but they differ in the spatio-temporal paths between them.
The main focus of the paper was the generalized L\'evy walk, which consists of flight episodes, where the velocities of the flights deterministically depend on the heavy-tailed distributed flight durations.
We investigated this model in full analytical detail, provided descriptive explanations for all the observed phenomena, and compared our findings with previous results for the other models that follow from the general model.
We found that the resulting types of anomalous diffusion based on the time-lag dependence of the ensemble-averaged squared displacement only depend on the statistics of the turning points,
whereas the spatio-temporal paths between them control whether the ensemble-averaged squared displacement may diverge or not for certain parameter ranges.
Furthermore, we derived the time-lag dependencies of the time-averaged squared displacement of the generalized L\'evy walk
and compared them with the corresponding results for the space-time coupled L\'evy flight that consists of jumps and waiting times.
While the time-averaged squared displacement of the space-time coupled L\'evy flight always increases linearly, the generalized L\'evy walk shows a richer spectrum of time-lag dependencies
indicating that the spatio-temporal paths between the turning points crucially influence the behavior of time averages.
For a certain range of parameters of the generalized L\'evy walk, the time-lag dependence of the ensemble-averaged squared displacement leads to subdiffusion, whereas the time-averaged squared displacement indicates superdiffusion.
We argued that this new kind of ergodicity breaking (``subdiffusion appearing as superdiffusion'') is general in so far as it should also be observed in almost all the other space-time coupled models that follow from the general model.
Whereas the subdiffusive behavior is caused by the fact that the second moment of the travelled distances with each elementary event diverges weaker than the mean event duration,
the superdiffusive behavior is caused by the continuous motion during long elementary events.
An investigation of the random nature of the time-averaged squared displacement based on the analytical derivation of the ergodicity breaking parameter revealed further surprising results.
We found that the fluctuations of the time-averaged squared displacements can become so large that the ergodicity breaking parameter diverges (``infinitely strong ergodicity breaking'') although the mean-squared displacements are finite.
Even more surprisingly, we argued that this can also happen for parameter ranges where the generalized L\'evy walk shows normal diffusion.
The reason is that, although the second moment of the travelled distances per flight and the mean flight duration are finite leading to normal diffusion,
the algebraically decaying distribution of flight durations can still cause very long flights
that lead to a heavy-tailed distribution of time-averaged squared displacements which is responsible for the divergence of the ergodicity breaking parameter.
All the observed surprising results and new kinds of ergodicity breaking are essential for the interpretation of time averages obtained, for instance, from single-particle tracking experiments.

\appendix

\section{\label{sec:0}Derivation of the propagator}

In this appendix, we derive the propagator $p(\mathbf{x},t)$ in terms of the probability distributions $\psi(\mathbf{x},t)$ and $W(\mathbf{x},t)$ defined in Eqs.~(\ref{eq:psi_x_t}) and (\ref{eq:W_x_t}) of the main text, respectively.
The propagator can be written as
\begin{equation}
\label{eq:p_x_t}
p(\mathbf{x},t)=\int\limits_{\mathbb{R}^d}\int\limits_0^tQ(\mathbf{x}-\mathbf{x}',t-t')\,W(\mathbf{x}',t')\,\text{d}t'\,\text{d}^d\mathbf{x}',
\end{equation}
where $Q(\mathbf{x},t)\,\text{d}^d\mathbf{x}\,\text{d}t$ is the probability that a random walker arrives with a completed flight at an infinitesimal volume $\text{d}^d\mathbf{x}$ around $\mathbf{x}$ in the time interval $[t,t+\text{d}t]$.
We can further write
\begin{equation}
\label{eq:Q_x_t}
Q(\mathbf{x},t)=\sum\limits_{n=0}^{\infty}Q_n(\mathbf{x},t),
\end{equation}
where $Q_n(\mathbf{x},t)$ is the probability density to reach position $\mathbf{x}$ at time $t$ after $n$ flights with $Q_0(\mathbf{x},t)=\delta(\mathbf{x})\,\delta(t)$ and
\begin{equation}
\label{eq:Q_n_x_t}
Q_n(\mathbf{x},t)=\int\limits_{\mathbb{R}^d}\int\limits_0^tQ_{n-1}(\mathbf{x}-\mathbf{x}',t-t')\,\psi(\mathbf{x}',t')\,\text{d}t'\,\text{d}^d\mathbf{x}'.
\end{equation}
A combined Fourier $\mathcal{F}\{f(\mathbf{x});\mathbf{x},\mathbf{k}\}=\int_{\mathbb{R}^d}f(\mathbf{x})\,e^{\text{i}\mathbf{k}\mathbf{x}}\,\text{d}^d\mathbf{x}$
and Laplace $\mathcal{L}\{g(t);t,s\}=\int_0^{\infty}g(t)\,e^{-st}\,\text{d}t$ transform \cite{schiff1999} of the convolutions, Eq.~(\ref{eq:p_x_t}) and Eq.~(\ref{eq:Q_n_x_t}),
together with Eq.~(\ref{eq:Q_x_t}) gives the Fourier and Laplace transform of the propagator according to Eq.~(\ref{eq:p_k_s}) of the main text,
\begin{equation}
\label{eq:p_k_s_}
p(\mathbf{k},s)=\frac{W(\mathbf{k},s)}{1-\psi(\mathbf{k},s)}.
\end{equation}

\section{\label{sec:A}Derivation of the ensemble-averaged squared displacement (EASD)}

The following derivation and the ones in Appendices \ref{sec:B} and \ref{sec:E} are among others the multidimensional extensions of the results obtained in \cite{albers2016}.

The ensemble-averaged squared displacement (EASD) is defined by Eq.~(\ref{eq:EASD_definition}) of the main text and, according to Eq.~(\ref{eq:EASD_propagator}), is equal to the second moment of the propagator $p(\mathbf{x},t)$
with the initial condition $p(\mathbf{x},t=0)=\delta(\mathbf{x})$.
Because the Fourier transform of the propagator is a moment-generating function, the Laplace transform of the EASD can be written as \cite{zumofen1993_2}
\begin{equation}
\label{eq:MSD_E_s}
\begin{split}
&\mathcal{L}\left\{\langle\Delta\mathbf{x}^2(\tau)\rangle_{\text{E}};\tau,s\right\}=\int_{\mathbb{R}^d}\mathbf{x}^2\,p(\mathbf{x},s)\,\text{d}^d\mathbf{x}\\[1ex]
&=\int_{\mathbb{R}^d}-\left.\frac{\partial^2}{\partial\mathbf{k}^2}\left(e^{\text{i}\mathbf{k}\mathbf{x}}\right)\right|_{\mathbf{k}=\mathbf{0}}\,p(\mathbf{x},s)\,\text{d}^d\mathbf{x}=-\left.\frac{\partial^2}{\partial\mathbf{k}^2}p(\mathbf{k},s)\right|_{\mathbf{k}=\mathbf{0}}\\[1ex]
&=-\left.\frac{\Delta_{\mathbf{k}}W(\mathbf{k},s)}{1-\psi(\mathbf{k},s)}\right|_{\mathbf{k}=\mathbf{0}}-\left.\frac{W(\mathbf{k},s)\Delta_{\mathbf{k}}\psi(\mathbf{k},s)}{[1-\psi(\mathbf{k},s)]^2}\right|_{\mathbf{k}=\mathbf{0}}.
\end{split}
\end{equation}
Note that $\nabla_{\mathbf{k}}\psi(\mathbf{k}=\mathbf{0},s)$ and $\nabla_{\mathbf{k}}W(\mathbf{k}=\mathbf{0},s)$ are equal to zero due to the spatial isotropy of the considered random walk.
The long-time behavior of $\langle\Delta\mathbf{x}^2(\tau)\rangle_{\text{E}}$ is determined by the small-$s$ behavior of $\mathcal{L}\left\{\langle\Delta\mathbf{x}^2(\tau)\rangle_{\text{E}};\tau,s\right\}$.
Therefore, and because of the Laplace operator with respect to $\mathbf{k}$ at the point $\mathbf{k}=\mathbf{0}$ in Eq.~(\ref{eq:MSD_E_s}),
we need the small-$s$ behavior and the Taylor expansion in powers of $\mathbf{k}$ up to the second order of the quantities $\psi(s)$, $\psi(\mathbf{k},s)$, and $W(\mathbf{k},s)$.
For $\psi(s)$, using the definition in Eq.~(\ref{eq:psi_t}), we find \cite{zumofen1993_2}
\begin{equation}
\label{eq:psi_s_1}
\psi(s)\overset{s\rightarrow0}{\simeq}
\begin{cases}
1-\Gamma(1-\gamma)(t_0s)^{\gamma},\,&0<\gamma<1\\
1-\frac{1}{\gamma-1}t_0s-\Gamma(1-\gamma)(t_0s)^{\gamma},\,&1<\gamma<2\\
1-\frac{1}{\gamma-1}t_0s+\frac{1}{(\gamma-1)(\gamma-2)}(t_0s)^2,\,&\gamma>2
\end{cases}.
\end{equation}
This formula can be obtained by using the Cauchy-Saalsch\"utz representation of the gamma function for negative arguments \cite{zumofen1990}.
For $\gamma>2$, the small-$s$ behavior in Eq.~(\ref{eq:psi_s_1}) is identical to the Taylor expansion of $\psi(s)$ in powers of $s$ up to the second order,
where the prefactors of $s$ and $s^2$ correspond to the first and second moment of the probability density function $\psi(t)$ of flight durations, respectively.
The zeroth order reproduces the normalization of the distribution $\psi(t)$.
For $1<\gamma<2$, the second moment of $\psi(t)$ diverges, and, therefore, the second-order term is replaced by a term of fractional order.
For $0<\gamma<1$, also the first moment of $\psi(t)$ diverges, and so the first-order term is replaced by the fractional-order term.
For $\gamma<0$, $\psi(t)$ is not normalizable and, thus, not a probability density function.
A Taylor expansion of $\psi(\mathbf{k},s)$ in powers of $\mathbf{k}$ up to the second order gives
\begin{equation}
\label{eq:psi_k_s}
\begin{split}
\psi(\mathbf{k},s)&\simeq\psi(s)-\frac{1}{2d}c^2\mathbf{k}^2I_1(s)\quad(\mathbf{k}\rightarrow\mathbf{0}),\\
I_1(s)&=\int_0^{\infty}t^{2\nu}\,\psi(t)\,e^{-st}\,\text{d}t,\\
I_1(s)&\overset{s\rightarrow0}{\simeq}
\begin{cases}
\frac{\Gamma(2\nu+1)\Gamma(\gamma-2\nu)}{\Gamma(\gamma)}t_0^{2\nu},\,&2\nu<\gamma\\
\gamma\Gamma(2\nu-\gamma)t_0^{\gamma}s^{\gamma-2\nu},\,&2\nu>\gamma
\end{cases},
\end{split}
\end{equation}
where the small-$s$ behavior of the integral $I_1(s)$ for $2\nu<\gamma$ is equal to the zeroth order of the corresponding Taylor expansion, i.e., $I_1(0)$.
For $2\nu>\gamma$, $I_1(0)$ diverges, and, therefore, the zeroth order is replaced by a fractional order, which can be obtained from the definition of $I_1(s)$ in Eq.~(\ref{eq:psi_k_s}) by using the definition of the gamma function.
Finally, according to the calculation in Eq.~(\ref{eq:psi_k_s}), the small $\mathbf{k}$ and $s$ expansion of $W(\mathbf{k},s)$ yields
\begin{equation}
\label{eq:W_k_s}
\begin{split}
W(\mathbf{k},s)&\simeq\frac{1-\psi(s)}{s}-\frac{1}{2d}c^2\mathbf{k}^2I_2(s)\quad(\mathbf{k}\rightarrow0),\\[1ex]
I_2(s)&\overset{s\rightarrow0}{\simeq}
\begin{cases}
\frac{\Gamma(2\nu+2)\Gamma(\gamma-2\nu-1)}{3\Gamma(\gamma)}t_0^{2\nu+1},\,&2\nu<\gamma-1\\[1ex]
\frac{\gamma\Gamma(2\nu+1-\gamma)}{\gamma+2-2\nu}t_0^{\gamma}s^{\gamma-2\nu-1},\,&2\nu>\gamma-1
\end{cases},
\end{split}
\end{equation}
where the double integral
\begin{equation}
\label{eq:I_2_s}
I_2(s)=\int\limits_0^{\infty}\int\limits_t^{\infty}t'^{2\nu-2}\,\psi(t')\,\text{d}t'\,t^2e^{-st}\,\text{d}t
\end{equation}
diverges for $2\nu\geq\gamma+2$.
Therefore, also the EASD, Eq.~(\ref{eq:MSD_E_s}), diverges under the same condition,
\begin{equation}
\label{eq:LW_MSD_E_tau_divergence}
\langle\Delta\mathbf{x}^2(\tau)\rangle_{\text{E}}=\infty\quad\text{if}\quad2\nu\geq\gamma+2.
\end{equation}
However, for $2\nu<\gamma+2$, by inserting Eqs.~(\ref{eq:psi_s_1}-\ref{eq:W_k_s}) in Eq.~(\ref{eq:MSD_E_s}) and performing the inverse Laplace transform, we obtain for the EASD with $0<\gamma<1$,
\begin{equation}
\label{eq:LW_MSD_E_tau_1}
\begin{split}
&\langle\Delta\mathbf{x}^2(\tau)\rangle_{\text{E}}\overset{\tau\rightarrow\infty}{\simeq}\\[1ex]
&\begin{cases}
c^2\frac{\Gamma(2\nu+1)\Gamma(\gamma-2\nu)}{\Gamma(1-\gamma)\Gamma(\gamma)\Gamma(\gamma+1)}t_0^{2\nu-\gamma}\tau^{\gamma},\,&0<2\nu<\gamma\\[1ex]
2c^2\frac{\gamma}{\gamma+2-2\nu}\frac{\Gamma(2\nu-\gamma)}{\Gamma(1-\gamma)\Gamma(2\nu+1)}\tau^{2\nu},\,&\gamma<2\nu<\gamma+2
\end{cases},
\end{split}
\end{equation}
and for $\gamma>1$, we get
\begin{equation}
\label{eq:LW_MSD_E_tau_2}
\begin{split}
&\langle\Delta\mathbf{x}^2(\tau)\rangle_{\text{E}}\overset{\tau\rightarrow\infty}{\simeq}\\[1ex]
&\begin{cases}
c^2\frac{\Gamma(2\nu+1)\Gamma(\gamma-2\nu)}{\Gamma(\gamma-1)}t_0^{2\nu-1}\tau,\,&0<2\nu<\gamma\\[1ex]
2c^2\frac{\gamma(\gamma-1)}{\gamma+2-2\nu}\frac{\Gamma(2\nu-\gamma)}{\Gamma(2\nu+2-\gamma)}t_0^{\gamma-1}\tau^{2\nu+1-\gamma},\,&\gamma<2\nu<\gamma+2
\end{cases}.
\end{split}
\end{equation}
The $\tau$-dependencies in Eq.~(\ref{eq:LW_MSD_E_tau_1}) and Eq.~(\ref{eq:LW_MSD_E_tau_2}) are the ones indicated in Fig.~\ref{fig:phase_diagrams} (a).

\section{\label{sec:B}Derivation of the ensemble average of the time-averaged squared displacement (EATASD)}

Our starting point for the analytical derivation is the Green-Kubo formula \cite{kubo1966,hansen2006,godec2013}
\begin{equation}
\label{eq:Green_Kubo_formula}
\langle\Delta\mathbf{x}^2(\tau)\rangle_{\text{T}}=2\int_0^{\tau}(\tau-t)\,C_{\mathbf{v}}(t)\,\text{d}t,
\end{equation}
which relates the time-averaged squared displacement (TASD) with the autocorrelation function $C_{\mathbf{v}}(t)$ of the velocity process defined as time average,
\begin{equation}
\label{eq:C_v_t}
C_{\mathbf{v}}(t)=\frac{1}{T-t}\int_0^{T-t}\mathbf{v}(t')\mathbf{v}(t'+t)\,\text{d}t'.
\end{equation}
With the Green-Kubo formula, we can write for the ensemble average of the time-averaged squared displacement (EATASD)
\begin{equation}
\label{eq:Green_Kubo_formula_E}
\langle\langle\Delta\mathbf{x}^2(\tau)\rangle_{\text{T}}\rangle_{\text{E}}=2\int_0^{\tau}(\tau-t)\,\langle C_{\mathbf{v}}(t)\rangle_{\text{E}}\,\text{d}t.
\end{equation}
By taking the Laplace transform of Eq.~(\ref{eq:Green_Kubo_formula_E}) and by using the convolution theorem of the Laplace transform \cite{schiff1999}, we obtain
\begin{equation}
\label{eq:Green_Kubo_formula_E_LT}
\mathcal{L}\left\{\langle\langle\Delta\mathbf{x}^2(\tau)\rangle_{\text{T}}\rangle_{\text{E}};\tau,s\right\}=2\,\frac{1}{s^2}\,\mathcal{L}\left\{\langle C_{\mathbf{v}}(t)\rangle_{\text{E}};t,s\right\}.
\end{equation}
According to the definition in Eq.~(\ref{eq:C_v_t}), the ensemble average of the autocorrelation function $C_{\mathbf{v}}(t)$ of the velocity process defined as time average is given by
\begin{equation}
\label{eq:C_v_t_E}
\langle C_{\mathbf{v}}(t)\rangle_{\text{E}}\overset{T\gg t}{\simeq}\frac{1}{T}\int_0^T\langle\mathbf{v}(t')\mathbf{v}(t'+t)\rangle_{\text{E}}\,\text{d}t'.
\end{equation}
By using the properties of the Laplace transform \cite{schiff1999}, the Laplace transform of Eq.~(\ref{eq:C_v_t_E}) reads
\begin{equation}
\label{eq:C_v_t_E_LT}
\begin{split}
&\hspace{-3em}\mathcal{L}^2\left\{T\langle C_{\mathbf{v}}(t)\rangle_{\text{E}};t,s;T,u\right\}\\
&=\frac{1}{u}\,\mathcal{L}^2\left\{\langle\mathbf{v}(t')\mathbf{v}(t'+t)\rangle_{\text{E}};t,s;t',u\right\}.
\end{split}
\end{equation}
The expectation value on the right-hand side of Eq.~(\ref{eq:C_v_t_E_LT}) can be expressed by the probability density $p(\mathbf{v}',t';\mathbf{v},t'+t)$
that the velocity of the process is equal to $\mathbf{v}'$ at time $t'$ and equal to $\mathbf{v}$ at a later time $t'+t$,
\begin{equation}
\label{eq:v_v'_E_1}
\langle\mathbf{v}(t')\mathbf{v}(t'+t)\rangle_{\text{E}}=\int_{\mathbb{R}^d}\int_{\mathbb{R}^d}\mathbf{v}'\mathbf{v}\,p(\mathbf{v}',t';\mathbf{v},t'+t)\,\text{d}^d\mathbf{v}'\,\text{d}^d\mathbf{v},
\end{equation}
where $p(\mathbf{v}',t;\mathbf{v},t'+t)$ is normalized with respect to $\mathbf{v}'$ and $\mathbf{v}$.
For this probability density, we can write
\begin{equation}
\label{eq:p_v_v'}
\begin{split}
&p(\mathbf{v}',t';\mathbf{v},t'+t)\\
&=p^=(\mathbf{v};t',t'+t)\,\delta(\mathbf{v}-\mathbf{v}')+p^{\neq}(\mathbf{v}',t';\mathbf{v},t'+t),
\end{split}
\end{equation}
where $p^=(\mathbf{v};t',t'+t)$ denotes the probability density that the two instants of time $t'$ and $t'+t$ belong to the same flight of velocity $\mathbf{v}$,
and $p^{\neq}(\mathbf{v}',t';\mathbf{v},t'+t)$ denotes the probability density that the two instants of time belong to different flights of velocity $\mathbf{v}'$ and $\mathbf{v}$, respectively.
Note that the normalization condition for $p^=(\mathbf{v};t',t'+t)$ reads $\lim_{t\rightarrow0}\int p^=(\mathbf{v};t',t'+t)\,\text{d}^d\mathbf{v}=1$.
Due to the isotropy of the generalized L\'evy walk, i.e., a flight with velocity $\mathbf{v}$ has the same probability as a flight with velocity $-\mathbf{v}$,
\begin{equation}
\label{eq:p^neq_v_v'}
\begin{split}
&p^{\neq}(\mathbf{v}',t';\mathbf{v},t'+t)=p^{\neq}(\mathbf{v}',t';-\mathbf{v},t'+t)\\
&=p^{\neq}(-\mathbf{v}',t';\mathbf{v},t'+t)=p^{\neq}(-\mathbf{v}',t';-\mathbf{v},t'+t).
\end{split}
\end{equation}
By inserting Eq.~(\ref{eq:p_v_v'}) into Eq.~(\ref{eq:v_v'_E_1}) and by using Eq.~(\ref{eq:p^neq_v_v'}), we get
\begin{equation}
\label{eq:v_v'_E_2}
\begin{split}
\langle\mathbf{v}(t')\mathbf{v}(t'+t)\rangle_{\text{E}}&=\int_{\mathbb{R}^d}\mathbf{v}^2\,p^=(\mathbf{v};t',t'+t)\,\text{d}^d\mathbf{v}\\
&=\int_0^{\infty}|\mathbf{v}|^2\,p^=(|\mathbf{v}|;t',t'+t)\,\text{d}|\mathbf{v}|\\
&=\int_0^{\infty}c^2t_f^{2\nu-2}\,p^=(t_f;t',t'+t)\,\text{d}t_f,
\end{split}
\end{equation}
where we also used the deterministic dependence of the absolute value $|\mathbf{v}|$ of the flight velocities on the corresponding flight durations $t_f$.
Therefore, the calculation of the expectation value on the left-hand side of Eq.~(\ref{eq:v_v'_E_2}) reduces to the problem of finding an analytical expression for the probability density $p^=(t_f;t',t'+t)$
that the two instants of time $t'$ and $t'+t$ belong to the same flight of duration $t_f$.
An appropriate ansatz for solving this problem is
\begin{equation}
\label{eq:p^eq_t_f_t_t'}
p^=(t_f;t',t'+t)=\sum\limits_{n=1}^{\infty}p^=_n(t_f;t',t'+t),
\end{equation}
where $p^=_n(t_f;t',t'+t)$ denotes the probability density that the two instants of time belong to the $n^{\text{th}}$ flight of duration $t_f$.
In order to find an analytical expression for $p^=_n(t_f;t',t'+t)$, we use methods introduced by Godr\`eche and Luck in \cite{godreche2001}, where similar distributions of renewal processes were investigated
(these methods were also applied, e.g., in \cite{barkai2007}).
To do so, we write
\begin{equation}
\label{eq:p^eq_n_t_f_t_t'}
\begin{split}
p^=_n(t_f,t',t'+t)=\left\langle\delta(t_f-T_n)\,I\left(\sum\limits_{i=1}^{n-1}T_i<t'<\sum\limits_{i=1}^nT_i\right)\right.\\
\left.\times\,I\left(\sum\limits_{i=1}^{n-1}T_i<t'+t<\sum\limits_{i=1}^nT_i\right)\right\rangle_{\text{E}},
\end{split}
\end{equation}
where $I(\dots)$ is a so-called indicator function, which is equal to unity if the condition in parentheses is true and zero otherwise.
$T_i$ denotes the duration of the $i^{\text{th}}$ flight of a trajectory of the generalized L\'evy walk.
The indicator functions in Eq.~(\ref{eq:p^eq_n_t_f_t_t'}) account for the above-mentioned condition that the two instants of time $t'$ and $t'+t$ belong to the $n^{\text{th}}$ flight.
As usual, $\langle\dots\rangle_{\text{E}}$ denotes an ensemble average over all possible trajectories of the generalized L\'evy walk, i.e., an average over all possible sequences of flight durations,
\begin{equation}
\label{eq:ensemble_average}
\langle\dots\rangle_{\text{E}}=\int_0^{\infty}\dotsi\int_0^{\infty}\dots\,\psi(t_1)\dotsm\psi(t_n)\,\text{d}t_1\dotsm\text{d}t_n.
\end{equation}
A Laplace transform of $p^=_n(t_f;t',t'+t)$ with respect to $t$ can easily be calculated by using the definition of the indicator function,
\begin{equation}
\label{eq:p^eq_n_t_f_s_t'}
\begin{split}
&\mathcal{L}\left\{p^=_n(t_f,t',t'+t);t,s\right\}=\Biggl\langle\delta(t_f-T_n)\\
&\left.\times\,I\left(\sum\limits_{i=1}^{n-1}T_i<t'<\sum\limits_{i=1}^nT_i\right)\frac{1}{s}\left(1-e^{-s\left(\sum\limits_{i=1}^nT_i-t'\right)}\right)\right\rangle_{\text{E}}.
\end{split}
\end{equation}
A further Laplace transform with respect to $t'$ results in
\begin{equation}
\label{eq:p^eq_n_t_f_s_u}
\begin{split}
&\mathcal{L}^2\left\{p^=_n(t_f,t',t'+t);t,s;t',u\right\}=\\[1ex]
&\frac{1}{s}\left\langle\delta(t_f-T_n)\int_{\sum\limits_{i=1}^{n-1}T_i}^{\sum\limits_{i=1}^nT_i}\left(1-e^{-s\left(\sum\limits_{i=1}^nT_i-t'\right)}\right)e^{-ut'}\,\text{d}t'\right\rangle_{\text{E}}.
\end{split}
\end{equation}
The integral on the right-hand side of Eq.~(\ref{eq:p^eq_n_t_f_s_u}) can straightforwardly be calculated,
\begin{equation}
\label{eq:integral}
\begin{split}
&\int_{\sum\limits_{i=1}^{n-1}T_i}^{\sum\limits_{i=1}^nT_i}\left(1-e^{-s\left(\sum\limits_{i=1}^nT_i-t'\right)}\right)e^{-ut'}\,\text{d}t'\\[1ex]
&=\frac{1}{u}\left(e^{-u\sum\limits_{i=1}^{n-1}T_i}-e^{-u\sum\limits_{i=1}^nT_i}\right)\\
&\quad-\frac{1}{s-u}e^{-s\sum\limits_{i=1}^nT_i}\left(e^{(s-u)\sum\limits_{i=1}^nT_i}-e^{(s-u)\sum\limits_{i=1}^{n-1}T_i}\right)\\[1ex]
&=\left(\frac{1}{u}+\frac{1}{s-u}e^{-sT_n}\right)e^{-u\sum\limits_{i=1}^{n-1}T_i}+\frac{s}{u(u-s)}e^{-u\sum\limits_{i=1}^nT_i}.
\end{split}
\end{equation}
By inserting Eq.~(\ref{eq:integral}) into Eq.~(\ref{eq:p^eq_n_t_f_s_u}) and by combining the result with the Laplace transform of Eq.~(\ref{eq:v_v'_E_2}) and Eq.~(\ref{eq:p^eq_t_f_t_t'}), we obtain
\begin{equation}
\label{eq:v_v'_E_LT_1}
\begin{split}
&\mathcal{L}^2\left\{\langle\mathbf{v}(t')\mathbf{v}(t'+t)\rangle_{\text{E}};t,s;t',u\right\}\\[1ex]
&=\int_0^{\infty}c^2t_f^{2\nu-2}\,\mathcal{L}^2\left\{p^=(t_f;t',t'+t);t,s;t',u\right\}\,\text{d}t_f\\[1ex]
&=\sum\limits_{n=1}^{\infty}\int_0^{\infty}c^2t_f^{2\nu-2}\,\mathcal{L}^2\left\{p^=_n(t_f;t',t'+t);t,s;t',u\right\}\,\text{d}t_f\\[1ex]
&=\sum\limits_{n=1}^{\infty}\int_0^{\infty}c^2t_f^{2\nu-2}\,\frac{1}{s}\left\langle\delta(t_f-T_n)\left[\left(\frac{1}{u}+\frac{1}{s-u}e^{-sT_n}\right)\right.\right.\\
&\hspace{6em}\times\left.\left.e^{-u\sum\limits_{i=1}^{n-1}T_i}+\frac{s}{u(u-s)}e^{-u\sum\limits_{i=1}^nT_i}\right]\right\rangle_{\text{E}}\text{d}t_f\\[1ex]
&=\sum\limits_{n=1}^{\infty}\frac{1}{s}\left\langle c^2T_n^{2\nu-2}\left[\left(\frac{1}{u}+\frac{1}{s-u}e^{-sT_n}\right)e^{-u\sum\limits_{i=1}^{n-1}T_i}\right.\right.\\
&\hspace{10em}+\left.\left.\frac{s}{u(u-s)}e^{-u\sum\limits_{i=1}^nT_i}\right]\right\rangle_{\text{E}}.
\end{split}
\end{equation}
We now can evaluate the ensemble average in Eq.~(\ref{eq:v_v'_E_LT_1}) by using Eq.~(\ref{eq:ensemble_average}) and the definition of the Laplace transform,
\begin{equation}
\label{eq:v_v'_E_LT_2}
\begin{split}
&\mathcal{L}^2\left\{\langle\mathbf{v}(t')\mathbf{v}(t'+t)\rangle_{\text{E}};t,s;t',u\right\}=\\[1ex]
&c^2\sum\limits_{n=1}^{\infty}\left[\frac{\langle T^{2\nu-2}\rangle\psi^{n-1}(u)}{su}+\frac{\mathcal{L}\left\{t^{2\nu-2}\psi(t);t,s\right\}\psi^{n-1}(u)}{s(s-u)}\right.\\
&\hspace{6em}+\left.\frac{\mathcal{L}\left\{t^{2\nu-2}\psi(t);t,u\right\}\psi^{n-1}(u)}{u(u-s)}\right].
\end{split}
\end{equation}
The sum in Eq.~(\ref{eq:v_v'_E_LT_2}) can be calculated by using the geometric series,
\begin{equation}
\label{eq:v_v'_E_LT_3}
\begin{split}
&\mathcal{L}^2\left\{\langle\mathbf{v}(t')\mathbf{v}(t'+t)\rangle_{\text{E}};t,s;t',u\right\}=\frac{c^2}{1-\psi(u)}\times\\[1ex]
&\left[\frac{\langle T^{2\nu-2}\rangle}{su}+\frac{\mathcal{L}\left\{t^{2\nu-2}\psi(t);t,s\right\}}{s(s-u)}+\frac{\mathcal{L}\left\{t^{2\nu-2}\psi(t);t,u\right\}}{u(u-s)}\right].
\end{split}
\end{equation}
Note that this equation for the special case $\nu=1$ has already been found for the standard L\'evy walk with constant flight velocity in \cite{froemberg2013_2}.
By inserting Eq.~(\ref{eq:v_v'_E_LT_3}) into Eq.~(\ref{eq:C_v_t_E_LT}) and combining the result with Eq.~(\ref{eq:Green_Kubo_formula_E_LT}),
we obtain the double Laplace transform of $T\langle\langle\Delta\mathbf{x}^2(\tau)\rangle_{\text{T}}\rangle_{\text{E}}$,
\begin{equation}
\label{eq:MSD_T_E_s_u}
\begin{split}
&\mathcal{L}^2\left\{T\langle\langle\Delta\mathbf{x}^2(\tau)\rangle_{\text{T}}\rangle_{\text{E}};\tau,s;T,u\right\}=\frac{2c^2}{s^2u(1-\psi(u))}\times\\[1ex]
&\left[\frac{\langle T^{2\nu-2}\rangle}{su}+\frac{\mathcal{L}\left\{t^{2\nu-2}\psi(t);t,s\right\}}{s(s-u)}+\frac{\mathcal{L}\left\{t^{2\nu-2}\psi(t);t,u\right\}}{u(u-s)}\right].
\end{split}
\end{equation}
The fractional moment appearing on the right-hand side of Eq.~(\ref{eq:MSD_T_E_s_u}) can be calculated by using the definition of the beta function and its connection to the gamma function,
\begin{equation}
\label{eq:T^2v-2_E}
\begin{split}
&\langle T^{2\nu-2}\rangle=\int_0^{\infty}t^{2\nu-2}\,\psi(t)\,\text{d}t\\[1ex]
&\hspace{1em}=\frac{\Gamma(2\nu-1)\Gamma(\gamma+2-2\nu)}{\Gamma(\gamma)}t_0^{2\nu-2}\quad\text{if}\quad1<2\nu<\gamma+2.
\end{split}
\end{equation}
The small-$s$ behavior of the Laplace transform $\mathcal{L}\left\{t^{2\nu-2}\psi(t);t,s\right\}$ can be obtained by using the Cauchy-Saalsch\"utz representation of the gamma function,
\begin{equation}
\label{eq:t^2v-2_psi_t_LT}
\begin{split}
&\mathcal{L}\left\{t^{2\nu-2}\psi(t);t,s\right\}\overset{s\rightarrow0}{\simeq}\langle T^{2\nu-2}\rangle-\langle T^{2\nu-1}\rangle s+\frac{1}{2}\langle T^{2\nu}\rangle s^2\\[1ex]
&\hspace{11em}+\gamma\Gamma(2\nu-\gamma-2)t_0^{\gamma}s^{\gamma+2-2\nu},\\[2ex]
&\langle T^{2\nu-1}\rangle=\frac{\Gamma(2\nu)\Gamma(\gamma+1-2\nu)}{\Gamma(\gamma)}t_0^{2\nu-1},\\[1ex]
&\langle T^{2\nu}\rangle=\frac{\Gamma(2\nu+1)\Gamma(\gamma-2\nu)}{\Gamma(\gamma)}t_0^{2\nu}.
\end{split}
\end{equation}
The fractional moment in Eq.~(\ref{eq:T^2v-2_E}) diverges for $2\nu\geq\gamma+2$.
Therefore, also the EATASD, Eq.~(\ref{eq:MSD_T_E_s_u}), diverges under the same condition,
\begin{equation}
\label{eq:LW_MSD_T_E_tau_divergence}
\langle\langle\Delta\mathbf{x}^2(\tau)\rangle_{\text{T}}\rangle_{\text{E}}=\infty\quad\text{if}\quad2\nu\geq\gamma+2.
\end{equation}
Interestingly, the fractional moment in Eq.~(\ref{eq:T^2v-2_E}) also diverges for $2\nu\leq1$.
This divergence, however, is compensated by the Laplace transform $\mathcal{L}\left\{t^{2\nu-2}\psi(t);t,s\right\}=\int_0^{\infty}t^{2\nu-2}\,\psi(t)\,e^{-st}\,\text{d}t$,
which contains exactly the same divergence at the lower bound of integration.
In order to see that these divergences cancel out, one only has to bring the fractions in the square brackets on the right-hand side of Eq.~(\ref{eq:MSD_T_E_s_u}) to the common denominator.
Therefore, for $2\nu<\gamma+2$, the EATASD is finite.
By inserting Eq.~(\ref{eq:T^2v-2_E}) and Eq.~(\ref{eq:t^2v-2_psi_t_LT}) into Eq.~(\ref{eq:MSD_T_E_s_u}),
we obtain the small-$s$ and the small-$u$ behavior of the Laplace transform of $T\langle\langle\Delta\mathbf{x}^2(\tau)\rangle_{\text{T}}\rangle_{\text{E}}$,
from which the large-$\tau$ and the large-$T$ behavior of $\langle\langle\Delta\mathbf{x}^2(\tau)\rangle_{\text{T}}\rangle_{\text{E}}$ can be calculated by an inverse Laplace transform.
Additionally, we consider the asymptotic behavior for $s\gg u$ in the Laplace space, which corresponds to the asymptotic behavior for $T\gg\tau$ in the original space.
For $0<\gamma<1$, we get
\begin{widetext}
\begin{equation}
\label{eq:MSD_T_E_s_u_1}
\begin{split}
&\hspace{-3em}\mathcal{L}^2\left\{T\langle\langle\Delta\mathbf{x}^2(\tau)\rangle_{\text{T}}\rangle_{\text{E}};\tau,s;T,u\right\}\overset{(s,u)\rightarrow(0,0)}{\simeq}\\[1ex]
&\begin{cases}
\frac{c^2\langle t^{2\nu}\rangle_{\psi(t)}}{\Gamma(1-\gamma)t_0^{\gamma}}s^{-2}u^{-\gamma-1},\,&0<2\nu<\gamma\\[1ex]
\frac{2c^2\gamma\Gamma(2\nu-\gamma-2)}{\Gamma(1-\gamma)s^2u^{\gamma+1}}\frac{s^{\gamma+1-2\nu}-u^{\gamma+1-2\nu}}{s-u}\overset{s\gg u}{\simeq}\frac{2c^2\gamma\Gamma(2\nu-\gamma-2)}{\Gamma(1-\gamma)}s^{\gamma-2\nu-2}u^{-\gamma-1},\,&\gamma<2\nu<\gamma+1\\[1ex]
\frac{2c^2\gamma\Gamma(2\nu-\gamma-2)}{\Gamma(1-\gamma)s^2u^{\gamma+1}}\frac{s^{\gamma+1-2\nu}-u^{\gamma+1-2\nu}}{s-u}\overset{s\gg u}{\simeq}-\frac{2c^2\gamma\Gamma(2\nu-\gamma-2)}{\Gamma(1-\gamma)}s^{-3}u^{-2\nu},\,&\gamma+1<2\nu<\gamma+2
\end{cases}.
\end{split}
\end{equation}
After an inverse Laplace transform of Eq.~(\ref{eq:MSD_T_E_s_u_1}), we obtain
\begin{equation}
\label{eq:LW_MSD_T_E_tau_1}
\begin{split}
\langle\langle\Delta\mathbf{x}^2(\tau)\rangle_{\text{T}}\rangle_{\text{E}}\overset{1\ll\tau\ll T}{\simeq}
\begin{cases}
c^2\frac{\Gamma(2\nu+1)\Gamma(\gamma-2\nu)}{\Gamma(1-\gamma)\Gamma(\gamma)\Gamma(\gamma+1)}t_0^{-\gamma}T^{\gamma-1}\tau,\,&0<2\nu<\gamma\\[1ex]
2c^2\frac{\Gamma(2\nu-\gamma-2)}{\Gamma(1-\gamma)\Gamma(\gamma)\Gamma(2\nu+2-\gamma)}T^{\gamma-1}\tau^{2\nu+1-\gamma},\,&\gamma<2\nu<\gamma+1\\[1ex]
-c^2\gamma\frac{\Gamma(2\nu-\gamma-2)}{\Gamma(1-\gamma)\Gamma(2\nu)}T^{2\nu-2}\tau^2,\,&\gamma+1<2\nu<\gamma+2
\end{cases}.
\end{split}
\end{equation}
Accordingly, for $\gamma>1$, the Laplace transform of $T\langle\langle\Delta\mathbf{x}^2(\tau)\rangle_{\text{T}}\rangle_{\text{E}}$ reads
\begin{equation}
\label{eq:MSD_T_E_s_u_2}
\begin{split}
&\hspace{-3em}\mathcal{L}^2\left\{T\langle\langle\Delta\mathbf{x}^2(\tau)\rangle_{\text{T}}\rangle_{\text{E}};\tau,s;T,u\right\}\overset{(s,u)\rightarrow(0,0)}{\simeq}\\[1ex]
&\begin{cases}
\frac{c^2(\gamma-1)\langle t^{2\nu}\rangle_{\psi(t)}}{t_0}s^{-2}u^{-2},\,&0<2\nu<\gamma\\[1ex]
\frac{2c^2\gamma(\gamma-1)\Gamma(2\nu-\gamma-2)t_0^{\gamma-1}}{s^2u^2}\frac{s^{\gamma+1-2\nu}-u^{\gamma+1-2\nu}}{s-u}\overset{s\gg u}{\simeq}\frac{2c^2\gamma(\gamma-1)\Gamma(2\nu-\gamma-2)}{t_0^{1-\gamma}s^{2\nu+2-\gamma}u^2},\,&\gamma<2\nu<\gamma+1\\[1ex]
\frac{2c^2\gamma(\gamma-1)\Gamma(2\nu-\gamma-2)t_0^{\gamma-1}}{s^2u^2}\frac{s^{\gamma+1-2\nu}-u^{\gamma+1-2\nu}}{s-u}\overset{s\gg u}{\simeq}-\frac{2c^2\gamma(\gamma-1)\Gamma(2\nu-\gamma-2)}{t_0^{1-\gamma}s^3u^{2\nu+1-\gamma}},\,&\gamma+1<2\nu<\gamma+2
\end{cases},
\end{split}
\end{equation}
and in the time domain, we get
\begin{equation}
\label{eq:LW_MSD_T_E_tau_2}
\begin{split}
\langle\langle\Delta\mathbf{x}^2(\tau)\rangle_{\text{T}}\rangle_{\text{E}}\overset{1\ll\tau\ll T}{\simeq}
\begin{cases}
c^2\frac{\Gamma(2\nu+1)\Gamma(\gamma-2\nu)}{\Gamma(\gamma-1)}t_0^{2\nu-1}\tau,\,&0<2\nu<\gamma\\[1ex]
2c^2\gamma(\gamma-1)\frac{\Gamma(2\nu-\gamma-2)}{\Gamma(2\nu+2-\gamma)}t_0^{\gamma-1}\tau^{2\nu+1-\gamma},\,&\gamma<2\nu<\gamma+1\\[1ex]
-c^2\gamma(\gamma-1)\frac{\Gamma(2\nu-\gamma-2)}{\Gamma(2\nu+1-\gamma)}t_0^{\gamma-1}T^{2\nu-\gamma-1}\tau^2,\,&\gamma+1<2\nu<\gamma+2
\end{cases}.
\end{split}
\end{equation}
\end{widetext}
The $T$- and $\tau$-dependencies in Eq.~(\ref{eq:LW_MSD_T_E_tau_1}) and Eq.~(\ref{eq:LW_MSD_T_E_tau_2}) are the ones of Fig.~\ref{fig:phase_diagrams} (b).

\section{\label{sec:C}Derivation of the contribution of the completed flights to the ensemble average of the time-averaged squared displacement (EATASD)}

The derivations in Appendices \ref{sec:C} and \ref{sec:D} provide a simpler, more intuitive understanding of the results given in Eq.~(\ref{eq:LW_MSD_T_E_tau_1}) and Eq.~(\ref{eq:LW_MSD_T_E_tau_2}), respectively.
In this appendix, we derive the contribution of the completed flights of the generalized L\'evy walk to the ensemble average of the time-averaged squared displacement (EATASD).
This contribution corresponds to the first summand on the right-hand side of Eq.~(\ref{eq:EATASD_contributions}).
In order to calculate the expectation value, we need the joint probability distribution $p_N(t_1,\dots,t_N;T)$ for the occurrence of $N$ completed flights up to measurement time $T$ with durations $t_1$ to $t_N$.
This distribution was derived in \cite{godreche2001} and captures the full statistical information about the number and the durations of the completed flights,
\begin{equation}
\label{eq:joint_probability_density}
\begin{split}
&p_N(t_1,\dots,t_N;T)\\
&=\prod\limits_{i=1}^N\psi(t_i)\hspace{0em}\int\limits_{T-\sum\limits_{j=1}^Nt_j}^{\infty}\hspace{-1em}\psi(t')\,\text{d}t'\hspace{1em}\Theta\left(T-\sum\limits_{j=1}^Nt_j\right),
\end{split}
\end{equation}
where $\Theta(\dots)$ denotes the Heaviside step function.
In the following, we concentrate on the contribution to the EATASD coming from flights of durations $T_i$ longer than the time-lag $\tau$
because it can be expected that the long flights mainly determine the asymptotic time-lag dependence of the EATASD.
If the time window $[t,t+\tau]$ lies completely within these flights, the corresponding squared displacement is given by $[cT_i^{\nu-1}\tau]^2$.
Therefore, we can write for the time integral over such a flight,
\begin{equation}
\label{eq:time_integral_1}
\int_{t_{i-1}}^{t_i-\tau}[\mathbf{x}(t+\tau)-\mathbf{x}(t)]^2\,\text{d}t=[cT_i^{\nu-1}\tau]^2(T_i-\tau)=:f(T_i),
\end{equation}
which we abbreviate as $f(T_i)$ in the following.
The expectation value of the random sum of these time integrals can now be calculated with the distribution in Eq.~(\ref{eq:joint_probability_density}),
\begin{widetext}
\begin{equation}
\label{eq:EATASD_contribution_1_1}
\begin{split}
\left\langle\sum\limits_{i=1}^{N_T}f(T_i)\right\rangle_{\text{E}}&=\sum\limits_{N=1}^{\infty}\left[\int_{\tau}^{\infty}\dotsi\int_{\tau}^{\infty}\left(\sum\limits_{i=1}^Nf(t_i)\right)p_N(t_1,\dots,t_N;T)\,\text{d}t_1\dotsm\text{d}t_N\right]\\
&=\sum\limits_{N=1}^{\infty}N\left[\int_{\tau}^{\infty}\dotsi\int_{\tau}^{\infty}f(t_1)\,p_N(t_1,\dots,t_N;T)\,\text{d}t_1\dotsm\text{d}t_N\right]\\
&=\sum\limits_{N=1}^{\infty}N\left[\int_{\tau}^{\infty}f(t_1)\,p_N(t_1;T)\,\text{d}t_1\right].
\end{split}
\end{equation}
\end{widetext}
In the second line of this calculation, we used that the $N$-fold integral of each summand $f(t_i)$ gives the same result due to the symmetry of the distribution $p_N(t_1,\dots,t_N;T)$ in Eq.~(\ref{eq:joint_probability_density}).
The integration over durations $t_2$ to $t_N$ of this distribution leads to the marginal distribution $p_N(t_1;T)$ of having $N$ completed flights up to measurement time $T$ with the duration of the first flight equal to $t_1$.
This marginal distribution was also derived in \cite{godreche2001},
\begin{equation}
\label{eq:p_N_t1_T}
p_N(t_1;T)=\psi(t_1)\,p_{N-1}(T-t_1)\,\Theta(T-t_1),
\end{equation}
where $p_{N-1}(T-t_1)$ denotes the probability for the occurrence of $N-1$ completed flights in the time $T-t_1$.
Inserting Eq.~(\ref{eq:p_N_t1_T}) into the third line of Eq.~(\ref{eq:EATASD_contribution_1_1}) results in
\begin{equation}
\label{eq:EATASD_contribution_1_2}
\begin{split}
&\left\langle\sum\limits_{i=1}^{N_T}f(T_i)\right\rangle_{\text{E}}\\
&=\sum\limits_{N=1}^{\infty}N\left[\int_{\tau}^Tf(t_1)\,\psi(t_1)\,p_{N-1}(T-t_1)\,\text{d}t_1\right]\\
&=\sum\limits_{N=0}^{\infty}(N+1)\left[\int_{\tau}^Tf(t_1)\,\psi(t_1)\,p_N(T-t_1)\,\text{d}t_1\right]\\
&=\int_{\tau}^Tf(t_1)\,\psi(t_1)\,\langle N_{T-t_1}\rangle_{\text{E}}\,\text{d}t_1+\int_{\tau}^Tf(t_1)\,\psi(t_1)\,\text{d}t_1.
\end{split}
\end{equation}
An asymptotic analysis shows that the first integral $J$ in the third line of Eq.~(\ref{eq:EATASD_contribution_1_2}) dominates the second one.

In order to calculate the asymptotic behavior of the first integral, we need the asymptotic behavior of the mean number $\langle N_T\rangle_{\text{E}}$ of completed flights up to measurement time $T$.
From the renewal theory \cite{godreche2001}, it is well known that the Laplace transform of the mean value $\langle N_T\rangle_{\text{E}}$ is given by
\begin{equation}
\label{eq:N_s}
\mathcal{L}\{\langle N_T\rangle_{\text{E}};T,s\}=\frac{\psi(s)}{s[1-\psi(s)]}.
\end{equation}
By using the small-$s$ behavior of the Laplace transform $\psi(s)$ of the flight time distribution in Eq.~(\ref{eq:psi_s_1}), we get
\begin{equation}
\label{eq:N_T}
\langle N_T\rangle_{\text{E}}\overset{T\rightarrow\infty}{\simeq}\begin{cases}\frac{T^{\gamma}}{\Gamma(1-\gamma)\Gamma(1+\gamma)t_0^{\gamma}},\,&\gamma<1\\[1ex]\frac{\gamma-1}{t_0}T,\,&\gamma>1\end{cases}.
\end{equation}
There are simple interpretations for these results.
For $\gamma>1$, where the mean flight duration is finite, the mean number of completed flights should be asymptotically equal to the total measurement time divided by the mean flight duration, i.e.,
$\langle N_T\rangle_{\text{E}}\simeq T/\langle T_i\rangle=(\gamma-1)/t_0\,T$.
For $\gamma<1$, where the mean flight duration diverges, one can use the generalized central limit theorem, i.e.,
$T\approx\sum_{i=1}^{N_T}T_i\sim N_T^{1/\gamma}\Rightarrow N_T\sim T^{\gamma}$.

Inserting the definition in Eq.~(\ref{eq:time_integral_1}), the asymptotic behavior of the flight time distribution $\psi(t)\sim t^{-\gamma-1}$,
and the asymptotic behavior of $\langle N_T\rangle_{\text{E}}$ for $\gamma<1$ into the first integral in the third line of Eq.~(\ref{eq:EATASD_contribution_1_2}) leads to
\begin{equation}
\label{eq:I_1_1}
J\sim\tau^2\int_{\tau}^Tt_1^{2\nu-\gamma-3}\,(t_1-\tau)\,(T-t_1)^{\gamma}\,\text{d}t_1.
\end{equation}
The asymptotic behavior of this integral for $T\gg\tau$ is given by
\begin{equation}
\label{eq:I_1_1_asymptotic}
J\overset{T\gg\tau}{\sim}\begin{cases}T^{\gamma}\,\tau^{1+2\nu-\gamma},\,&2\nu<\gamma+1\\T^{2\nu-1}\,\tau^2,\,&2\nu>\gamma+1\end{cases}.
\end{equation}
Dividing these asymptotics by the measurement time $T$ according to Eq.~(\ref{eq:EATASD_contributions}) results in the correct asymptotic behavior of the EATASD for $0<\gamma<1$
(see Fig.~\ref{fig:phase_diagrams} (b) and Eq.~(\ref{eq:LW_MSD_T_E_tau_1})).
For $\gamma>1$, we can write for the first integral in the third line of Eq.~(\ref{eq:EATASD_contribution_1_2})
\begin{equation}
\label{eq:I_1_2}
J\sim\tau^2\int_{\tau}^Tt_1^{2\nu-\gamma-3}\,(t_1-\tau)\,(T-t_1)\,\text{d}t_1.
\end{equation}
The asymptotic behavior for $T\gg\tau$ reads
\begin{equation}
\label{eq:I_1_2_asymptotic}
J\overset{T\gg\tau}{\sim}\begin{cases}T\,\tau^{1+2\nu-\gamma},\,&2\nu<\gamma+1\\T^{2\nu-\gamma}\,\tau^2,\,&2\nu>\gamma+1\end{cases}.
\end{equation}
Again, dividing these asymptotics by the measurement time $T$ leads to the correct asymptotic behavior of the EATASD for $\gamma>1$ (see Fig.~\ref{fig:phase_diagrams} (b) and Eq.~(\ref{eq:LW_MSD_T_E_tau_2})).

We would like to point out that one also has to evaluate the contributions to the EATASD where the time window $[t,t+\tau]$ contains turning points marking the end of the flights and the beginning of the next flights
and the related contribution coming from flights of duration smaller than $\tau$.
Because from the random walk theory it is known that a sequence of short flights leads to normal diffusion with a linear time-lag dependence of the mean-squared displacement,
it can be expected that these contributions to the EATASD lead to the normal diffusive behavior observed for $2\nu<\gamma$ in the phase diagram of the EATASD in Fig.~\ref{fig:phase_diagrams} (b)
as given by Eq.~(\ref{eq:LW_MSD_T_E_tau_1}) and Eq.~(\ref{eq:LW_MSD_T_E_tau_2}).
Note that for $2\nu<\gamma$, a linear time-lag dependence dominates the asymptotic behavior found for $2\nu<\gamma+1$ in Eq.~(\ref{eq:I_1_1_asymptotic}) and Eq.~(\ref{eq:I_1_2_asymptotic}).

\section{\label{sec:D}Derivation of the contribution of the last incomplete flight to the ensemble average of the time-averaged squared displacement (EATASD)}

\begin{figure}
\includegraphics[width=\linewidth]{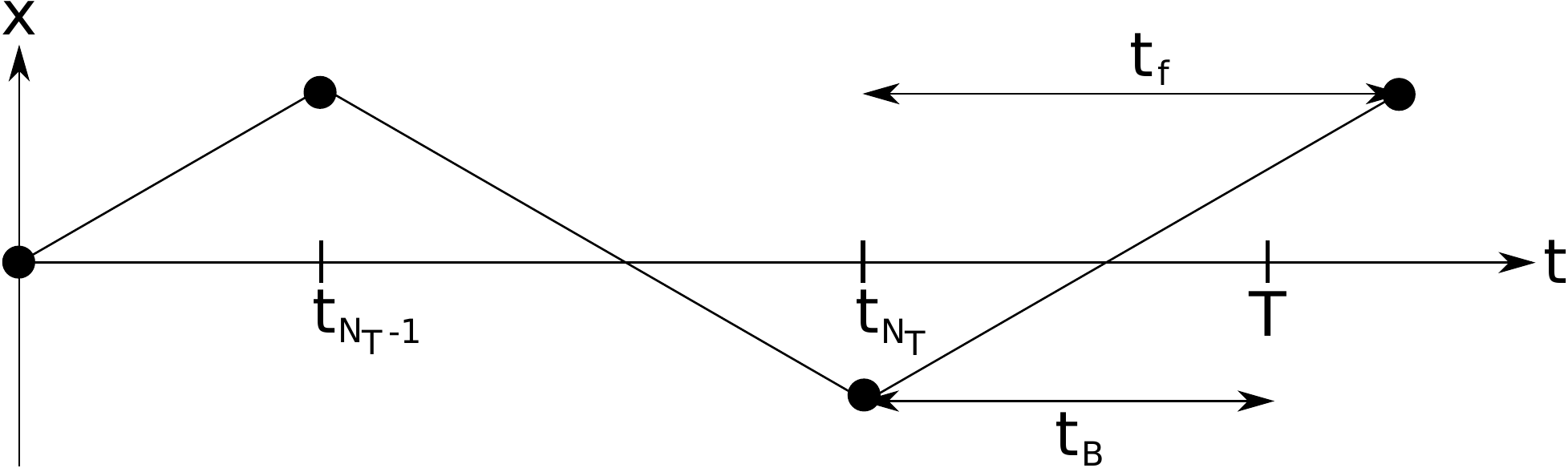}
\caption{\label{fig:explanation2}
Schematic representation of one realization of a generalized L\'evy walk with the current flight time $t_f$ and the backward recurrence time $t_B$ with respect to the instant of time $T$.}
\end{figure}

In this appendix, we calculate the contribution of the last incomplete flight of a generalized L\'evy walk to the ensemble average of the time-averaged squared displacement (EATASD) in Eq.~(\ref{eq:EATASD_contributions}).
If the duration $t_f$ of the last incomplete flight is larger than the time lag $\tau$, a squared displacement within this flight is given by $[\mathbf{x}(t+\tau)-\mathbf{x}(t)]^2=[ct_f^{\nu-1}\tau]^2$.
The corresponding time integral over the last incomplete flight reads
\begin{equation}
\label{eq:time_integral_2}
\int_{t_{N_T}}^T[\mathbf{x}(t+\tau)-\mathbf{x}(t)]^2\,\text{d}t=[ct_f^{\nu-1}\tau]^2\,t_B,
\end{equation}
where $t_B$ is the so-called backward recurrence time, see Fig.~\ref{fig:explanation2}.
The average over all possible last incomplete flights longer than the time lag $\tau$ gives
\begin{equation}
\label{eq:EATASD_contribution_2}
\begin{split}
&\left\langle\int_{t_{N_T}}^T[\mathbf{x}(t+\tau)-\mathbf{x}(t)]^2\,\text{d}t\right\rangle_{\text{E}}\\
&=\int_{\tau}^T\int_{t_B}^{\infty}[ct_f^{\nu-1}\tau]^2\,t_B\,p_T(t_f,t_B)\,\text{d}t_f\,\text{d}t_B,
\end{split}
\end{equation}
where $p_T(t_f,t_B)$ is the probability density for the occurrence of a flight of duration $t_f$ with backward recurrence time $t_B$ at instant of time $T$.
Note that, of course, the duration $t_f$ of the last incomplete flight must be larger than the backward recurrence time $t_B$, which cannot be longer than the measurement time $T$ by definition.

In order to continue the calculation, we need the distribution $p_T(t_f,t_B)$ whose triple Laplace transform is derived at the end of this appendix,
\begin{equation}
\label{eq:p_z_s_u_2}
p_T(t_f,t_B)\,\overset{\mathcal{L}^3}{\longrightarrow}\,p_z(s,u)=\frac{1}{u+z}\,\frac{\psi(s)-\psi(s+u+z)}{1-\psi(z)}.
\end{equation}
From this equation one can obtain, for instance, the double Laplace transform of the distribution $p_T(t_B)$ of the backward recurrence time, which is well known from renewal theory \cite{godreche2001},
\begin{equation}
\label{eq:p_T_tB}
\begin{split}
p_T(t_B)&=\int_0^{\infty}p_T(t_f,t_B)\,\text{d}t_f\\
\overset{\mathcal{L}^2}{\longrightarrow}\,p_z(u)&=\lim\limits_{s\rightarrow0}p_z(s,u)=\frac{1}{u+z}\,\frac{1-\psi(u+z)}{1-\psi(z)}.
\end{split}
\end{equation}
Performing the inverse Laplace transform of Eq.~(\ref{eq:p_z_s_u_2}) with respect to the Laplace variable $s$ results in
\begin{equation}
\label{eq:p_z_tf_u}
p_z(s,u)\,\underset{s\rightarrow t_f}{\overset{\mathcal{L}^{-1}}{\longrightarrow}}\,p_z(t_f,u)=\psi(t_f)\,\frac{1-e^{-(u+z)t_f}}{(u+z)[1-\psi(z)]}.
\end{equation}
Another inverse Laplace transform with respect to the Laplace variable $u$ leads to
\begin{equation}
\label{eq:p_z_tf_tB}
p_z(t_f,u)\,\underset{u\rightarrow t_B}{\overset{\mathcal{L}^{-1}}{\longrightarrow}}\,p_z(t_f,t_B)=\psi(t_f)\,\frac{e^{-zt_B}}{1-\psi(z)}\,\Theta(t_f-t_B),
\end{equation}
where $\Theta(\dots)$ is the Heaviside step function.
Performing the last inverse Laplace transform with respect to the Laplace variable $z$ yields the distribution $p_T(t_f,t_B)$ in the time domain,
\begin{equation}
\label{eq:p_T_tf_tB_2}
\begin{split}
&p_z(t_f,t_B)\,\underset{z\rightarrow T}{\overset{\mathcal{L}^{-1}}{\longrightarrow}}\,p_T(t_f,t_B)=\\
&\psi(t_f)\,\mathcal{L}^{-1}\left[\frac{1}{1-\psi(z)};z,T-t_B\right]\,\Theta(T-t_B)\,\Theta(t_f-t_B).
\end{split}
\end{equation}
This result can be inserted in Eq.~(\ref{eq:EATASD_contribution_2}) in order to obtain the contribution of the last incomplete flight to the EATASD according to Eq.~(\ref{eq:EATASD_contributions}) of the main text,
\begin{equation}
\label{eq:EATASD_contribution_2_0}
\begin{split}
&\frac{1}{T}\left\langle\int_{t_{N_T}}^T[\mathbf{x}(t+\tau)-\mathbf{x}(t)]^2\,\text{d}t\right\rangle_{\text{E}}\\
&=\frac{1}{T}\int_{\tau}^T\int_{t_B}^{\infty}[ct_f^{\nu-1}\tau]^2\,t_B\,\psi(t_f)\\
&\hspace{8em}\times\mathcal{L}^{-1}\left[\frac{1}{1-\psi(z)};z,T-t_B\right]\,\text{d}t_f\,\text{d}t_B\\
&\sim\frac{\tau^2}{T}\int_{\tau}^Tt_B^{2\nu-\gamma-1}\,\mathcal{L}^{-1}\left[\frac{1}{1-\psi(z)};z,T-t_B\right]\,\text{d}t_B.
\end{split}
\end{equation}
Interestingly, the inner integral in the second line of Eq.~(\ref{eq:EATASD_contribution_2_0}) diverges for $2\nu\geq\gamma+2$ providing again the correct condition for the divergence of the EATASD.
In order to evaluate the inverse Laplace transform in the last line of Eq.~(\ref{eq:EATASD_contribution_2_0}), we use the asymptotic behavior of the Laplace transform of the distribution of the flight durations in Eq.~(\ref{eq:psi_s_1}).
For $\gamma<1$, we obtain
\begin{equation}
\label{eq:EATASD_contribution_2_1}
\sim\frac{\tau^2}{T}\int_{\tau}^Tt_B^{2\nu-\gamma-1}\,(T-t_B)^{\gamma-1}\,\text{d}t_B\overset{T\gg\tau}{\sim}T^{2\nu-2}\tau^2,
\end{equation}
and for $\gamma>1$, we get
\begin{equation}
\label{eq:EATASD_contribution_2_2}
\sim\frac{\tau^2}{T}\int_{\tau}^Tt_B^{2\nu-\gamma-1}\,\text{d}t_B\overset{T\gg\tau}{\sim}T^{2\nu-\gamma-1}\tau^2.
\end{equation}
We can see that the contribution of the last incomplete flight of the generalized L\'evy walk to the EATASD leads to the quadratic time-lag dependencies
that were obtained rigorously in Eq.~(\ref{eq:LW_MSD_T_E_tau_1}) and Eq.~(\ref{eq:LW_MSD_T_E_tau_2}) as shown in the phase diagram for the EATASD in Fig.~\ref{fig:phase_diagrams} (b) of the main text.
Note that also the dependence on the total measurement time $T$ is reproduced correctly.

\subsection{Derivation of $p_z(s,u)$}

In the last part of this appendix, we derive the triple Laplace transform $p_z(s,u)$ of the distribution $p_T(t_f,t_B)$.
As an ansatz, we can write
\begin{equation}
\label{eq:p_T_tf_tB_1}
p_T(t_f,t_B)=\sum\limits_{n=1}^{\infty}p_T^n(t_f,t_B),
\end{equation}
where $p_T^n(t_f,t_B)$ is the probability that the instant of time $T$ belongs to the $n^{\text{th}}$ flight with flight duration $t_f$ and backward recurrence time $t_B$ with respect to $T$.
In order to find an analytical expression for this probability, we use methods of Godr\'eche and Luck \cite{godreche2001} similar to the derivation in the previous appendix.
Therefore, we have
\begin{widetext}
\begin{equation}
\label{eq:p^n_T_tf_tB}
p_T^n(t_f,t_B)=\left\langle\delta(t_f-T_n)\,\delta\left(t_B-\left(T-\sum\limits_{i=1}^{n-1}T_i\right)\right)I\left(\sum\limits_{i=1}^{n-1}T_i<T<\sum\limits_{i=1}^nT_i\right)\right\rangle_{\text{E}},
\end{equation}
where $I(\dots)$ is again the indicator.
The indicator function ensures that the instant of time $T$ belongs to the $n^{\text{th}}$ flight.
The double Laplace transform of Eq.~(\ref{eq:p^n_T_tf_tB}) with respect to $t_f$ and $t_B$ can easily be calculated due to the delta functions,
\begin{equation}
\label{eq:p^n_T_s_u}
p_T^n(t_f,t_B)\,\underset{t_B\rightarrow u}{\underset{t_f\rightarrow s}{\overset{\mathcal{L}^2}{\longrightarrow}}}\,p_T^n(s,u)=\left\langle e^{-sT_n}\,e^{-u\left(T-\sum\limits_{i=1}^{n-1}T_i\right)}I\left(\sum\limits_{i=1}^{n-1}T_i<T<\sum\limits_{i=1}^nT_i\right)\right\rangle_{\text{E}}.
\end{equation}
A further Laplace transform with respect to $T$ results in
\begin{equation}
\label{eq:p^n_z_s_u_1}
\begin{split}
p_T^n(s,u)\,\underset{T\rightarrow z}{\overset{\mathcal{L}}{\longrightarrow}}\,p_z^n(s,u)&=\left\langle e^{-sT_n}\,e^{u\sum\limits_{i=1}^{n-1}T_i}\frac{1}{u+z}\left(e^{-(u+z)\sum\limits_{i=1}^{n-1}T_i}-e^{-(u+z)\sum\limits_{i=1}^nT_i}\right)\right\rangle_{\text{E}}\\
&=\frac{1}{u+z}\left\langle e^{-sT_n-z\sum\limits_{i=1}^{n-1}T_i}-e^{-(s+u+z)T_n-z\sum\limits_{i=1}^{n-1}T_i}\right\rangle_{\text{E}}.
\end{split}
\end{equation}
The ensemble average in Eq.~(\ref{eq:p^n_z_s_u_1}) has to be performed according to Eq.~(\ref{eq:ensemble_average}) of Appendix \ref{sec:B}.
Using the definition of the Laplace transform $\psi(s)$ of the distribution $\psi(t)$ of flight durations $T_i$, we get
\begin{equation}
\label{eq:p^n_z_s_u_2}
p_z^n(s,u)=\frac{1}{u+z}\left[\psi(s)-\psi(s+u+z)\right]\psi^{n-1}(z).
\end{equation}
Inserting this result in the Laplace transform of Eq.~(\ref{eq:p_T_tf_tB_1}) and performing the sum using the geometric series, we obtain the triple Laplace transform of the distribution $p_T(t_f,t_B)$,
\begin{equation}
\label{eq:p_z_s_u_1}
p_z(s,u)=\sum\limits_{n=1}^{\infty}p_z^n(s,u)=\frac{1}{u+z}\,\frac{\psi(s)-\psi(s+u+z)}{1-\psi(z)}.
\end{equation}
\end{widetext}

\section{\label{sec:E}Derivation of the ergodicity breaking (EB) parameter}

The ergodicity breaking (EB) parameter is defined in Eq.~(\ref{eq:EB_tau}) of the main text as variance of the random variable $\widehat{\xi}(\tau)$ specified in Eq.~(\ref{eq:xi_tau}).
According to Eq.~(\ref{eq:xi_xi*}), the random variable $\widehat{\xi}(\tau)$ is equal in distribution to the random variable $\xi^*$ in the limit $\tau\rightarrow0$.
If the diffusion exponents obtained from all time-averaged squared displacements (TASDs) coincide, $\widehat{\xi}(\tau)$ does not depend on $\tau$, and the equality in distribution also holds for all values of $\tau$.
Because it is difficult to obtain an analytical expression for the distribution of the random variable $\xi^*$ for the generalized L\'evy walk,
we concentrate on finding an analytical expression for the variance of the random variable $\xi^*$, i.e., the EB parameter in the limit $\tau\rightarrow0$.
For the different regions of the two-dimensional parameter space, we will infer the distribution of the random variable $\widehat{\xi}(\tau)$ from the corresponding formula of the EB parameter.
The first moment of the random variable $\xi^*$ is equal to unity, and, therefore, we can write for the EB parameter
\begin{equation}
\label{eq:lim_tau_0_EB_tau}
\lim\limits_{\tau\rightarrow0}\text{EB}(\tau)=\frac{\left\langle\left[\int_0^T\mathbf{v}^2(t)\,\text{d}t\right]^2\right\rangle_{\text{E}}}{\left\langle\int_0^T\mathbf{v}^2(t)\,\text{d}t\right\rangle_{\text{E}}^2}-1.
\end{equation}
The time integral of the squared velocity in Eq.~(\ref{eq:lim_tau_0_EB_tau}) can be interpreted as a biased L\'evy walk in one spatial dimension with modified definitions of the distributions $\psi(x,t)$ and $W(x,t)$, namely
\begin{equation}
\label{eq:psi_x_t_}
\psi(x,t)=\delta(x-c^2t^{2\nu-1})\,\psi(t)
\end{equation}
and
\begin{equation}
\label{eq:W_x_t_}
W(x,t)=\int_t^{\infty}\delta(x-c^2t'^{2\nu-2}t)\,\psi(t')\,\text{d}t'.
\end{equation}
Then, the analytical treatment of the numerator and the denominator on the right-hand side of Eq.~(\ref{eq:lim_tau_0_EB_tau}) reduces to the calculation of the second and the first moment of the propagator of this biased L\'evy walk,
respectively.
However, this calculation is very extensive.
Instead, we here use the methods from Appendix \ref{sec:B} for the analytical treatment of the numerator and the denominator on the right-hand side of Eq.~(\ref{eq:lim_tau_0_EB_tau}).
The Laplace transform of the square root of the denominator is given by
\begin{equation}
\label{eq:int_0^T_v^2_t_E_LT_1}
\mathcal{L}\left\{\int_0^T\langle\mathbf{v}^2(t)\rangle_{\text{E}}\,\text{d}t;T,s\right\}=\frac{\mathcal{L}\left\{\langle\mathbf{v}^2(t)\rangle_{\text{E}};t,s\right\}}{s}.
\end{equation}
The Laplace transform on the right-hand side of Eq.~(\ref{eq:int_0^T_v^2_t_E_LT_1}) can easily be calculated from Eq.~(\ref{eq:v_v'_E_LT_3}),
\begin{equation}
\label{eq:v^2_t_E_LT}
\begin{split}
\mathcal{L}\left\{\langle\mathbf{v}^2(t)\rangle_{\text{E}};t,s\right\}&=\left.\mathcal{L}\left\{\lim\limits_{t\rightarrow0}\langle\mathbf{v}(t')\mathbf{v}(t'+t)\rangle_{\text{E}};t',u\right\}\right|_{u=s}\\[1ex]
&\hspace{-3em}=\left.\lim\limits_{s\rightarrow\infty}s\,\mathcal{L}^2\left\{\langle\mathbf{v}(t')\mathbf{v}(t'+t)\rangle_{\text{E}};t,s;t',u\right\}\right|_{u=s}\\[1ex]
&\hspace{-3em}=c^2\frac{\langle T^{2\nu-2}\rangle-\mathcal{L}\left\{t^{2\nu-2}\psi(t);t,s\right\}}{s(1-\psi(s))}.
\end{split}
\end{equation}
If we insert Eq.~(\ref{eq:v^2_t_E_LT}) into Eq.~(\ref{eq:int_0^T_v^2_t_E_LT_1}), we obtain
\begin{equation}
\label{eq:int_0^T_v^2_t_E_LT_2}
\begin{split}
&\mathcal{L}\left\{\int_0^T\langle\mathbf{v}^2(t)\rangle_{\text{E}}\,\text{d}t;T,s\right\}\\[1ex]
&\hspace{3em}=c^2\frac{\langle T^{2\nu-2}\rangle-\mathcal{L}\left\{t^{2\nu-2}\psi(t);t,s\right\}}{s^2(1-\psi(s))}.
\end{split}
\end{equation}
We now treat the numerator on the right-hand side of Eq.~(\ref{eq:lim_tau_0_EB_tau}),
\begin{equation}
\label{eq:int_0^T_v^2_t^2_E}
\left\langle\left[\int_0^T\mathbf{v}^2(t)\,\text{d}t\right]^2\right\rangle_{\text{E}}=\int_0^T\int_0^T\langle\mathbf{v}^2(t)\mathbf{v}^2(t')\rangle_{\text{E}}\,\text{d}t\,\text{d}t'.
\end{equation}
Because the Laplace transform of the right-hand side of Eq.~(\ref{eq:int_0^T_v^2_t^2_E}) with respect to $T$ is difficult to calculate,
we set the upper bounds of integration equal to $T_1$ and $T_2$ and calculate the corresponding double Laplace transform,
\begin{equation}
\label{eq:int_0^T_v^2_t^2_E_LT_1}
\begin{split}
&\mathcal{L}^2\left\{\int_0^{T_1}\int_0^{T_2}\langle\mathbf{v}^2(t)\mathbf{v}^2(t')\rangle_{\text{E}}\,\text{d}t\,\text{d}t';T_1,s;T_2,u\right\}\\[1ex]
&=\frac{\mathcal{L}^2\left\{\langle\mathbf{v}^2(t)\mathbf{v}^2(t')\rangle_{\text{E}};t,s;t',u\right\}}{su}.
\end{split}
\end{equation}
After an inverse double Laplace transform at the end of our derivation, we set $T_1$ and $T_2$ equal to $T$.
The analytical treatment of the Laplace transform of the correlation function of squared velocities in Eq.~(\ref{eq:int_0^T_v^2_t^2_E_LT_1})
is analogous to the one of the Laplace transform of the correlation function of nonsquared velocities in Appendix \ref{sec:B}.

We can write
\begin{equation}
\label{eq:v^2_v'^2_E}
\begin{split}
\langle\mathbf{v}^2(t)\mathbf{v}^2(t')\rangle_{\text{E}}&=\int_{\mathbb{R}^d}\int_{\mathbb{R}^d}\mathbf{v}^2\mathbf{v}'^2\,p(\mathbf{v},t;\mathbf{v}',t')\,\text{d}^d\mathbf{v}\,\text{d}^d\mathbf{v}'\\[1ex]
&\hspace{-5em}=\int_0^{\infty}\int_0^{\infty}c^2t_f^{2\nu-2}c^2t_f'^{2\nu-2}\,p(t_f,t;t_f',t')\,\text{d}t_f\,\text{d}t_f',
\end{split}
\end{equation}
where $p(\mathbf{v},t;\mathbf{v}',t')$ and $p(t_f,t;t_f',t')$ denote the probability densities of having a flight of velocity $\mathbf{v}$ / duration $t_f$ at time $t$
and a flight of velocity $\mathbf{v}'$ / duration $t_f'$ at time $t'$.
An appropriate ansatz for the latter is
\begin{equation}
\label{eq:p_tf_t_tf'_t'}
p(t_f,t;t_f',t')=\sum\limits_{n,m=1}^{\infty}p_{nm}(t_f,t;t_f',t'),
\end{equation}
where $p_{nm}(t_f,t;t_f',t')$ is the probability density of having the $n^{\text{th}}$ flight of duration $t_f$ at time $t$ and the $m^{\text{th}}$ flight of duration $t_f'$ at time $t'$.
Again, as in Appendix \ref{sec:B}, we use the methods introduced by Godr\`eche and Luck in \cite{godreche2001} in order to write down an equation for the probability density $p_{nm}(t_f,t;t_f',t')$,
\begin{equation}
\label{eq:p_nm_tf_t_tf'_t'}
\begin{split}
&p_{nm}(t_f,t;t_f',t')=\Biggl\langle\delta(t_f-T_n)\,\delta(t_f'-T_m)\,\times\\[1ex]
&\left.I\left(\sum\limits_{i=1}^{n-1}T_i<t<\sum\limits_{i=1}^nT_i\right)\,I\left(\sum\limits_{i=1}^{m-1}T_i<t'<\sum\limits_{i=1}^mT_i\right)\right\rangle_{\text{E}},
\end{split}
\end{equation}
where $I(\dots)$ are again the indicator functions, which account for the above-mentioned condition that the two instants of time $t$ and $t'$ belong to the $n^{\text{th}}$ and the $m^{\text{th}}$ flight, respectively.
$T_i$ denotes the duration of the $i^{\text{th}}$ flight, and the ensemble average, Eq.~(\ref{eq:ensemble_average}), takes all possible sequences of flight durations into account.
The Laplace transform of Eq.~(\ref{eq:v^2_v'^2_E}) combined with Eq.~(\ref{eq:p_tf_t_tf'_t'}) and Eq.~(\ref{eq:p_nm_tf_t_tf'_t'}) yields
\begin{widetext}
\begin{equation}
\label{eq:v^2_v'^2_E_LT}
\begin{split}
&\hspace{-3em}\mathcal{L}^2\left\{\langle\mathbf{v}^2(t)\mathbf{v}^2(t')\rangle_{\text{E}};t,s;t',u\right\}\\[2ex]
&=\int_0^{\infty}\int_0^{\infty}c^2t_f^{2\nu-2}c^2t_f'^{2\nu-2}\,\mathcal{L}^2\left\{p(t_f,t;t_f',t');t,s;t',u\right\}\,\text{d}t_f\,\text{d}t_f'\\[1ex]
&=\sum\limits_{n,m=1}^{\infty}\int_0^{\infty}\int_0^{\infty}c^2t_f^{2\nu-2}c^2t_f'^{2\nu-2}\,\mathcal{L}^2\left\{p_{nm}(t_f,t;t_f',t');t,s;t',u\right\}\,\text{d}t_f\,\text{d}t_f'\\[1ex]
&=\sum\limits_{n,m=1}^{\infty}\left\langle c^2T_n^{2\nu-2}c^2T_m^{2\nu-2}\,\frac{1}{s}\left(e^{-s\sum\limits_{i=1}^{n-1}T_i}-e^{-s\sum\limits_{i=1}^nT_i}\right)\frac{1}{u}\left(e^{-u\sum\limits_{i=1}^{m-1}T_i}-e^{-u\sum\limits_{i=1}^mT_i}\right)\right\rangle_{\text{E}}.
\end{split}
\end{equation}
Inserting Eq.~(\ref{eq:v^2_v'^2_E_LT}) into Eq.~(\ref{eq:int_0^T_v^2_t^2_E_LT_1}) leads to
\begin{equation}
\label{eq:int_0^T_v^2_t^2_E_LT_2}
\begin{split}
&\hspace{-3em}\mathcal{L}^2\left\{\int_0^{T_1}\int_0^{T_2}\langle\mathbf{v}^2(t)\mathbf{v}^2(t')\rangle_{\text{E}}\,\text{d}t\,\text{d}t';T_1,s;T_2,u\right\}\\[1ex]
&=\frac{c^4}{s^2u^2}\sum\limits_{n,m=1}^{\infty}\left\langle T_n^{2\nu-2}T_m^{2\nu-2}\left(e^{-s\sum\limits_{i=1}^{n-1}T_i}-e^{-s\sum\limits_{i=1}^nT_i}\right)\left(e^{-u\sum\limits_{i=1}^{m-1}T_i}-e^{-u\sum\limits_{i=1}^mT_i}\right)\right\rangle_{\text{E}}.
\end{split}
\end{equation}
The double sum on the right-hand side of Eq.~(\ref{eq:int_0^T_v^2_t^2_E_LT_2}) can be split into three terms,
\begin{equation}
\label{eq:double_sum_1}
\sum\limits_{n,m=1}^{\infty}=\sum\limits_{n=1}^{\infty}\sum\limits_{m=n}^n+\sum\limits_{m=1}^{\infty}\sum\limits_{n=m+1}^{\infty}+\sum\limits_{n=1}^{\infty}\sum\limits_{m=n+1}^{\infty},
\end{equation}
where for the first one $n=m$, for the second one $n>m$, and for the third one $n<m$.
Evaluating the ensemble average, Eq.~(\ref{eq:ensemble_average}), in Eq.~(\ref{eq:int_0^T_v^2_t^2_E_LT_2}) with respect to the first double sum on the right-hand side of Eq.~(\ref{eq:double_sum_1}) results in
\begin{equation}
\label{eq:n=m}
\begin{split}
&\sum\limits_{n=1}^{\infty}\left\langle T_n^{4\nu-4}\left(e^{-(s+u)\sum\limits_{i=1}^{n-1}T_i}+e^{-(s+u)\sum\limits_{i=1}^nT_i}-e^{-(s+u)\sum\limits_{i=1}^{n-1}T_i-uT_n}-e^{-(s+u)\sum\limits_{i=1}^{n-1}T_i-sT_n}\right)\right\rangle_{\text{E}}\\[1ex]
&=\frac{\langle T^{4\nu-4}\rangle+\mathcal{L}\left\{t^{4\nu-4}\psi(t)\right\}(s+u)-\mathcal{L}\left\{t^{4\nu-4}\psi(t)\right\}(u)-\mathcal{L}\left\{t^{4\nu-4}\psi(t)\right\}(s)}{1-\psi(s+u)}.
\end{split}
\end{equation}
Correspondingly, with respect to the second double sum for which $n>m$, we get
\begin{equation}
\label{eq:n>m}
\begin{split}
&\sum\limits_{m=1}^{\infty}\sum\limits_{n=m+1}^{\infty}\left\langle T_n^{2\nu-2}T_m^{2\nu-2}\left(e^{-s\sum\limits_{i=1}^{n-1}T_i}-e^{-s\sum\limits_{i=1}^nT_i}\right)\left(e^{-u\sum\limits_{i=1}^{m-1}T_i}-e^{-u\sum\limits_{i=1}^mT_i}\right)\right\rangle_{\text{E}}\\[1ex]
&=\sum\limits_{m=1}^{\infty}\sum\limits_{n=m+1}^{\infty}\psi^{m-1}(s+u)\,\psi^{n-m-1}(s)\left[\mathcal{L}\left\{t^{2\nu-2}\psi(t)\right\}(s)\,\langle T^{2\nu-2}\rangle\right.\\[1ex]
&\quad+\mathcal{L}\left\{t^{2\nu-2}\psi(t)\right\}(s+u)\,\mathcal{L}\left\{t^{2\nu-2}\psi(t)\right\}(s)-\mathcal{L}\left\{t^{2\nu-2}\psi(t)\right\}(s+u)\,\langle T^{2\nu-2}\rangle\\[1ex]
&\quad\left.-\mathcal{L}\left\{t^{2\nu-2}\psi(t)\right\}(s)\,\mathcal{L}\left\{t^{2\nu-2}\psi(t)\right\}(s)\right],
\end{split}
\end{equation}
where the double sum on the right-hand side of Eq.~(\ref{eq:n>m}) can easily be calculated by using the geometric series,
\begin{equation}
\label{eq:double_sum_2}
\sum\limits_{m=1}^{\infty}\sum\limits_{n=m+1}^{\infty}\psi^{m-1}(s+u)\,\psi^{n-m-1}(s)=\frac{1}{(1-\psi(s))(1-\psi(s+u))}.
\end{equation}
The third double sum for which $n<m$ gives the same result as in Eq.~(\ref{eq:n>m}) with $s$ and $u$ interchanged.
By combining Eq.~(\ref{eq:n=m}), Eq.~(\ref{eq:n>m}), and Eq.~(\ref{eq:double_sum_2}), we obtain our final result
\begin{equation}
\label{eq:int_0^T_v^2_t^2_E_LT_3}
\begin{split}
&\hspace{-3em}\mathcal{L}^2\left\{\int_0^{T_1}\int_0^{T_2}\langle\mathbf{v}^2(t)\mathbf{v}^2(t')\rangle_{\text{E}}\,\text{d}t\,\text{d}t';T_1,s;T_2,u\right\}\\[1ex]
&=\frac{c^4}{s^2u^2}\left[\frac{\langle T^{4\nu-4}\rangle+\mathcal{L}\left\{t^{4\nu-4}\psi(t)\right\}(s+u)-\mathcal{L}\left\{t^{4\nu-4}\psi(t)\right\}(s)-\mathcal{L}\left\{t^{4\nu-4}\psi(t)\right\}(u)}{1-\psi(s+u)}\right.\\[1ex]
&\quad+\frac{\left[\mathcal{L}\left\{t^{2\nu-2}\psi(t)\right\}(s+u)-\mathcal{L}\left\{t^{2\nu-2}\psi(t)\right\}(s)\right]\left[\mathcal{L}\left\{t^{2\nu-2}\psi(t)\right\}(s)-\langle T^{2\nu-2}\rangle\right]}{(1-\psi(s))(1-\psi(s+u))}\\[1ex]
&\quad\left.+\frac{\left[\mathcal{L}\left\{t^{2\nu-2}\psi(t)\right\}(s+u)-\mathcal{L}\left\{t^{2\nu-2}\psi(t)\right\}(u)\right]\left[\mathcal{L}\left\{t^{2\nu-2}\psi(t)\right\}(u)-\langle T^{2\nu-2}\rangle\right]}{(1-\psi(u))(1-\psi(s+u))}\right].
\end{split}
\end{equation}
\end{widetext}
The fractional moment appearing on the right-hand side of Eq.(\ref{eq:int_0^T_v^2_t^2_E_LT_3}) can be calculated by using the definition of the beta function and its connection to the gamma function,
\begin{equation}
\label{eq:T^4v-4_E}
\begin{split}
&\langle T^{4\nu-4}\rangle=\int_0^{\infty}t^{4\nu-4}\,\psi(t)\,\text{d}t\\[1ex]
&=\frac{\Gamma(4\nu-3)\Gamma(\gamma+4-4\nu)}{\Gamma(\gamma)}t_0^{4\nu-4}\quad\text{if}\quad\frac{3}{2}<2\nu<\frac{\gamma}{2}+2.
\end{split}
\end{equation}
The small-$s$ behavior of the Laplace transform $\mathcal{L}\left\{t^{4\nu-4}\psi(t);t,s\right\}$ can be obtained by using the Cauchy-Saalsch\"utz representation of the gamma function,
\begin{equation}
\label{eq:t^4v-4_psi_t_LT}
\begin{split}
&\mathcal{L}\left\{t^{4\nu-4}\psi(t);t,s\right\}\overset{s\rightarrow0}{\simeq}\langle T^{4\nu-4}\rangle-\langle T^{4\nu-3}\rangle s+\frac{1}{2}\langle T^{4\nu-2}\rangle s^2\\[1ex]
&\hspace{11em}+\gamma\Gamma(4\nu-\gamma-4)t_0^{\gamma}s^{\gamma+4-4\nu},\\[2ex]
&\langle T^{4\nu-3}\rangle=\frac{\Gamma(4\nu-2)\Gamma(\gamma+3-4\nu)}{\Gamma(\gamma)}t_0^{4\nu-3},\\[1ex]
&\langle T^{4\nu-2}\rangle=\frac{\Gamma(4\nu-1)\Gamma(\gamma+2-4\nu)}{\Gamma(\gamma)}t_0^{4\nu-2}.
\end{split}
\end{equation}
The fractional moment in Eq.~(\ref{eq:T^4v-4_E}) diverges for $2\nu\geq\gamma/2+2$.
Therefore, also the EB parameter diverges under the same condition,
\begin{equation}
\label{eq:LW_EB_tau_divergence}
\text{EB}(\tau)=\infty\quad\text{if}\quad2\nu\geq\frac{\gamma}{2}+2.
\end{equation}
The fractional moment in Eq.~(\ref{eq:T^4v-4_E}) also diverges for $2\nu\leq3/2$, but, similar to the discussion in Appendix \ref{sec:B},
this divergence is compensated by the Laplace transform $\mathcal{L}\left\{t^{4\nu-4}\psi(t);t,s\right\}=\int_0^{\infty}t^{4\nu-4}\,\psi(t)\,e^{-st}\,\text{d}t$,
which contains exactly the same divergence at the lower bound of integration.
These divergences cancel out in Eq.~(\ref{eq:int_0^T_v^2_t^2_E_LT_3}).
Therefore, for $2\nu<\gamma/2+2$, the EB parameter is finite.
The small-$s$ and the small-$u$ behavior of its Laplace transform can be obtained by inserting Eq.~(\ref{eq:T^4v-4_E}) and Eq.~(\ref{eq:t^4v-4_psi_t_LT}) into Eq.~(\ref{eq:int_0^T_v^2_t^2_E_LT_3})
and by inserting Eq.~(\ref{eq:T^2v-2_E}) and Eq.~(\ref{eq:t^2v-2_psi_t_LT}) into Eq.~(\ref{eq:int_0^T_v^2_t_E_LT_2}) and Eq.~(\ref{eq:int_0^T_v^2_t^2_E_LT_3}) and by combining the results with Eq.~(\ref{eq:lim_tau_0_EB_tau}).
Note that the Laplace variables $s$ and $u$ in Eq.~(\ref{eq:int_0^T_v^2_t^2_E_LT_3}) belong to the variables $T_1$ and $T_2$ in the time domain, where $T_1$ and $T_2$ have to be set equal to $T$.
Remember that we introduced these two variables $T_1$ and $T_2$ because it was too difficult to find the Laplace transform of the right-hand side of Eq.~(\ref{eq:int_0^T_v^2_t^2_E}) with respect to $T$
(see also Eq.~(\ref{eq:int_0^T_v^2_t^2_E_LT_1})).
Furthermore, for the inverse double Laplace transform of Eq.~(\ref{eq:int_0^T_v^2_t^2_E_LT_3}), we need the following relation
\begin{equation}
\label{eq:inverse_double_Laplace_transform}
\begin{split}
&\mathcal{L}^{-2}\left\{s^{-\alpha}u^{-\beta}(s+u)^{-\gamma};s,T;u,T\right\}\\[1ex]
&=\frac{\Gamma(\alpha+\beta-1)}{\Gamma(\alpha)\Gamma(\beta)\Gamma(\alpha+\beta+\gamma-1)}T^{\alpha+\beta+\gamma-2},
\end{split}
\end{equation}
which can be obtained by using the convolution theorem of the Laplace transform.
In the following, we list our analytical results for the EB parameter.
To do so, we have to distinguish different cases:

For $0<\gamma<1$ and $0<2\nu<\gamma/2+1$, we obtain
\begin{widetext}
\begin{equation}
\label{eq:EB_s_u_1}
\begin{split}
\mathcal{L}\left\{\left\langle\int_0^T\mathbf{v}^2(t)\,\text{d}t\right\rangle_{\text{E}};T,s\right\}&\simeq c^2\frac{\Gamma(2\nu)\Gamma(\gamma+1-2\nu)}{\Gamma(1-\gamma)\Gamma(\gamma)}t_0^{2\nu-\gamma-1}s^{-\gamma-1}\quad(s\rightarrow0),\\[1ex]
\left\langle\int_0^T\mathbf{v}^2(t)\,\text{d}t\right\rangle_{\text{E}}&\simeq c^2\frac{\Gamma(2\nu)\Gamma(\gamma+1-2\nu)}{\Gamma(1-\gamma)\Gamma(\gamma)\Gamma(\gamma+1)}t_0^{2\nu-\gamma-1}T^{\gamma}\quad(T\rightarrow\infty),\\[1ex]
\mathcal{L}\left\{\left\langle\left[\int_0^T\mathbf{v}^2(t)\,\text{d}t\right]^2\right\rangle_{\text{E}};T,s,u\right\}&\simeq\frac{c^4}{s^2u^2}\left[\frac{\Gamma(4\nu-1)\Gamma(\gamma+2-4\nu)}{\Gamma(1-\gamma)\Gamma(\gamma)}t_0^{4\nu-\gamma-2}su(s+u)^{-\gamma}\right.\\[1ex]
&\quad\left.+2\frac{\Gamma^2(2\nu)\Gamma^2(\gamma+1-2\nu)}{\Gamma^2(1-\gamma)\Gamma^2(\gamma)}t_0^{4\nu-2\gamma-2}s^{1-\gamma}u(s+u)^{-\gamma}\right],\\[1ex]
\left\langle\left[\int_0^T\mathbf{v}^2(t)\,\text{d}t\right]^2\right\rangle_{\text{E}}&\simeq2c^4\frac{\Gamma^2(2\nu)\Gamma^2(\gamma+1-2\nu)}{\Gamma^2(1-\gamma)\Gamma^2(\gamma)\Gamma(2\gamma+1)}t_0^{4\nu-2\gamma-2}T^{2\gamma}\quad(T\rightarrow\infty),
\end{split}
\end{equation}
which leads to
\begin{equation}
\label{eq:EB_1}
\lim\limits_{T\rightarrow\infty}\text{EB}=2\frac{\Gamma^2(\gamma+1)}{\Gamma(2\gamma+1)}-1.
\end{equation}
The case $0<\gamma<1$ and $\gamma/2+1<2\nu<\gamma+1$ leads to the same asymptotic result.
This is the result for region $\text{M}_{\gamma}$ in Fig.~\ref{fig:phase_diagrams} (c).

For $0<\gamma<1$ and $\gamma+1<2\nu<\gamma/2+2$, i.e., for region $\text{A}_{\gamma,\nu}$ in Fig.~\ref{fig:phase_diagrams} (c), we get
\begin{equation}
\label{eq:EB_s_u_2}
\begin{split}
\mathcal{L}\left\{\left\langle\int_0^T\mathbf{v}^2(t)\,\text{d}t\right\rangle_{\text{E}};T,s\right\}&\simeq-c^2\gamma\frac{\Gamma(2\nu-\gamma-2)}{\Gamma(1-\gamma)}s^{-2\nu}\quad(s\rightarrow0),\\[1ex]
\left\langle\int_0^T\mathbf{v}^2(t)\,\text{d}t\right\rangle_{\text{E}}&\simeq-c^2\gamma\frac{\Gamma(2\nu-\gamma-2)}{\Gamma(1-\gamma)\Gamma(2\nu)}T^{2\nu-1}\quad(T\rightarrow\infty),\\[1ex]
\mathcal{L}\left\{\left\langle\left[\int_0^T\mathbf{v}^2(t)\,\text{d}t\right]^2\right\rangle_{\text{E}};T,s,u\right\}&\simeq\frac{c^4}{s^2u^2}\left[\gamma\frac{\Gamma(4\nu-\gamma-4)}{\Gamma(1-\gamma)}\frac{(s+u)^{\gamma+4-4\nu}-s^{\gamma+4-4\nu}-u^{\gamma+4-4\nu}}{(s+u)^{\gamma}}\right.\\[1ex]
&\quad\left.+2\gamma^2\frac{\Gamma^2(2\nu-\gamma-2)}{\Gamma^2(1-\gamma)}\frac{\left((s+u)^{\gamma+2-2\nu}-s^{\gamma+2-2\nu}\right)s^{\gamma+2-2\nu}}{s^{\gamma}(s+u)^{\gamma}}\right],\\[1ex]
\left\langle\left[\int_0^T\mathbf{v}^2(t)\,\text{d}t\right]^2\right\rangle_{\text{E}}&\simeq2c^4\left(\gamma^2(\gamma+2-2\nu)\frac{\Gamma^2(2\nu-\gamma-2)}{\Gamma^2(1-\gamma)\Gamma(4\nu-1)}\right.\\[1ex]
&\quad\left.+\gamma(\gamma+3-4\nu)\frac{\Gamma(4\nu-\gamma-4)}{\Gamma(1-\gamma)\Gamma(4\nu-1)}\right)T^{4\nu-2}\quad(T\rightarrow\infty),
\end{split}
\end{equation}
which results in
\begin{equation}
\label{eq:EB_2}
\lim\limits_{T\rightarrow\infty}\text{EB}=2\frac{\Gamma^2(2\nu)}{\Gamma(4\nu-1)}(\gamma+2-2\nu)\left(1+\frac{\gamma+2-2\nu}{\gamma+4-4\nu}\frac{\Gamma(1-\gamma)\Gamma(4\nu-\gamma-2)}{\gamma\,\Gamma^2(2\nu-\gamma-1)}\right)-1.
\end{equation}

Furthermore, for $\gamma>1$ and $0<2\nu<\gamma/2+1$, we obtain
\begin{equation}
\label{eq:EB_s_u_3}
\begin{split}
\mathcal{L}\left\{\left\langle\int_0^T\mathbf{v}^2(t)\,\text{d}t\right\rangle_{\text{E}};T,s\right\}&\simeq c^2\frac{\Gamma(2\nu)\Gamma(\gamma+1-2\nu)}{\Gamma(\gamma-1)}t_0^{2\nu-2}s^{-2}\quad(s\rightarrow0),\\[1ex]
\left\langle\int_0^T\mathbf{v}^2(t)\,\text{d}t\right\rangle_{\text{E}}&\simeq c^2\frac{\Gamma(2\nu)\Gamma(\gamma+1-2\nu)}{\Gamma(\gamma-1)}t_0^{2\nu-2}T\quad(T\rightarrow\infty),\\[1ex]
\mathcal{L}\left\{\left\langle\left[\int_0^T\mathbf{v}^2(t)\,\text{d}t\right]^2\right\rangle_{\text{E}};T,s,u\right\}&\simeq\frac{c^4}{s^2u^2}\left[\frac{\Gamma(4\nu-1)\Gamma(\gamma+2-4\nu)}{\Gamma(\gamma-1)}t_0^{4\nu-3}su(s+u)^{-1}\right.\\[1ex]
&\quad\left.+2\frac{\Gamma^2(2\nu)\Gamma^2(\gamma+1-2\nu)}{\Gamma^2(\gamma-1)}t_0^{4\nu-4}u(s+u)^{-1}\right],\\[1ex]
\left\langle\left[\int_0^T\mathbf{v}^2(t)\,\text{d}t\right]^2\right\rangle_{\text{E}}&\simeq c^4\frac{\Gamma^2(2\nu)\Gamma^2(\gamma+1-2\nu)}{\Gamma^2(\gamma-1)}t_0^{4\nu-4}T^2\quad(T\rightarrow\infty),
\end{split}
\end{equation}
which yields
\begin{equation}
\label{eq:EB_3}
\lim\limits_{T\rightarrow\infty}\text{EB}=0
\end{equation}
corresponding to ergodicity in the lower part of region $\text{D}$ of the phase diagram in Fig.~\ref{fig:phase_diagrams} (c).
Additionally, for $\gamma>1$ and $\gamma/2+1<2\nu<\gamma+1$, we get
\begin{equation}
\label{eq:EB_s_u_4}
\begin{split}
\mathcal{L}\left\{\left\langle\int_0^T\mathbf{v}^2(t)\,\text{d}t\right\rangle_{\text{E}};T,s\right\}&\simeq c^2\frac{\Gamma(2\nu)\Gamma(\gamma+1-2\nu)}{\Gamma(\gamma-1)}t_0^{2\nu-2}s^{-2}\quad(s\rightarrow0),\\[1ex]
\left\langle\int_0^T\mathbf{v}^2(t)\,\text{d}t\right\rangle_{\text{E}}&\simeq c^2\frac{\Gamma(2\nu)\Gamma(\gamma+1-2\nu)}{\Gamma(\gamma-1)}t_0^{2\nu-2}T\quad(T\rightarrow\infty),\\[1ex]
\mathcal{L}\left\{\left\langle\left[\int_0^T\mathbf{v}^2(t)\,\text{d}t\right]^2\right\rangle_{\text{E}};T,s,u\right\}&\simeq\frac{c^4}{s^2u^2}\left[\frac{\gamma(\gamma-1)t_0^{\gamma-1}}{\Gamma^{-1}(4\nu-\gamma-4)}\frac{(s+u)^{\gamma+4-4\nu}-s^{\gamma+4-4\nu}-u^{\gamma+4-4\nu}}{s+u}\right.\\[1ex]
&\quad\left.+2\frac{\Gamma^2(2\nu)\Gamma^2(\gamma+1-2\nu)}{\Gamma^2(\gamma-1)}t_0^{4\nu-4}u(s+u)^{-1}\right],\\[1ex]
\left\langle\left[\int_0^T\mathbf{v}^2(t)\,\text{d}t\right]^2\right\rangle_{\text{E}}&\simeq2c^4\gamma(\gamma-1)(\gamma+3-4\nu)\frac{\Gamma(4\nu-\gamma-4)}{\Gamma(4\nu-\gamma)}t_0^{\gamma-1}T^{4\nu-\gamma-1}\\[1ex]
&\quad+c^4\frac{\Gamma^2(2\nu)\Gamma^2(\gamma+1-2\nu)}{\Gamma^2(\gamma-1)}t_0^{4\nu-4}T^2\quad(T\rightarrow\infty),
\end{split}
\end{equation}
which leads to
\begin{equation}
\label{eq:EB_4}
\text{EB}\simeq
\begin{cases}
0,\quad&\frac{\gamma}{2}+1<2\nu<\frac{\gamma}{2}+\frac{3}{2}\\[1ex]
\frac{2}{\gamma+4-4\nu}\frac{\Gamma(\gamma-1)\Gamma(\gamma+1)\Gamma(4\nu-\gamma-2)}{\Gamma^2(2\nu)\Gamma^2(\gamma+1-2\nu)\Gamma(4\nu-\gamma)}t_0^{\gamma+3-4\nu}T^{4\nu-\gamma-3},\quad&\frac{\gamma}{2}+\frac{3}{2}<2\nu<\gamma+1
\end{cases}
\quad(T\rightarrow\infty),
\end{equation}
where the upper line corresponds to ergodicity in the upper part of region $\text{D}$, and the lower line correponds to the behavior in region $\text{B}_{\gamma,\nu}$ in Fig.~\ref{fig:phase_diagrams} (c).
Note, however, that the last line in Eq.~(\ref{eq:EB_4}) is valid only below the line of EB-divergence $2\nu=\gamma/2+2$ given by Eq.~(\ref{eq:LW_EB_tau_divergence}).

So far, the identification of the dominant term in Eq.~(\ref{eq:int_0^T_v^2_t^2_E_LT_3}) revealed that the EB parameter asymptotically goes to zero in sector $D$.
If one is interested in the details of this transition to ergodicity, i.e., the dependence of the EB parameter on the total measurement time $T$, one has to incorporate additional terms in the asymptotic analysis.
Because of the complicated structure of Eq.~(\ref{eq:int_0^T_v^2_t^2_E_LT_3}), this leads to additional cases that have to be distinguished.
In Fig.~\ref{fig:sector_D}, we summarize the dependence of the EB parameter on the total measurement time $T$ for sector $D$.

\begin{figure}
\includegraphics[width=\linewidth]{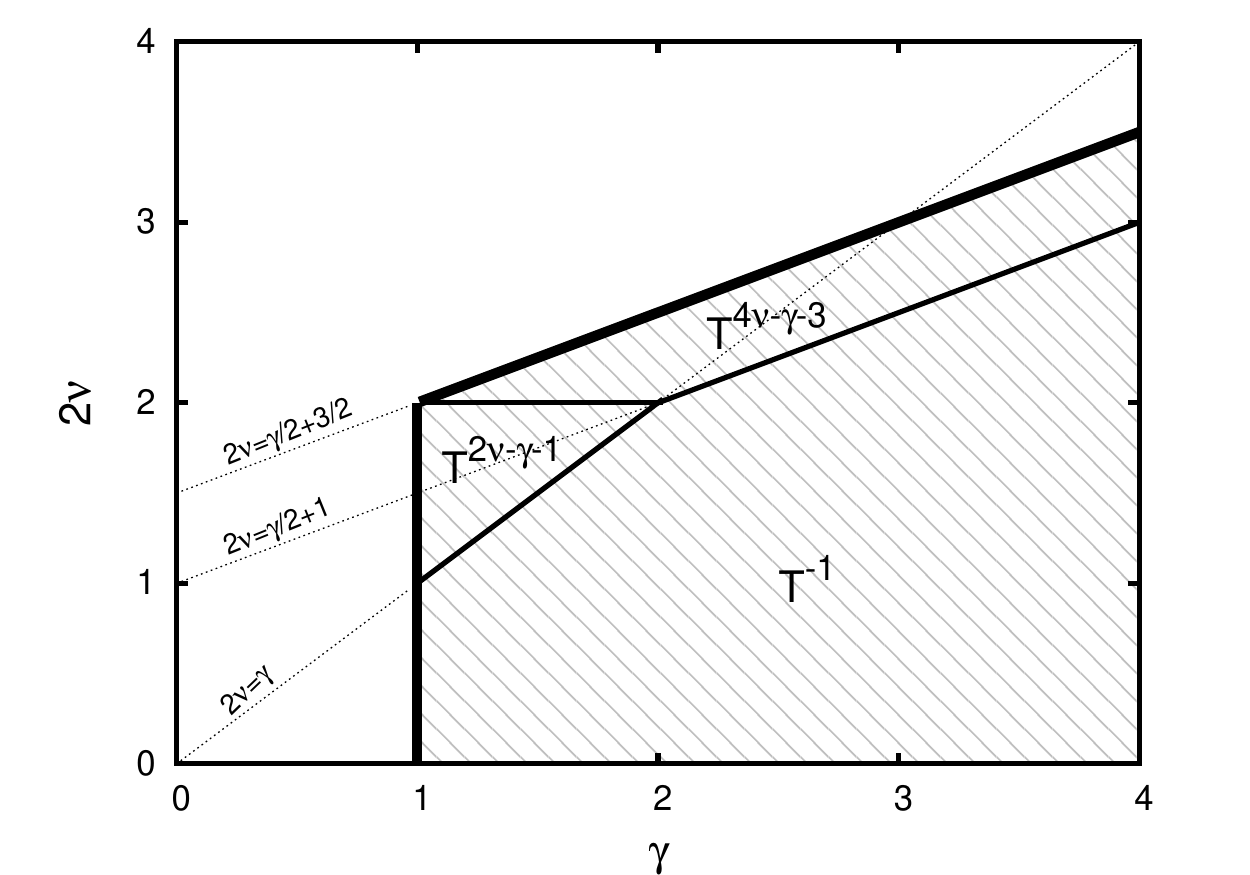}
\caption{\label{fig:sector_D}
Enlargement of the ergodic sector $D$ of the phase diagram of the EB parameter shown in Fig.~\ref{fig:phase_diagrams} (c) in the main text.
This sector, where the EB parameter asymptotically goes to zero as the measurement time $T$ goes to infinity, is represented by the grey shaded region bounded by thick black lines.
The thin black lines divide sector $D$ in three regions with different power-law dependencies of the EB parameter on the measurement time $T$.
Note that all exponents in this sector are negative.
The thin dotted lines serve as a guide to the eye.}
\end{figure}

Finally, for $\gamma>1$ and $\gamma+1<2\nu<\gamma/2+2$, i.e., for region $\text{C}_{\gamma,\nu}$ in Fig.~\ref{fig:phase_diagrams} (c), we obtain
\begin{equation}
\label{eq:EB_s_u_5}
\begin{split}
\mathcal{L}\left\{\left\langle\int_0^T\mathbf{v}^2(t)\,\text{d}t\right\rangle_{\text{E}};T,s\right\}&\simeq-c^2\gamma(\gamma-1)\Gamma(2\nu-\gamma-2)t_0^{\gamma-1}s^{\gamma-2\nu-1}\quad(s\rightarrow0),\\[1ex]
\left\langle\int_0^T\mathbf{v}^2(t)\,\text{d}t\right\rangle_{\text{E}}&\simeq-c^2\gamma(\gamma-1)\frac{\Gamma(2\nu-\gamma-2)}{\Gamma(2\nu+1-\gamma)}t_0^{\gamma-1}T^{2\nu-\gamma}\quad(T\rightarrow\infty),\\[1ex]
\mathcal{L}\left\{\left\langle\left[\int_0^T\mathbf{v}^2(t)\,\text{d}t\right]^2\right\rangle_{\text{E}};T,s,u\right\}&\simeq\frac{c^4}{s^2u^2}\left[\frac{\gamma(\gamma-1)t_0^{\gamma-1}}{\Gamma^{-1}(4\nu-\gamma-4)}\frac{(s+u)^{\gamma+4-4\nu}-s^{\gamma+4-4\nu}-u^{\gamma+4-4\nu}}{s+u}\right.\\[1ex]
&\quad\left.+\frac{2\gamma^2(\gamma-1)^2t_0^{2\gamma-2}}{\Gamma^{-2}(2\nu-\gamma-2)}\frac{\left((s+u)^{\gamma+2-2\nu}-s^{\gamma+2-2\nu}\right)s^{\gamma+2-2\nu}}{s(s+u)}\right],\\[1ex]
\left\langle\left[\int_0^T\mathbf{v}^2(t)\,\text{d}t\right]^2\right\rangle_{\text{E}}&\simeq2c^4\gamma(\gamma-1)(\gamma+3-4\nu)\frac{\Gamma(4\nu-\gamma-4)}{\Gamma(4\nu-\gamma)}t_0^{\gamma-1}T^{4\nu-\gamma-1},
\end{split}
\end{equation}
which results in
\begin{equation}
\label{eq:EB_5}
\text{EB}\simeq\frac{2(\gamma+3-4\nu)}{\gamma(\gamma-1)}\frac{\Gamma(4\nu-\gamma-4)\Gamma^2(2\nu+1-\gamma)}{\Gamma(4\nu-\gamma)\Gamma^2(2\nu-\gamma-2)}t_0^{1-\gamma}T^{\gamma-1}\quad(T\rightarrow\infty).
\end{equation}
\end{widetext}

Note that in sectors $\text{B}_{\gamma,\nu}$ and $\text{C}_{\gamma,\nu}$ in Fig.~\ref{fig:phase_diagrams} (c), the EB parameter is finite for every measurement time $T$,
whereas the EB parameter diverges for every finite $T$ in sector $\infty$ for $2\nu\geq\gamma/2+2$.\\[2ex]

\begin{acknowledgments}
The authors gratefully acknowledge funding by the Deutsche Forschungsgemeinschaft (DFG, German Research Foundation) - 438881351.
\end{acknowledgments}

\bibliography{references}

\end{document}